\pdfoutput=1
\documentclass[12pt,a4paper]{article}

\usepackage[utf8]{inputenc}    % Encoding
\usepackage[T1]{fontenc}       % Font encoding
\usepackage{lmodern}           % Latin Modern font
\usepackage{amsmath,amssymb}   % Math symbols
\usepackage{graphicx}          % Figures
\usepackage{cite}              % Better citation handling
\usepackage{hyperref}          % Clickable refs
\usepackage{geometry}          % Page layout
\usepackage{authblk}           % Authors & affiliations
\usepackage{mathrsfs}
\usepackage{float}

\title{\textbf{Holographic Tessellations Tensor Networks $\{p,q\}$ from Entangled Quantum Tetrahedron in Loop Quantum Gravity}}

\author[1]{Deden M. Akbar\thanks{Corresponding author: \texttt{deden442@guru.sma.belajar.id}}}
\author[1,2]{Freddy P. Zen\thanks{Email: \texttt{fpzen@fi.itb.ac.id}}}
\author[1,2]{Jusak S. Kosasih\thanks{Email: \texttt{jusak@fi.itb.ac.id}}}

\affil[1]{\textit{Theoretical High Energy Physics Group, Department of Physics, Institut Teknologi Bandung, Jl. Ganesha 10, Bandung 40132, Indonesia}}
\affil[2]{\textit{Indonesia Center for Theoretical and Mathematical Physics (ICTMP), Institut Teknologi Bandung, Jl. Ganesha 10, Bandung 40132, Indonesia}}

\date{} % leave empty for arXiv

\begin{document}

\maketitle

\begin{abstract}
The relationship between spin network states in loop quantum gravity and bulk–boundary duality is further explored using the tessellated holographic tensor network. We develop a holographic tensor network construction based on quantum entanglement between tetrahedra in loop quantum gravity (LQG). This construction relates four-valent intertwiner spaces and Bell states dressed with $SU(2)$ holonomies to effective tensors on hyperbolic $\{p,q\}$ tessellations. For the $\{5,4\}$ tessellation, six tetrahedra are assembled into a single effective cell with five network legs and one logical index. Provided that the microscopic transfer map has full row rank, a right inverse and a completion on the kernel space are employed to construct an isometric embedding, yielding an effective tensor proportional to a perfect tensor. The construction is generalized to the $p=2k-1$ family, with local dimensions supporting absolutely maximally entangled states. On finite networks satisfying a radial orientation condition, an appropriate normalization produces an isometric bulk--boundary map. For cuts satisfying the isometry condition on both sides, the boundary entropy decomposes into a link contribution $\sum_{\ell}\log(2j_{\ell}+1)$ and the bulk entropy, thereby providing a discrete realization of the quantum Ryu--Takayanagi structure. Applications to the black hole interior are formulated via postselection-based non-isometric encoding. Assuming additional Haar statistics, this framework is used to analyze the preservation of overlaps on average and the competition between entropy branches underlying the Page Curve. These results establish a constructive link between the geometric degrees of freedom of LQG and holographic codes, while clarifying the conditions required to connect them to geometric area in LQG, the location of the quantum extremal surface, and evaporation of black hole.
\end{abstract}

\section{Introduction}
%Start writing your introduction here. Reference papers like this~\cite{Einstein1916}.

The holographic principle has become one of the important ideas in efforts to understand the relationship among quantum gravity, quantum field theory, and quantum information. In general, this principle states that the physical information associated with a gravitational region can be represented by degrees of freedom defined on the boundary of that region. The most concrete realization of this idea is provided by the anti-de Sitter/conformal field theory (AdS/CFT) correspondence, which relates a gravitational theory in an asymptotically AdS spacetime of dimension $d+1$ to a nongravitational $d$-dimensional conformal field theory on the boundary. The consistency of its symmetry structure is reflected in the equivalence between the isometry group of AdS and the conformal group of the CFT, namely $SO(d,2)$. In a broader sense, holographic ideas have also been developed beyond cases with exact conformal symmetry \cite{ref1,ref2,ref3,ref4,ref5}.

One of the most fundamental results in the holographic correspondence is the relation between bulk geometry and the entanglement structure on the boundary. This relation is formulated quantitatively through holographic entanglement entropy and is crystallized in the Ryu--Takayanagi (RT) formula. For a region $A$ on the boundary, the formula relates the entanglement entropy to the area of a minimal surface $A_{\min}$ in the bulk anchored on the boundary of $A$ \cite{ref6,ref7,ref8}
\begin{equation}
S(A)=\frac{\mathrm{Area}(A_{\min})}{4G_N},
\label{eq:RT}
\end{equation}
where $G_N$ is Newton's gravitational constant. This formula provides a direct relation between an information-theoretic quantity on the boundary and a geometric quantity in the bulk. In a gravitational context, the same area-law structure is also closely related to the Bekenstein--Hawking entropy. Because of its universality in a wide class of holographic configurations, the relation between entanglement and geometric area has become one of the main signatures of the emergence of semiclassical geometry from quantum gravity \cite{ref6,ref7,ref8}.

Nevertheless, realizing the structure of AdS/CFT explicitly in the language of quantum information is not straightforward. Continuum field theories generally possess infinitely many degrees of freedom, whereas quantum-information formulations are typically more transparent in finite-dimensional Hilbert spaces. At the same time, one of the characteristic features of holographic duality is bulk reconstruction, namely the ability to reconstruct bulk information or bulk operators from data available on the boundary. The redundancy of this reconstruction exhibits a structure closely resembling quantum error correction: a bulk operator may admit several different boundary representations without losing its physical information \cite{ref4,ref5,ref9,ref11}.

This motivation leads naturally to the development of holographic quantum error-correcting codes, in which bulk degrees of freedom are treated as logical degrees of freedom and boundary degrees of freedom as physical degrees of freedom. An important realization was introduced by Pastawski, Yoshida, Harlow, and Preskill through the HaPPY code \cite{ref10}. In this model, perfect tensors are placed on a hyperbolic tessellation so that the entire tensor network defines a bulk-to-boundary encoding map. For the pentagon code, the basic building block is related to the five-qubit quantum error-correcting code and can be represented by a six-leg tensor: one leg plays the role of a logical bulk index, while the remaining five legs are network indices that can be contracted with neighboring tensors. The resulting network is a discretization of hyperbolic space and provides an explicit realization of quantum-error-correcting structure in holography \cite{ref9,ref10,ref11}.

In the discrete HaPPY-code model, bulk reconstruction can be understood through tensor pushing or, geometrically, through the greedy procedure \cite{ref10,ref66}. Starting from a boundary region, tensors with a sufficient number of legs contained in the already reconstructed region can be absorbed progressively into the wedge. This procedure produces a discrete entanglement wedge that determines which part of the bulk can be reconstructed from the chosen boundary region. The HaPPY code therefore provides an explicit example of how redundancy in bulk reconstruction arises as a quantum-error-correcting property of a holographic tensor network \cite{ref10,ref11,ref66}.

An important generalization of the RT formula is obtained when bulk quantum effects are included. Faulkner, Lewkowycz, and Maldacena showed that the leading-order quantum correction to the boundary entanglement entropy involves the entanglement entropy of quantum fields in the bulk. For a region $A$, the generalized entropy takes the form \cite{ref12,ref13}
\begin{equation}
S(A)
=\frac{\mathrm{Area}(A_{\min})}{4G_N}
+S_{\mathrm{bulk}}(A_b),
\label{eq:generalized_entropy}
\end{equation}
where $A_b$ denotes the bulk region bounded by the relevant extremal surface and the boundary region, while $S_{\mathrm{bulk}}(A_b)$ is the entanglement entropy of the effective quantum fields in that region. The first term is the classical geometric contribution, whereas the second term represents the quantum correction arising from entanglement of the bulk degrees of freedom. The ultraviolet divergence appearing in the bulk entanglement entropy must be understood together with the renormalization of gravitational parameters or an appropriate ultraviolet regularization in quantum gravity \cite{ref12,ref13}.

An analogous structure arises naturally in holographic tensor networks. For a boundary region $R$, the entropy of the tensor-network model can be written schematically as \cite{ref10,ref14,ref50}
\begin{equation}
S(\rho_A)
=|\gamma_c|\log D
+S\!\left(\rho^{\mathrm{bulk}}_{E_A}\right),
\label{eq:TN_generalized_entropy}
\end{equation}
where $\gamma_c$ is the minimal cut separating the boundary region $A$ from its complement and simultaneously defining the entanglement wedge $E_A$. The parameter $D$ denotes the bond dimension of the tensor network. The first term is the discrete analogue of the area term, whereas the second term originates from entanglement of the bulk degrees of freedom inside the wedge. Thus, a tensor-network realization not only models bulk reconstruction but also provides a discrete representation of the relation between geometry and quantum entanglement \cite{ref10,ref12,ref14,ref50}.

The HaPPY code remains a toy model, and several of its limitations become important when the model is extended to states involving black holes \cite{ref10,ref11,ref16}. An infinite-dimensional extension of the HaPPY code has also been investigated through a sequence of stabilizer codes whose Hilbert spaces are related by isometric embeddings. Although this construction allows entanglement-wedge reconstruction and bulk dynamics to be consistently defined, the resulting boundary Hamiltonian does not generate the long-range entanglement expected of a conformal field theory, thereby revealing an additional limitation of the HaPPY code as a complete model of AdS/CFT \cite{ref16}.

The isometric encoding of the HaPPY code naturally describes a code subspace of the boundary Hilbert space. Pastawski et al. showed that boundary states outside the semiclassical code subspace need not be interpreted merely as small perturbations around empty AdS; a significant fraction may instead correspond to black-hole states. In the tensor-network representation, a black hole can be modeled by excising tensors from an interior region of the network, leaving dangling legs on an inner boundary that act as black-hole degrees of freedom \cite{ref10,ref15,ref16}.

The problem becomes more fundamental when effective-field-theory degrees of freedom in the black-hole interior are included. The number of effective interior degrees of freedom can exceed the number of fundamental black-hole degrees of freedom determined by the horizon entropy. In such a situation, the bulk-to-boundary encoding cannot remain an isometry on the entire Hilbert space. This motivates the development of non-isometric holographic codes. Akers, Engelhardt, Harlow, Penington, and Vardhan showed that non-isometric encoding can be used to describe the black-hole interior while reproducing a number of important properties, including quantum-extremal-surface structure, Page-curve behavior, and state-dependent reconstruction. Subsequent developments incorporated hyperbolic tessellations so that the interior acquires a notion of locality absent in the simplest qudit model. In this approach, the relative positions of interior degrees of freedom can be represented by the discrete geometric structure of the tensor network \cite{ref15,ref16,ref17,ref18}.

These developments show that tensor networks provide an effective language for connecting geometry, entanglement, and quantum error correction. However, tensor-network holography and canonical approaches to quantum gravity begin from fundamentally different constructions. The question then arises whether building blocks of quantum geometry in a more fundamental theory can microscopically generate the structure required by holographic tensor networks \cite{ref5,ref10,ref14,ref15,ref16,ref17,ref45,ref46,ref47,ref48,ref49,ref50,ref51,ref52}.

Loop quantum gravity (LQG) is one of the nonperturbative and background - independent approaches to quantum gravity. In its canonical formulation, general relativity is expressed in terms of a pair of variables with an $SU(2)$ gauge structure~\cite{ref19,ref20}. LQG originates from the canonical formulation of classical four-dimensional gravity as a dynamical theory of gauge connections. In this formulation, the gravitational phase space has a structure analogous to that of an $SU(2)$ gauge theory, and the canonical variables represent the geometry of three-dimensional space. Quantization of this phase space leads to a Hilbert-space formulation with geometric operators acting on it. Promoting the canonical variables to operators gives a quantization of three-dimensional spatial geometry~\cite{ref45}. The Hilbert space admits eigenstates of geometric operators known as spin-network states. The spin-network formulation is based on cylindrical functions satisfying the quantum Gauss constraint in LQG. Among the most important geometric operators are the area and volume operators. Acting on spin-network states, these operators possess discrete eigenvalues, implying that the quantum geometry of area and volume is fundamentally discrete at the Planck scale. In this work, we employ the dual formulation of three-dimensional geometry constructed from nodes and edges in the spin-network description, commonly referred to as a dual triangulation. In this dual picture, nodes are associated with geometric objects interpreted as polyhedra, with the tetrahedron as the simplest structure, whereas edges are associated with faces shared by neighboring polyhedra. The area eigenvalues of the polyhedral faces and the volume eigenvalues of the polyhedra are determined by the corresponding area and volume operators \cite{ref19,ref20,ref21,ref35,ref36,ref61}.

Spin networks also admit a useful geometric interpretation. On a four-valent graph, a node can be associated with a quantum tetrahedron, while links represent adjacency relations between quantum polyhedra. The spin carried by a link determines the area of the corresponding face, whereas the intertwiner at a node encodes information about the shape and volume of the quantum polyhedron. A spin network can therefore be viewed as a collection of quantum polyhedra connected through shared faces \cite{ref21,ref22,ref23,ref24,ref25,ref26}.

The relation among spin networks, tensor networks, and holography has been investigated from several directions. Exact holographic mappings based on LQG have shown that spin-network states can be used to define bulk-to-boundary maps, relating bulk spin-network states to their boundary description through an exact holographic mapping in a tensor-network framework \cite{ref45,ref51}. In the work of Chen and Livine, the presence of spatial boundaries naturally leads to boundary quantum states that provide quantum boundary conditions for bulk fields, thereby defining a bulk/boundary map as a generalization of holographic duality from an LQG perspective \cite{ref62}. The relation between holographic duality and bulk/boundary maps has also been investigated in quantum gravity using group-field-theory formalisms \cite{ref46,ref47,ref48,ref49,ref63}. Recent studies on LQG and quantum information have further identified holographic features, such as entanglement entropy, within the structure of spin-network states \cite{ref35}. Other works have shown that spin-network structures can be represented explicitly as tensor networks and even implemented through quantum circuits and quantum-simulation platforms \cite{ref36,ref37,ref38,ref64,ref65}. In particular, recent developments show that $SU(2)$ holonomies and intertwiners can be mapped to efficient tensor-network representations, opening a direct connection among spin-network quantum geometry, quantum simulation, and holographic bulk-boundary maps \cite{ref35,ref36,ref37,ref38,ref45,ref47,ref51,ref62,ref63,ref64,ref65}.

Although LQG and holographic tensor networks originate from different constructions, they share several structural similarities. Both represent physical systems in terms of interconnected discrete units carrying local algebraic data. In LQG, these data are representations and intertwiners, whereas in HaPPY-like codes they are encoded in local tensors and bond structures. More importantly, entanglement plays a central role in both formulations. In tensor-network holography, the entanglement pattern determines effective geometry and entanglement wedges. In LQG, entanglement between quantum polyhedra can provide the geometric correlations required to glue neighboring regions of quantum space \cite{ref5,ref10,ref14,ref31,ref32,ref45,ref46,ref47,ref48,ref49,ref50,ref51,ref52}.

An important example of the latter mechanism is provided by Bell-network states. In this construction, spin-network states are modified using squeezed-vacuum states so that the intertwiners at neighboring nodes become entangled. These correlations induce geometric conditions on adjacent polyhedra, in particular a back-to-back relation between the normals of shared faces. Entanglement therefore does not merely act as an abstract correlation but also plays a geometric role: quantum polyhedra that were initially uncorrelated can be glued into a more coherent quantum geometry \cite{ref31,ref32,ref33,ref34}.

In this work, we use Bell spin-network states with an explicit holonomy representation to construct face-to-face correlations between quantum tetrahedra. This correlation structure glues a tetrahedron to its neighbor through entanglement. Entanglement among tetrahedra generates an inter-tetrahedral structure that can subsequently be extended into a complete holographic model. In particular, we investigate whether the entanglement pattern of Bell spin-network states can be reorganized into a structure satisfying the essential properties of holographic encoding, including isometry, tensor-network connectivity, and holographic entanglement entropy. In contrast to Bell-network constructions that focus primarily on geometric correlations between neighboring polyhedra, our aim is to use entangled tetrahedra as microscopic building blocks of a tessellated tensor network. We first show how a collection of entangled tetrahedra can be organized into a structure corresponding to the pentagonal building block of the HaPPY code. In this contruction, we define a logical embedding tensor $\mathcal{N}$ acting on the specific amplitudes of the holonomy-dressed Bell spin network, yielding the same dimension as the perfect tensor in the HaPPY code. The construction is then generalized to a $p$-gonal structure on a regular tessellation $\{p,q\}$ \cite{ref10,ref31,ref32,ref33,ref38}.

We then extend this local construction to the entire hyperbolic disk with specific class. For hyperbolic tessellations satisfying the condition \cite{ref10,ref14,ref50,ref52,ref59,ref60}
\begin{equation}
(p-2)(q-2)>4,
\label{eq:hyperbolic_condition}
\end{equation}
We construct a holonomy dressed Bell spin network whose connectivity follows a specific hyperbolic $\{p,q\}$ tessellation within the family of HaPPY code tessellations. In this way, each building block of the holographic tessellation tensor network acquires a representation in terms of entangled quantum tetrahedra. The construction provides an explicit dictionary between LQG degrees of freedom and the tensor structure of the holographic network \cite{ref10,ref14,ref31,ref38,ref45,ref46,ref47,ref48,ref49,ref50,ref51,ref52}. Based on this construction, we define a bulk-to-boundary encoding operator for the tessellation network. For this purpose, we introduce two conditions: (i) the geometric hyperbolic structure condition, and (ii) the admissible microscopic transfer map condition. These two conditions are necessary because the expressive rank in the holonomy-Bell LQG model is sufficiently large to allow the selection of a logical embedding that realizes the HaPPY code perfect tensor. We rigorously show that, under these two conditions, the effective tensor constructed via this formulation is identically equal to a specific perfect tensor—which, in the context of the HaPPY code, corresponds to a six-legged effective perfect tensor. The isometric property of this map is then analyzed. We show that the resulting global encoding for whole tessellation is isometry even in the large-spin limit. This analysis provides a direct criterion for the emergence of holographic encoding from the entanglement structure among quantum tetrahedra with coarse graining approach which is introduced here~\cite{ref27,ref28,ref29,ref30,ref31,ref32,ref45,ref47}.

We subsequently calculate the holographic entanglement entropy of boundary subregions. In the relevant limit, the result takes the form of a generalized entropy consisting of a minimal-cut area contribution and a bulk-entanglement contribution~\cite{ref10,ref12,ref13,ref14}. Thus, the Bell-spin-network construction reproduces the discrete structure of the quantum Ryu-Takayanagi relation that also appears in holographic quantum error-correcting codes. This result indicates that the relation between area and entanglement in holographic tensor networks can acquire a microscopic interpretation in terms of entangled quantum geometry \cite{ref32,ref45,ref47,ref48,ref49,ref50}.

The framework also allows an extension to models of the black-hole interior. By using tessellated interior tensor networks, the number of bulk inputs relative to output legs can be varied so that the encoding need not remain isometric. In this regime, the tetrahedral construction provides an LQG analogue of an inner non-isometric code~\cite{ref16}. We analyze the change in the encoding properties and relate it to the qudit-counting description of black-hole evaporation and to changes in the generalized entropy that produce Page-curve behavior \cite{ref15,ref16,ref17,ref18}.

The paper is organized as follows. In Sec.~II, we introduce regular tessellation tensor networks $\{p,q\}$, which provide the discrete geometric structure for the holographic construction, and review the formulation of Bell spin-network states together with an explicit representation of $SU(2)$ holonomies. We also define the bulk and boundary Hilbert spaces used throughout the analysis. In Sec.~III, we show how entangled quantum tetrahedra can be assembled into the pentagonal building block of the HaPPY code and then generalize the construction to $p$-gonal building blocks on tessellations $\{p,q\}$ with the same family structure of HaPPY code $\{5,4\}$. The isometry condition and its relation to a perfect-tensor-like structure are then analyzed explicitly. In Sec.~IV, We apply the same framework for the global bulk-to-boundary encoding operator, and compute the holographic entanglement entropy to a tessellated black-hole interior in order to study the transition to non-isometric encoding and the entropy evolution in an evaporating black-hole model. Finally, in Sec.~V, we summarize the main results, discuss the limitations of the construction, and outline possible extensions toward bulk dynamics and more general holographic tensor-network models.

\section{Tessellation Tensor Network $\{p,q\}$ and Bell spin-network states in LQG}
\label{Basic theory}

In this section, we introduce a tessellation tensor network, a class of holographic tensor-network constructions inspired by the HaPPY code and the PHEVA framework, organized on a hyperbolic tessellation $\{p,q\}$ \cite{ref16}. We then discuss holonomy-dressed Bell spin-network states, in which spin-network states are modified using the formalism of squeezed spin networks \cite{ref31,ref33,ref34} and supplemented by an explicit Wigner holonomy matrix, following the representation developed in Ref.~\cite{ref38}. We introduce a class of states that combines spin-network holonomy amplitudes with the Bell-network gluing structure in order to construct entangled states of neighboring quantum tetrahedra.

\subsection{Tessellation Tensor Network $\{p,q\}$}
\label{subsec:tessellation-pq}

A two-dimensional regular tessellation is specified by the Schl\"afli symbol $\{p,q\}$, where $p$ denotes the number of sides of each polygon and $q$ denotes the number of polygons meeting at each vertex. In holographic tensor networks, hyperbolic tessellations provide a discrete geometric scaffold that organizes tensor connectivity. The HaPPY code uses such a structure to construct an isometric bulk-to-boundary encoding from perfect tensors \cite{ref5,ref10}. By contrast, the non-isometric PHEVA model describes a map between effective black-hole interior degrees of freedom and fundamental black-hole degrees of freedom \cite{ref15,ref17}. Bueller \textit{et al.} subsequently incorporated hyperbolic tessellations into a PHEVA-inspired interior tensor-network construction in order to introduce an explicit notion of locality for the effective interior degrees of freedom \cite{ref16}.

A tessellation may be characterized geometrically through its discrete curvature. In a two-dimensional regular tessellation, $q$ polygons meet at each vertex, producing an angular deviation $\delta$ from $2\pi$. This angular deficit or excess defines the tessellation curvature through
\begin{equation}
R_{\mathrm{tess}}
 = \frac{\delta}{2\pi}
 = \frac{2p-qp+2q}{2p}.
\label{eq:tess-curvature}
\end{equation}
Positive, vanishing, and negative curvature correspond, respectively, to $R_{\mathrm{tess}}>0$, $R_{\mathrm{tess}}=0$, and $R_{\mathrm{tess}}<0$. Equation~\eqref{eq:tess-curvature} therefore implies the classification
\begin{equation}
(p-2)(q-2)>4,
\label{eq:hyperbolic-condition}
\end{equation}
for a negatively curved, or hyperbolic, tessellation. Conversely, $(p-2)(q-2)<4$ corresponds to positive curvature, while $(p-2)(q-2)=4$ corresponds to a flat tessellation.

The spacetime geometry represented by the tessellation is characterized by comparison with the Ricci scalar of the spacetime under consideration. In the present work, the relevant geometries are negatively curved spacetimes, such as AdS and spatial slices of AdS--Schwarzschild spacetime. Accordingly, both AdS and AdS--Schwarzschild geometries can be represented by hyperbolic tessellations satisfying Eq.~\eqref{eq:tess-curvature}, or equivalently
\begin{equation}
q>\frac{2p}{p-2}.
\end{equation}

\begin{figure}[t]
    \centering
    \includegraphics[width=0.75\linewidth]{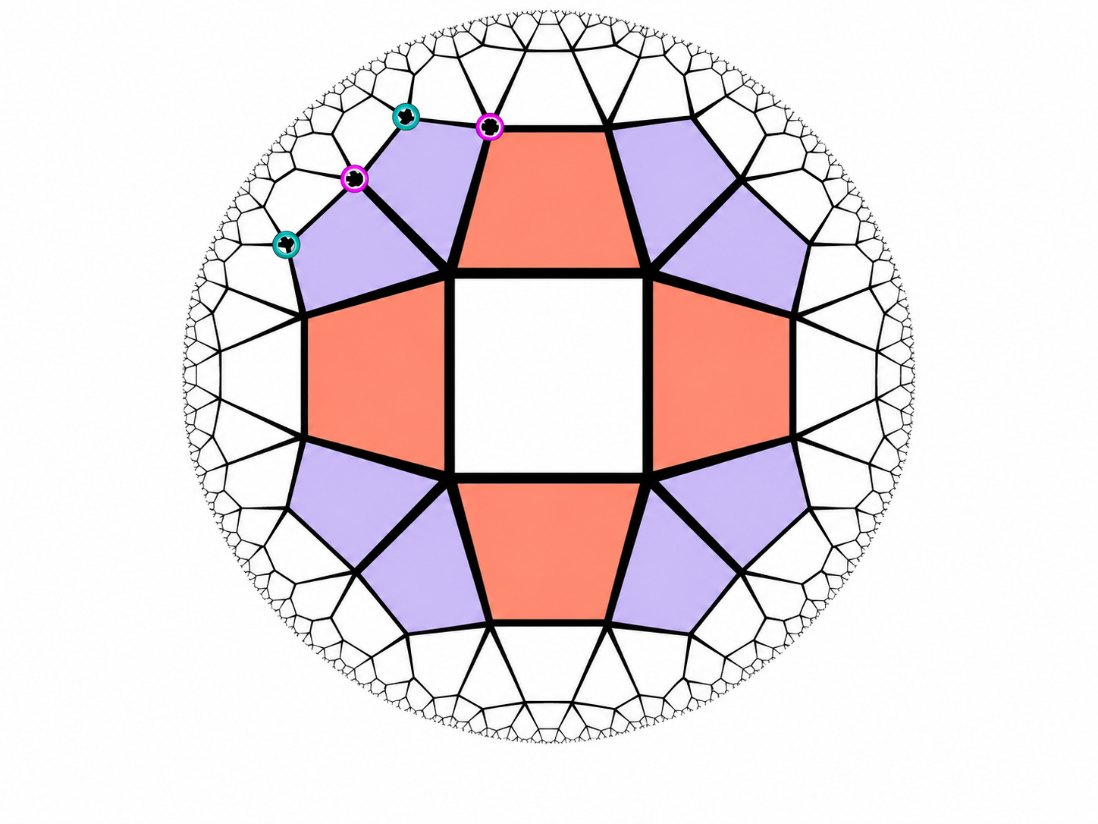}
    \caption{\small A depiction of a $\{4,5\}$ tessellation. The cells in row 1 are colored according to the type used for counting: edge cells are shown in
    coral, while vertex cells are shown in light
    purple. Two types of vertices are identified on the exterior of row 1: vertices circled in magenta are
    shared by two cells in row 1, while vertices
    circled in cyan are unshared.}
    \label{fig:tessellation45}
\end{figure}

Figure~\ref{fig:tessellation45} illustrates the two-dimensional hyperbolic tessellation $\{4,5\}$: each polygonal cell is a square ($p=4$), and five squares meet at each vertex ($q=5$). The parametrization of polygons into edge and vertex cells on successive rows of the tessellation is implemented through the procedure known as \emph{vertex inflation} \cite{ref15,ref16}. In Fig.~\ref{fig:tessellation45}, the central polygon is defined as the first row, while row $n$ consists of all cells that are absent from the preceding row and share a vertex with row $n-1$.

For the tessellation shown in Fig.~\ref{fig:tessellation45}, the numbers of edge and vertex cells are denoted by $e_n$ and $v_n$, respectively. The initial vector is taken to be
\begin{equation}
\begin{pmatrix}
e_{n-1}\\
v_{n-1}
\end{pmatrix}
=
\begin{pmatrix}
p\\
p(q-3)
\end{pmatrix}.
\label{eq:initial-vector}
\end{equation}
For an arbitrary tessellation $\{p,q\}$, the radial evolution of the cell numbers is written as
\begin{equation}
\begin{pmatrix}
e_n\\
v_n
\end{pmatrix}
=
M^{n-1}
\begin{pmatrix}
e_{n-1}\\
v_{n-1}
\end{pmatrix},
\label{eq:transfer-recursion}
\end{equation}
where
\begin{equation}
M=
\begin{pmatrix}
a & b\\
c & d
\end{pmatrix}.
\label{eq:M-general}
\end{equation}

The elements in the first row of $M$, namely $a$ and $b$, count the contribution of each cell in row $n-1$ to the number of edge cells in row $n$. An edge cell in row $n-1$ has three edges that are already occupied: one edge connecting it to row $n-2$ and two edges shared with neighboring cells in the same row. It therefore has $p-3$ free edges, which gives $a=p-3$. By contrast, a vertex cell has two edges already used by its two neighboring cells in row $n-1$, leaving $p-2$ free edges and hence $b=p-2$.

The second row of $M$ counts the contribution to the number of vertex cells in row $n$. Two types of vertices occur on the outer boundary of row $n-1$: \emph{shared vertices}, which are shared by two neighboring cells, and \emph{unshared vertices}, which belong to only one cell. Each shared vertex produces $q-4$ new vertex cells because, among the $q$ cells that can meet at a vertex, four positions are already occupied by two cells in row $n-1$ and two edge cells in row $n$. Each unshared vertex instead generates $q-3$ new vertex cells.

An edge cell in row $n-1$ contains $p-4$ unshared vertices. Including the contribution from one shared vertex, the number of newly generated vertex cells is
$c=(p-4)(q-3)+(q-4)$. A vertex cell contains $p-3$ unshared vertices, which together with the shared-vertex contribution gives $d=(p-3)(q-3)+(q-4)$. The transfer matrix controlling the growth of the number of cells from one row to the next is therefore
\begin{equation}
M=
\begin{pmatrix}
p-3 & p-2\\
(p-4)(q-3)+(q-4) & (p-3)(q-3)+(q-4)
\end{pmatrix}.
\label{eq:transfer-matrix}
\end{equation}

In the holographic setting, the spectrum of the transfer matrix $M$ in Eq.~\eqref{eq:transfer-matrix} provides a direct relation between the combinatorial geometry of the tessellation and the encoding capacity of the tensor network. Because $M\in SL(2,\mathbb{Z})$ is unimodular, the discrete radial evolution between successive rows is controlled by a reciprocal pair of eigenvalues. In the description of an AdS black hole as a hyperbolic tessellation, effective degrees of freedom are assigned to bulk cells, while fundamental degrees of freedom are associated with edges on the horizon. The spectrum of $M$ therefore directly controls the relation between the growth of the bulk geometry and the Hilbert-space capacity of the boundary. The common growth exponent in the limit $n\gg1$ provides a discrete representation of holographic scaling: increasing geometric complexity in the bulk is accompanied by an increasing information capacity of the boundary that encodes it.

\subsection{Holonomy-Dressed Bell Spin-Network States: Bell-Network States with Explicit $SU(2)$ Wigner Holonomy Matrices}
\label{subsec:holonomy-bell}

In loop quantum gravity (LQG), spatial geometry is quantized \cite{ref19,ref20,ref21,ref22,ref61}. Spin-network states form an orthogonal basis describing the quantum geometry of space in LQG. The spatial structure is represented by links and nodes; in the dual geometric interpretation, a node is associated with a quantum polyhedron, whose simplest relevant realization is a tetrahedron. The corresponding tetrahedral structure satisfies the Gauss constraint of LQG \cite{ref19}. In this sense, the discrete spatial structure of LQG can be understood in terms of collections of three-dimensional Euclidean polyhedra whose quantum geometry is encoded by spin-network states.

\begin{figure}[t]
    \centering
    \includegraphics[width=0.5\columnwidth]{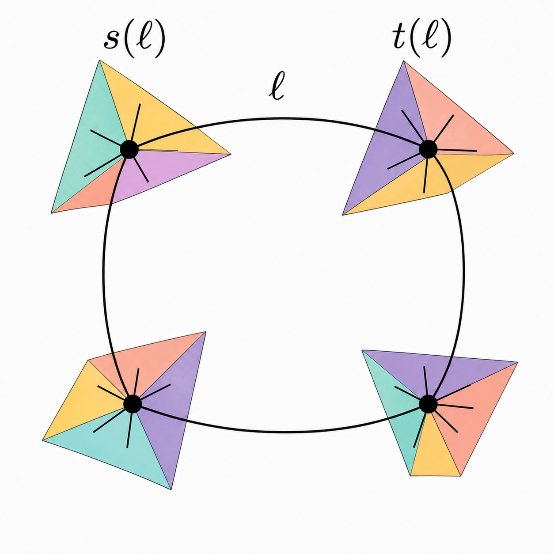}
    \caption{\small Neighboring polyhedra have adjacent faces with the same area but different shapes.}
    \label{fig:polyhedra-gluing}
\end{figure}

To describe the gluing of tetrahedral faces in LQG, the relevant faces are required to satisfy a back-to-back relation. Entanglement between polyhedra associated with neighboring nodes correlates their geometric fluctuations such that the back-to-back condition emerges semiclassically. This condition is written as
\begin{equation}
n_{s(l)}=- n_{t(l)},
\label{eq:back-to-back}
\end{equation}
where $n$ denotes the normal vector to the polyhedral faces being glued and the indices $s(l)$ and $t(l)$ label the source and target of the link $l$, as illustrated in Fig.~\ref{fig:polyhedra-gluing}.

We now consider correlations between spin-network nodes that can be interpreted as entanglement between quantum polyhedra \cite{ref31}. In the conventional spin-network basis, states associated with different nodes are essentially tensor products of intertwiner states, so geometric correlations between distinct nodes do not appear explicitly. To introduce such correlations, Bayta\c{s} \textit{et al.} \cite{ref31}, within the squeezed-spin-network framework \cite{ref31,ref33,ref34}, introduced Bell-network states. These states are constructed by placing Bell-type entanglement on the links connecting neighboring polyhedra and subsequently projecting onto the gauge-invariant subspace at each node. In the semiclassical limit, this construction generates geometric correlations that lead to a back-to-back relation between the normals of the two facing surfaces.

Inspired by the tensor-network representation of spin networks developed in Ref.~\cite{ref38}, we extend the Bell-network construction by retaining an explicit Wigner holonomy matrix on every link. The holonomy $h_l\in SU(2)$ represents parallel transport between the local frames at the source and target of a link, while $R(h_l)\in SO(3)$ denotes its vector representation. The back-to-back condition~\eqref{eq:back-to-back} can therefore be written in the gauge-covariant form
\begin{equation}
R(h_l) n_{s(l)}=- n_{t(l)}.
\label{eq:gauge-covariant-back-to-back}
\end{equation}
Equation~\eqref{eq:gauge-covariant-back-to-back} states that the face normals at the source and target become antiparallel after the normal at one endpoint is parallel transported into the local frame of the other endpoint.

The gluing of quantum polyhedra is implemented by entangling neighboring polyhedra using the squeezed Bell-state formalism introduced in Refs.~\cite{ref31,ref32,ref33}. We write
\begin{equation}
\left|B,\alpha_l\right\rangle
=
\left(1-|\alpha_l|^2\right)
\exp\!\left(\alpha_l\mathrm{\Phi}_l^{\dagger}\right)
|0\rangle_{s(l)}|0\rangle_{t(l)},
\label{eq:squeezed-bell}
\end{equation}
where the creation operators $a_{s(l)}^{\dagger A}$ and $a_{t(l)}^{\dagger A}$, with $A, B = 1, 2$ is a spinor index, we define a Bell state of the link $\ l$, parameter $\alpha \in \mathbb{C}$ and we define
\begin{equation}
\mathrm{\Phi}_l^{\dagger}
=\epsilon_{AB}
 a_{s(l)}^{\dagger A}
 a_{t(l)}^{\dagger B},
\end{equation}
with $\epsilon_{00}=\epsilon_{11}=0$, $\epsilon_{01}=1$, and $\epsilon_{10}=-1$. Hence
\begin{equation}
\mathrm{\Phi}_l^{\dagger}
=a_{s(l)}^{\dagger 0}a_{t(l)}^{\dagger 1}
-a_{s(l)}^{\dagger 1}a_{t(l)}^{\dagger 0}.
\end{equation}
Expanding the exponential gives
\begin{equation}
\exp(\alpha_l\mathrm{\Phi}_l^{\dagger})
=
\sum_{2j_l=0}^{\infty}
\frac{\alpha_l^{2j_l}}{(2j_l)!}
(\mathrm{\Phi}_l^{\dagger})^{2j_l},
\end{equation}
with
\begin{equation}
(\mathrm{\Phi}_l^{\dagger})^{2j_l}
=
\sum_{k=0}^{2j_l}
(-1)^k
\binom{2j_l}{k}
\left(a_{s(l)}^{\dagger0}a_{t(l)}^{\dagger1}\right)^{2j_l-k}
\left(a_{s(l)}^{\dagger1}a_{t(l)}^{\dagger0}\right)^k.
\end{equation}
The labels $j_l$ and $m$ denote, respectively, the spin and magnetic quantum number carried by the link $l$ connecting the two quantum polyhedra. Setting $k=j_l-m$, Eq.~\eqref{eq:squeezed-bell} becomes
\begin{align}
|B,\alpha_l\rangle
&=
(1-|\alpha_l|^2)
\sum_{2j_l=0}^{\infty}
\frac{\alpha_l^{2j_l}}{(2j_l)!}
\sum_{m=-j_l}^{j_l}
(-1)^{j_l-m}
\binom{2j_l}{j_l-m}
\nonumber\\
&\quad\times
(a_{s(l)}^{\dagger0})^{j_l+m}
(a_{s(l)}^{\dagger1})^{j_l-m}
(a_{t(l)}^{\dagger1})^{j_l+m}
(a_{t(l)}^{\dagger0})^{j_l-m}
|0\rangle_{s(l)}|0\rangle_{t(l)}.
\label{eq:squeezed-expanded}
\end{align}
Using the spin basis
\begin{align}
(a_{s(l)}^{\dagger0})^{j_l+m}
(a_{s(l)}^{\dagger1})^{j_l-m}|0\rangle_{s(l)}
&=
\sqrt{(j_l+m)!(j_l-m)!}
|j_l,m\rangle_{s(l)},\\
(a_{t(l)}^{\dagger1})^{j_l+m}
(a_{t(l)}^{\dagger0})^{j_l-m}|0\rangle_{t(l)}
&=
\sqrt{(j_l+m)!(j_l-m)!}
|j_l,-m\rangle_{t(l)},
\end{align}
we obtain
\begin{equation}
|B,\alpha_l\rangle
=
(1-|\alpha_l|^2)
\sum_{2j_l=0}^{\infty}
\alpha_l^{2j_l}
\sqrt{2j_l+1}
|B,j_l\rangle,
\label{eq:bell-spin-expansion}
\end{equation}
where
\begin{equation}
|B,j_l\rangle
=
\frac{1}{\sqrt{2j_l+1}}
\sum_{m=-j_l}^{j_l}
(-1)^{j_l-m}
|j_l,m\rangle_{s(l)}
|j_l,-m\rangle_{t(l)}.
\label{eq:fixed-spin-bell}
\end{equation}

We now dress the fixed-spin Bell state~\eqref{eq:fixed-spin-bell} with an explicit holonomy Wigner matrix $D(h_l)$. This distinguishes the present construction from the Bell-network formulation of Bayta\c{s} \textit{et al.} \cite{ref31}, in which the holonomy representation is not displayed explicitly in the Bell-network state. In that formulation, the squeezing parameter is written as $\alpha_l=|\alpha_l|e^{i\theta_l}$, with the phase $\theta_l$ associated with the corresponding link data. In our construction, $\alpha_l$ is retained in the form $\alpha_l=|\alpha_l|e^{i\theta_l}$ and is interpreted as an area-related parameter associated with the spin carried by the faces of the quantum polyhedra, whereas the holonomy operator is introduced separately through its Wigner-matrix representation. Equation~\eqref{eq:fixed-spin-bell} is therefore modified as
\begin{equation}
|B,j_l,h_l\rangle
=
(\widehat D_l(h_l)\otimes I)|B,j_l\rangle.
\label{eq:holonomy-dressed-def}
\end{equation}

The links of the spin networks in LQG are associated with SU(2) holonomies, which are unitary maps between the source and target Hilbert spaces~\cite{ref38}. The holonomy along a link $l$ is written as
\begin{equation}
h_l[A]
=
\mathcal P
\exp\!\left(\int_l A\right)
\in SU(2),
\label{eq:holonomy}
\end{equation}
where $A=A_a^i(x)\tau_i\,dx^a$ is the Ashtekar--Barbero connection, $\tau_i=-i\sigma_i/2$, $\sigma_i$ are the Pauli matrices, and $\mathcal P$ denotes path ordering. In the operator representation used in Eq.~\eqref{eq:holonomy-dressed-def}, inspired by \cite{ref38}, we define the holonomy operator as  
\begin{equation}
\widehat D_l
=
\sum_{m,m'}
D_{mm'}^{(j_l)}(h_l)
|j_l,m\rangle_{s(l)}
{}_{s(l)}\!\langle j_l,m'|.
\label{eq:holonomy-operator}
\end{equation}
The corresponding holonomy-dressed Bell state can then be written as
\begin{equation}
|B,j_l,h_l\rangle
=
\frac{1}{\sqrt{2j_l+1}}
\sum_{m,r,n=-j_l}^{j_l}
D_{mr}^{(j_l)}(h_l)
 g_{rn}^{(j_l)}
|j_l,m\rangle_{s(l)}
|j_l,n\rangle_{t(l)},
\label{eq:holonomy-dressed-bell}
\end{equation}
where
\begin{equation}
g_{nn'}^{(j_l)}
=(-1)^{j_l-n}\delta_{n,-n'}.
\end{equation}
For the trivial holonomy $h_l=\mathbb I$, one has $D^{(j_l)}(h_l)=\mathbb I$, and Eq.~\eqref{eq:holonomy-dressed-bell} reduces to Eq.~\eqref{eq:fixed-spin-bell}, namely the fixed-spin Bell-network state of Ref.~\cite{ref31}. Thus, the original fixed-spin Bell state is recovered as the trivial-holonomy sector of the present construction.

The full holonomy-dressed squeezed Bell spin-network state connecting the source $s(l)$ and target $t(l)$ quantum polyhedra is therefore
\begin{equation}
|B,\alpha_l,h_l\rangle
=
\left(1-|\alpha_l|^2\right)
\sum_{2j_l=0}^{\infty}
|\alpha_l|^{2j_l}
\sqrt{2j_l+1}
|B,j_l,h_l\rangle.
\label{eq:full-holonomy-dressed-link}
\end{equation}
Here $|B,j_l,h_l\rangle$ is given by Eq.~\eqref{eq:holonomy-dressed-bell}. In this construction, the holonomy does not change the amount of Bell entanglement at the level of the link state because it acts as a local unitary transformation on one endpoint. Its role is instead to change the basis in which source--target correlations are compared, thereby providing a more covariant geometric interpretation.

For the Bell-network state, the flux correlation can be expressed as
\begin{equation}
\left( J_{s(l)}+ J_{t(l)}\right)^2
|B,j_l\rangle=0,
\end{equation}
which implies the back-to-back condition~\eqref{eq:back-to-back}. Here $ J_{s(l)}$ and $ J_{t(l)}$ denote the bosonic flux operators associated with the two endpoints. The node constraint is written in the form \cite{ref31}
\begin{equation}
G_n
=
\sum_{i\in n} a_0 J_i\approx0,
\qquad
a_0=8\pi\gamma_{BI},
\end{equation}
where $(c=\hslash=G=1)$ and $\gamma_{BI}$ is the Barbero-Immirzi parameter. Since $\ J_{s(l)}$ and $ J_{t(l)}$ are in general defined with respect to different local frames, parallel transport of the source flux into the target frame leads to the condition
\begin{equation}
\left[
\ J_{s(l)}+R(h_l)\ J_{t(l)}
\right]^2
|B,j_l,h_l\rangle=0.
\label{eq:transported-flux}
\end{equation}
This condition implies the classical vector relation~\eqref{eq:gauge-covariant-back-to-back}. Consequently, gluing is no longer interpreted as an equality of normals in a specially chosen frame, but rather as back-to-back matching after $SU(2)$ parallel transport. The two objects $g^{(j_l)}$ and $D^{(j_l)}(h_l)$ therefore play distinct conceptual roles: $g^{(j_l)}$ implements the Bell correlation and orientation reversal, whereas $D^{(j_l)}(h_l)$ carries the parallel transport between local frames.

Following Ref.~\cite{ref31}, we now introduce a projection operator in order to apply Eq.~\eqref{eq:full-holonomy-dressed-link} to a graph $\Gamma$ consisting of nodes and links:
\begin{equation}
|\Gamma,B,\{\alpha_l\},\{h_l\}\rangle
=
P_{\Gamma}
|B,\{\alpha_l\},\{h_l\}\rangle.
\label{eq:graph-projection}
\end{equation}
The projector $P_{\Gamma}$ is defined through the outer products of the spin-network basis states,
\begin{equation}
P_{\Gamma}
=
\sum_{\{j_l\},\{i_n\}}
|\Gamma,\{i_n\},\{j_l\}\rangle
\langle\Gamma,\{i_n\},\{j_l\}|,
\end{equation}
where $i_n$ is an orthonormal basis of the intertwiner space
\begin{equation}
\mathcal I_n
=
\operatorname{Inv}_{SU(2)}
\left(
\bigotimes_{l\in n}V_{j_l}
\right).
\label{eq:intertwiner-space}
\end{equation}
The graph Hilbert space is therefore
\begin{equation}
\mathcal H_{\Gamma}
=
L^2\!\left(
\frac{SU(2)^L}{SU(2)^N}
\right)
=
\bigoplus_{\{j_l\}}
\left(
\bigotimes_{n\in V}\mathcal I_n
\right).
\end{equation}
This Hilbert space provides the vector space for the Bell spin-network states in Eq.~\eqref{eq:graph-projection}. The projection operator is essential because its action at the nodes implements the Gauss constraint.

Combining Eqs.~\eqref{eq:full-holonomy-dressed-link} and \eqref{eq:graph-projection}, we obtain
\begin{equation}
|\Gamma,B,\{\alpha_l\},\{h_l\}\rangle
=
\sum_{\{j_l\}}
\left[
\prod_l
(1-|\alpha_l|^2)
\alpha_l^{2j_l}
\sqrt{2j_l+1}
\right]
|\Gamma,B,\{j_l\},\{h_l\}\rangle,
\label{eq:graph-squeezed-state}
\end{equation}
where
\begin{equation}
|\Gamma,B,\{j_l\},\{h_l\}\rangle
=
\sum_{\{i_n\}}
A_{\Gamma}
(\{i_n\},\{j_l\},\{h_l\})
\bigotimes_n|i_n\rangle.
\label{eq:fixed-spin-graph-state}
\end{equation}
The amplitude is
\begin{equation}
M_{\Gamma}
(\{i_n\},\{j_l\},\{h_l\})
=
\sum_{\{m_l\}}
\prod_n
[i_n]^{\{m_l\}}
\prod_{l=(s,t)}
\left[
D^{(j_l)}(h_l)g^{(j_l)}
\right]_{m_{s(l)}m_{t(l)}}.
\label{eq:graph-amplitude}
\end{equation}
Equation~\eqref{eq:graph-amplitude} is a contraction of intertwiner tensors with explicit Wigner holonomy matrices on the graph $\Gamma$. The intertwiner tensor $[i_n]^{\{m_l\}}$ ensures that the magnetic degrees of freedom meeting at a node are combined into a gauge-invariant state, or equivalently satisfy the quantum closure/Gauss constraint. The tensor $g^{(j_l)}$ implements the Bell-singlet orientation reversal across the two sides of a shared face, while $D^{(j_l)}(h_l)$ carries the information associated with parallel transport between the local $SU(2)$ frames of two neighboring quantum polyhedra.

This construction therefore integrates Bell-network gluing and $SU(2)$ parallel transport into a single gauge-invariant spin-network amplitude. For $h_l=\mathbb I_l$, it reduces to the standard Bell-network state \cite{ref31,ref32}, whereas for nontrivial holonomy the Wigner $D$ matrices retain information about the relative local frames along the links \cite{ref19,ref38}. On open graphs, a subset of magnetic indices may be left uncontracted and interpreted as boundary legs, while the intertwiner degrees of freedom are retained as internal degrees of freedom. This structure provides a natural candidate for a bulk-to-boundary map in the spin-network framework \cite{ref38,ref45,ref62,ref63}.

In the following sections, we will show how the holonomy-dressed Bell spin-network states in Eqs.~\eqref{eq:graph-squeezed-state} and \eqref{eq:fixed-spin-graph-state} can be organized on hyperbolic tessellations $\{p,q\}$. We will analyze the conditions under which the corresponding encoding map becomes asymptotically isometric in the large-spin regime and examine the associated holographic entanglement entropy \cite{ref27,ref28,ref29,ref30,ref31,ref32,ref45,ref47,ref48,ref49}. For an open graph, uncontracted magnetic indices naturally define boundary legs, while intertwiner indices remain as internal or bulk degrees of freedom. This provides the basis for interpreting an LQG spin network as a bulk-to-boundary map \cite{ref36,ref38} and for treating networks of quantum tetrahedra as microscopic constituents of an effective holographic tensor network on a hyperbolic tessellation $\{p,q\}$ representing a discrete AdS geometry.

\section{Entangled Quantum Tetrahedra for Holographic Tessellation Tensor Networks $\{p,q\}$}

In this section, we construct a network of Bell-entangled quantum tetrahedra as a microscopic structure for a holographic tensor network defined on a hyperbolic tessellation. Tetrahedra are chosen as the polyhedra because the quantum-tetrahedron structure is the simplest four-valent intertwiner structure of a node that still satisfies gauge invariance, or equivalently the Gauss constraint in LQG, yielding the quantum closure condition for the fluxes associated with the four faces of the tetrahedron \cite{ref20,ref21}. Within this framework, the tetrahedron is a quantum-geometric unit that can be used as a microscopic constituent of an LQG-based tensor network \cite{ref36,ref37,ref38}. As an initial realization, we consider the regular hyperbolic tessellation $\{5,4\}$, namely the tessellation that also underlies the construction of the HaPPY code as a toy model of a holographic quantum error-correcting code \cite{ref4,ref5,ref10,ref16}. The purpose of this construction is not to identify a single $SU(2)$ intertwiner with a HaPPY perfect tensor. Instead, several gauge-invariant tetrahedral tensors are contracted and \emph{coarse-grained} into an \emph{effective cell tensor}, after which explicit conditions are determined under which the effective tensor can realize a HaPPY-type isometric encoding \cite{ref4,ref5,ref10,ref11}. More specifically, we consider the pentagonal structure in the hyperbolic tessellation $\{5,4\}$ organized into six mutually entangled quantum-tetrahedron structures. We then compute the generalized entropy or quantum Ryu--Takayanagi formula using the holonomy-dressed Bell spin-network states (Eqs.~\eqref{eq:graph-squeezed-state}--\eqref{eq:graph-amplitude}) on the graph identified with the entire spatial slice of the holographic tessellation $\{5,4\}$.

At the end of this section, we construct a theorem stating that the quantum - tetrahedron entanglement structure described by holonomy - dressed Bell spin-network states constitutes microscopically an effective holographic tensor network within a hyperbolic tessellation $\{p,q\}$, thereby producing two properties that must be possessed by a holographic tensor network representing holographic duality or bulk--boundary duality \cite{ref9,ref10,ref11,ref12,ref13}. The first is referred to as the bulk--boundary duality encoding operator, associated with the construction of an encoding operator inspired by the amplitude of the holonomy - dressed Bell spin-network states (Eq.~\eqref{eq:graph-amplitude}), for which the encoding operator has an isometric property at the local level for one polygonal cell in a holographic tessellation with a specified tiling $\{p,q\}$. This isometric property is important for bulk--boundary maps, such as the tensor-network structure of the HaPPY code, which is a toy model for a tensor-network representation of holographic duality, in this case AdS/CFT \cite{ref4,ref5,ref10}. The second concerns the geometric organization of boundary entanglement that yields the quantum Ryu--Takayanagi formula and generalized entropy when a particular subregion of a partial tessellation structure is considered; in this case, the partial tessellation structure is determined by the entanglement structure of quantum tetrahedra that replaces the polygonal structure in the hyperbolic tessellation tiling $\{p,q\}$. The symbols $p$ and $q$ are retained as the Schl\"afli parameters of the tessellation. In constructing this theorem, two conditions must be satisfied: (1) a tessellation-geometric-structure condition determined by the parameters $p$ and $q$ in the tessellation tiling $\{p,q\}$, referred to as the geometric and combinatorial compatibility condition; and (2) an admissibility condition associated with the microscopic transfer map, chosen such that it has full row rank for the perfect tensor.

\subsection{Microscopic Cell of Six Quantum Tetrahedra for the Tessellation $\{5,4\}$}

\subsubsection{Four-Valent Quantum Tetrahedra and the Fixed Microscopic Graph}

\begin{figure}[t]
    \centering
\includegraphics[width=0.5\columnwidth]{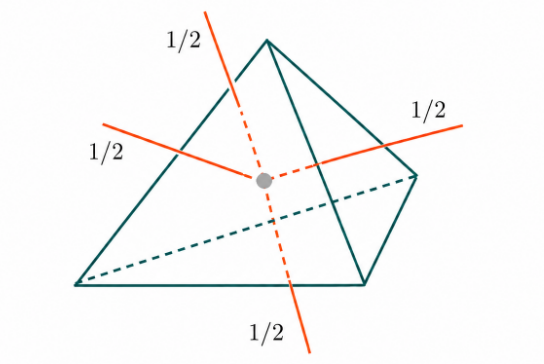}
    \caption{\small A four-valent node of the spin network and its triangulation as a quantum tetrahedron. Each link emerging from a face of the tetrahedron carries a holonomy-dressed Bell spin-network state with spin-$1/2$ that will entangle with other quantum tetrahedra, whereas the node carries an intertwiner state. Subsequently, the link and the node will be represented, respectively, as a boundary qubit and a logical qubit.}
    \label{fig:a four-valent node}
\end{figure}

Here we consider the hyperbolic tessellation $\{5,4\}$, which is the tessellation underlying the construction of the HaPPY code. We associate one pentagon with an abstract perfect tensor \cite{ref4,ref5,ref10}; in this construction, each pentagonal cell in the tessellation $\{5,4\}$ is realized microscopically by a network consisting of six mutually entangled quantum tetrahedra. Five tetrahedra are placed on the periphery of the cell and provide five inter-cell legs that are subsequently contracted with neighboring pentagonal cells, while one internal tetrahedron provides the logical degree of freedom (\emph{logical/bulk degree of freedom}). In the $j=1/2$ sector, each magnetic leg has dimension two and can be represented as a qubit \cite{ref36,ref37}, as shown in Fig.~\ref{fig:a four-valent node}, so that this construction provides a microscopic resolution of the effective six-leg tensor used in the tessellation $\{5,4\}$.

The entanglement across tetrahedral interfaces is constructed using the holonomy-dressed Bell spin-network states introduced in Eqs.~\eqref{eq:graph-squeezed-state}--\eqref{eq:graph-amplitude}. On each internal link $\ell$, there is a tensor $D^{\left( j_{l} \right)}\left( h_{l} \right)g^{\left( j_{l} \right)}$. After the Gauss projection at each node, these contractions produce a gauge-invariant state of mutually entangled quantum tetrahedra.

\begin{figure}[t]
    \centering
\includegraphics[width=0.9\columnwidth]{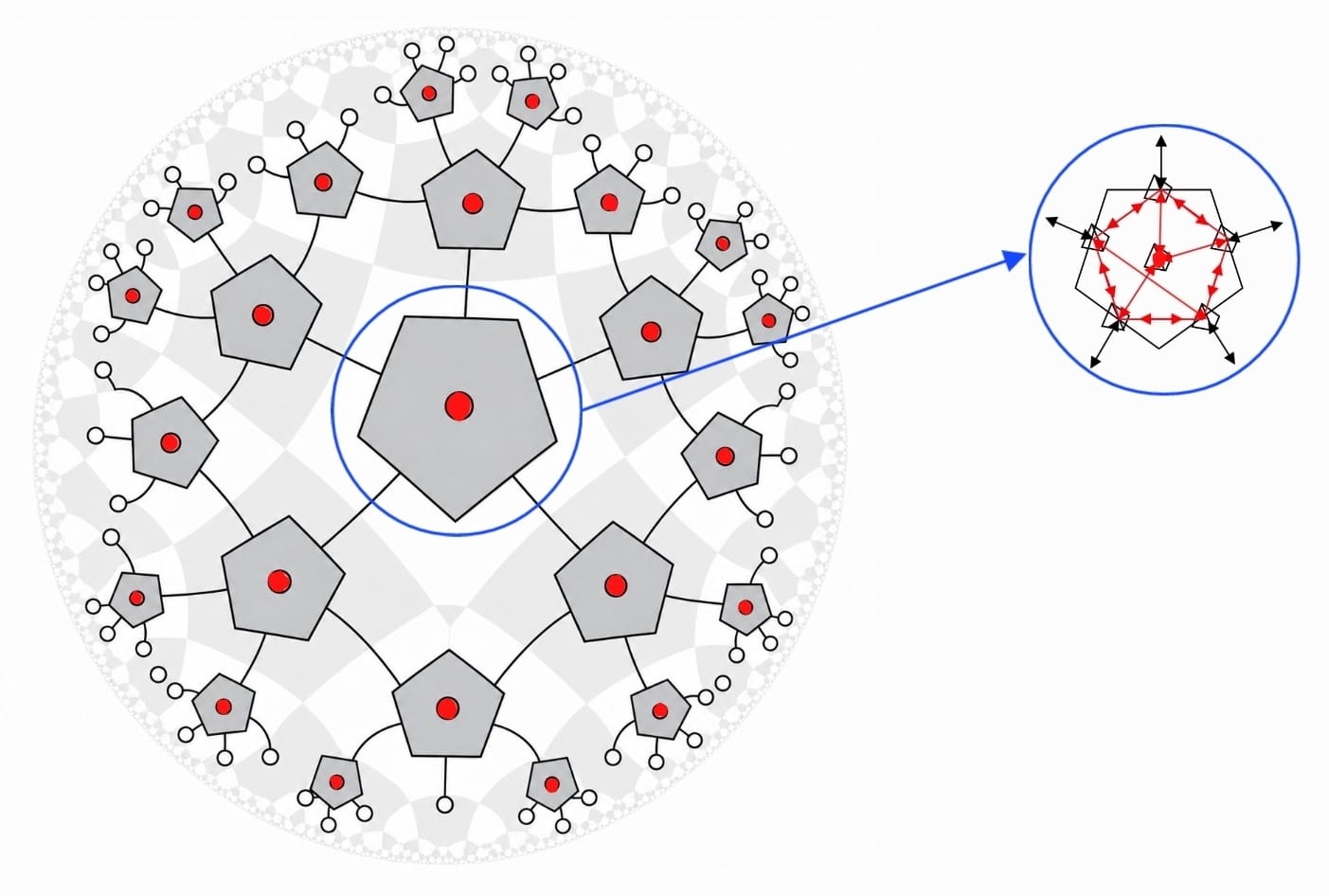}
    \caption{\small Tessellation with tiling $\{5,4\}$ representing perfect tensors in the HaPPY code, where the red dots represent logical bulk indices, while the white dots at the outer boundary represent physical boundary indices \cite{ref10}. The representation of a pentagon in the HaPPY code is defined as an entanglement structure of six quantum tetrahedra, where, for each quantum tetrahedron, every face satisfies the Gauss constraint and each link emerging from that face carries a holonomy-dressed Bell spin-network state. Consequently, one abstract perfect tensor on a pentagon is no longer regarded as fundamental ($\,{T^{b_{1}b_{2}b_{3}b_{4}b_{5}}}_{\mu}$), where $\mu$ is the logical bulk index and $b_{1},\ldots.,b_{5}$ are network-bond indices. Instead, the pentagon is ``opened'' or \emph{resolved} into a more microscopic structure consisting of several mutually entangled quantum tetrahedra. Thus, in our model we propose the relation $T_{b_{1},\ldots,b_{5};\ \mu\ }^{(HaPPY)}  \ \ \rightarrow \ \ T_{b_{1},\ldots,b_{5};\ \mu\ }^{(eff\ LQG)}$, where $T^{(eff\ LQG)}$ is not an elementary tensor, but an effective tensor obtained from the contraction of the quantum-tetrahedron network. For the tessellation $\{5,4\}$, one pentagonal cell is represented by six tetrahedra: five tetrahedra on the periphery and one internal tetrahedron.}
    \label{fig:Tessellation with tiling}
\end{figure}

Fig.~\ref{fig:Tessellation with tiling} shows how a pentagon in the tessellation tiling $\{5,4\}$ is represented as an entanglement structure among six quantum tetrahedra. Five tetrahedra are placed on the sides of the pentagon as an encoding from the bulk to the boundary, while one face of the quantum tetrahedron located at the center of the pentagon acts as the logical qubit. The red and black arrows indicate the entanglement carried by holonomy-dressed Bell spin-network states between quantum tetrahedra. The red arrows indicate the entanglement relations among quantum tetrahedra inside the pentagon, whereas the black arrows indicate the direction of entanglement out of the pentagon, thereby connecting the pentagon to neighboring pentagons.

In our model, the relation between the tensor structure of the HaPPY code and the effective tensor constructed from the entanglement of quantum tetrahedra can be written as
\begin{equation}
T_{b_{1},\ldots,b_{5};\ \mu\ }^{(HaPPY)}\ \ \ \  \rightarrow \ \ \ \ \ T_{b_{1},\ldots,b_{5};\ \mu\ }^{(eff\ LQG)}
\label{eq:s3-src29}
\end{equation}
Here $T^{(eff\ LQG)}$ is the composite/effective tensor obtained from the contraction of the quantum-tetrahedron network described by the amplitude of the holonomy-dressed Bell spin-network states (Eq.~\eqref{eq:graph-amplitude}). The tensor $T^{(eff\ LQG)}$ maps $T^{eff\ LQG} : \mathcal{H}_{logical}\  \rightarrow \ V_{j}^{\bigotimes 5}$. In our model, referring to the structure shown in Fig. 4, we assume that $T^{(eff\ LQG)}$ is not a perfect tensor. We now consider one pentagon in the hyperbolic tessellation $\{5,4\}$, for which we write the entanglement combinations as follows:

\begin{figure}[t]
    \centering
\includegraphics[width=0.45\columnwidth]{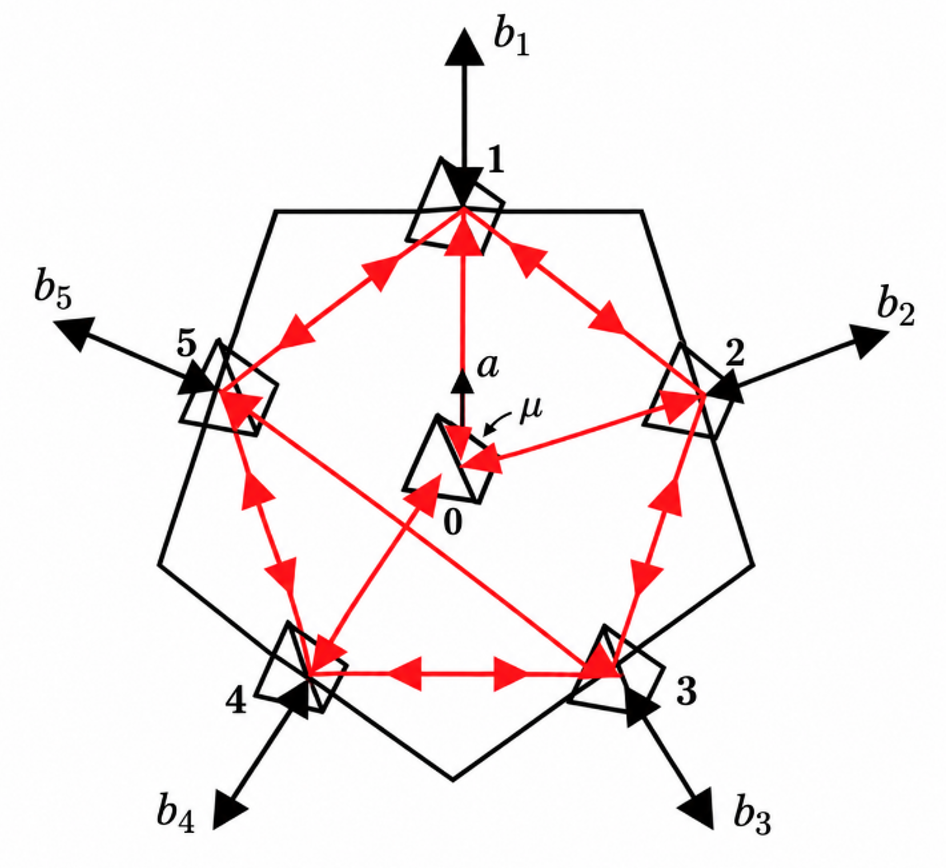}
    \caption{\small Microscopic realization of an effective pentagonal tensor from Bell-entangled quantum tetrahedra. Six four-valent quantum tetrahedra, labeled $v_{0}\  \rightarrow a$ and $v_{r}\  \rightarrow \ b_{r},\ \ r = 1,\ldots,\ 5$, are connected by nine internal holonomy-dressed Bell links (red) according to $L_{int} = \{(0,1),(0,2),(0,4),(1,2),(1,5),(2,3),(3,4),(3,5),(4,5)\}.$ The five open magnetic indices $b_{r} = b_{1},\ldots.,b_{5}$ (black outward arrows) define the effective inter-cell bond legs, while the remaining open magnetic index $a$ belongs to the microscopic input space.}
    \label{fig:microscopic realization}
\end{figure} 

The indices of the open legs $(v_{0},v_{r})$ are mapped to the effective tensor $T^{(eff\ LQG)}$ as follows:
\begin{equation}
\begin{gathered}
v_{0}\  \rightarrow a\\
v_{r}\  \rightarrow \ b_{r}\ ,\ \ \ \ r = 1,\ldots,\ 5
\end{gathered}
\label{eq:s3-src30}
\end{equation}
Thus, there are six open-leg indices for the six quantum tetrahedra: $v_{0}$ denotes the index on the tetrahedron at the center and represents the logical bulk qubit, while $v_{r}$ denotes the five peripheral tetrahedra on the sides of the pentagon. From Fig.~\ref{fig:microscopic realization}, it can be seen that the six quantum tetrahedra produce nine internal entanglement combinations among themselves to represent the microscopic constituent of one pentagon in the hyperbolic tessellation $\{5,4\}$.

One suitable three-regular graph on the six tetrahedra is the triangular-prism graph, with edge set
\begin{equation}
L_{int} = \{(0,1),(0,2),(0,4),(1,2),(1,5),(2,3),(3,4),(3,5),(4,5)\}.
\label{eq:s3-src31}
\end{equation}
The five open magnetic legs, $b_{1},\ldots,b_{5}$, become the effective inter-cell bond legs. The one remaining open magnetic leg, denoted by $a$, is part of the microscopic input space. The logical index $\mu$ is introduced through a separate embedding tensor $\mathcal{N}_{a,\mathbf{i};\mu}$ with $\mathbf{i} = \left( i_{0},\ldots,i_{5} \right)$. Therefore, in general, $a$ must not be identified directly with $\mu$. The special restriction $\mathcal{N}_{a,\mathbf{i};\mu} \propto \delta_{a\mu}$ would indeed make them identical, but the exact synthesis construction below uses a more general embedding. Each quantum tetrahedron, which is a four-valent node, leaves one open magnetic leg and uses its other three legs internally. Hence, the total number of magnetic half-edges is $4\times6=24$, or there are 24 spin-$1/2$ qubits per pentagonal cell; six of these are open, with five serving as network-bond indices or qubits entangled with qubits from neighboring pentagons, one serving as the logical qubit, and the remaining 18 forming internal half-edges that are mutually entangled within the pentagonal-cell structure. Equivalently, one internal link uses two half-edges or two qubits from the corresponding tetrahedra, yielding $L_{int}=9$ as shown in Eq.~\eqref{eq:s3-src31}.

\subsubsection{Microscopic Transfer Map}

For the case of one pentagon in the tessellation $\{5,4\}$, we propose that $T^{(eff\ LQG)}$ be represented in the amplitude of the holonomy-dressed spin-network states (Eq.~\eqref{eq:graph-amplitude}) as follows:
\begin{equation}
\begin{aligned}
&T_{b_{1},\ldots,b_{5};\,\mu}^{(eff\,LQG)}
\left(\mathbf{h}\right)
\\
&=
\sum_{a}
\sum_{i_{0},\ldots,i_{5}}
\mathcal{N}_{a\mu}^{i_{0},\ldots,i_{5}}
\sum_{\substack{
m_{01},m_{03},m_{05}\\
m_{10},m_{12},m_{15}\\
m_{21},m_{23},m_{24}\\
m_{30},m_{32},m_{34}\\
m_{42},m_{43},m_{45}\\
m_{50},m_{51},m_{45}
}}
\imath_{i_{0}}^{a m_{01},m_{03},m_{05}}
\\
&\qquad\times
\imath_{i_{1}}^{b_{1}m_{10},m_{12},m_{15}}
\imath_{i_{2}}^{b_{2}m_{21},m_{23},m_{24}}
\\
&\qquad\times
\imath_{i_{3}}^{b_{3}m_{30},m_{32},m_{34}}
\imath_{i_{4}}^{b_{4}m_{42},m_{43},m_{45}}
\\
&\qquad\times
\imath_{i_{5}}^{b_{5}m_{50},m_{51},m_{54}}
\\
&\qquad\times
L_{m_{10}m_{01}}^{(01)}
L_{m_{30}m_{03}}^{(03)}
L_{m_{50}m_{05}}^{(05)}
\\
&\qquad\times
L_{m_{21}m_{12}}^{(12)}
L_{m_{51}m_{15}}^{(15)}
L_{m_{32}m_{23}}^{(23)}
\\
&\qquad\times
L_{m_{42}m_{24}}^{(24)}
L_{m_{43}m_{34}}^{(34)}
L_{m_{54}m_{45}}^{(45)} .
\end{aligned}
\label{eq:s3-src32}
\end{equation}
where the indices $b_{1},\ldots,b_{5}$ denote the network-bond indices, while $\mu$ denotes the logical-qubit index. The tensor $\mathcal{N}_{a\mu}^{i_{0},\ldots,i_{5}}$ maps the code/logical degree of freedom of the effective index $\mu$ to the open leg $a$, which represents one qubit of the pentagon at the center. The next factor is the intertwiner tensor $(\imath_{k}^{m_{1}m_{2},m_{3},m_{4}})$ in a representation with four magnetic-leg indices, namely
\begin{equation}
\imath_{k}^{m_{1}m_{2},m_{3},m_{4}} = \sum_{M = - k}^{k}\frac{( - 1)^{k - M}}{\sqrt{2k + 1}}C_{jm_{1},jm_{2}}^{k,M}C_{jm_{3},jm_{4}}^{k, - M}
\label{eq:s3-src33}
\end{equation}
The Clebsch-Gordan coefficient $(C_{j_{1}m_{1},j_{2}m_{2}}^{k,M})$ is written as
\begin{equation}
C_{j_{1}m_{1},j_{2}m_{2}}^{k,M} = ( - 1)^{j_{1} - j_{2} + M}\sqrt{2J + 1}\begin{pmatrix}
j_{1} & j_{2} & J \\
m_{1} & m_{2} & - M
\end{pmatrix}
\label{eq:s3-src34}
\end{equation}
where $\begin{pmatrix} j_{1} & j_{2} & J \\ m_{1} & m_{2} & - M \end{pmatrix}$ is the Wigner $3j$ symbol. Reference \cite{ref38} uses a representation similar to Eq.~\eqref{eq:s3-src34}, namely a representation in terms of the Wigner $3j$ symbol. In addition, it should be noted that the intertwiner tensor in Eq.~\eqref{eq:s3-src33} satisfies the orthogonality property
\begin{equation}
\sum_{m_{1}m_{2},m_{3},m_{4}}^{}{\overline{\imath_{k^{'}}^{m_{1}m_{2},m_{3},m_{4}}}\imath_{k}^{m_{1}m_{2},m_{3},m_{4}}} = \delta_{k^{'}k}
\label{eq:s3-src35}
\end{equation}
The normalized link tensor appearing in Eq.~\eqref{eq:s3-src32} is known as the holonomy-dressed Bell link (see Eq.~\eqref{eq:holonomy-dressed-bell}); here we define it as

\begin{equation}
\begin{aligned}
L_{m_{w}m_{v}}^{(l)}
&=
\frac{1}{\sqrt{2j_{l}+1}}\,
D^{(j_{l})}\!\left(h_{l}\right)
g^{(j_{l})}
\\[1mm]
&=
\frac{(-1)^{j+m_{v}}}
{\sqrt{2j_{l}+1}}\,
D_{m_{w},-m_{v}}^{(j)}
\!\left(h_{l}\right)
\\[1mm]
&=
\frac{1}{\sqrt{2j_{l}+1}}\,
U_{m_{w},m_{v}}^{(j_{l})}
\!\left(h_{l}\right),
\qquad
h_{l}\in SU(2).
\end{aligned}
\label{eq:s3-src36}
\end{equation}
With the link index $l=(v,w)$ denoting an oriented internal link (see Eq.~\eqref{eq:s3-src31}), $U_{m_{w},m_{v}}^{\left( j_{l} \right)}\left( h_{l} \right) = D^{\left( j_{l} \right)}\left( h_{l} \right)g^{\left( j_{l} \right)} = \sqrt{2j_{l} + 1}L_{m_{w}m_{v}}^{(l)}$, where $g^{(j)}$ implements the Bell-singlet or orientation pairing, whereas $D^{(j)}\left( h_{e} \right)$ is the Wigner D-matrix that implements parallel transport between local $SU(2)$ frames \cite{ref31,ref38}. Since $D^{(j)}\left( h_{e} \right)$ and $g^{(j)}$ are both unitary, Eq.~\eqref{eq:s3-src55} satisfies the normalization condition
\begin{equation}
\sum_{m_{v}}^{}{U_{m_{w},m_{v}}^{\left( j_{l} \right)^{\dagger}}U_{m_{w},m_{v}}^{\left( j_{l} \right)}} = \delta_{m_{w}m_{w}^{'}}\delta_{m_{v}m_{v}^{'}}
\label{eq:s3-src37}
\end{equation}

\begin{equation}
\sum_{m_{v}}^{}{\overline{L_{m_{w}^{'}m_{v}^{'}}^{(l)}}L_{m_{w}m_{v}}^{(l)}} = \frac{\delta_{m_{w}m_{w}^{'}}\delta_{m_{v}m_{v}^{'}}}{{2j_{l} + 1}}
\label{eq:s3-src38}
\end{equation}
For $m_{w}^{'}=m_{w}$ and $m_{v}^{'}=m_{v}$, this normalization condition gives the density matrix
\begin{equation}
\rho_{w;v} = \frac{\mathbb{I}_{j_{l}}}{\sqrt{2j_{l} + 1}}
\label{eq:s3-src39}
\end{equation}
which is related to the entanglement entropy in LQG, $S_{cut}\left( h_{l} \right) = \log\left( 2j_{l} + 1 \right)$ \cite{ref32,ref35,ref63}. In our construction we use homogeneous spin $j_{l}$ with dimension $d_{l}=2j_{l}+1$, and each four-valent intertwiner state has

\begin{equation}
\dim{{Inv}_{SU(2)}(V_{j_{l}}^{\otimes 4}}) = d_{j_{l}}\
\label{eq:s3-src40}
\end{equation}

Next, define the matrix $M$ from Eq.~\eqref{eq:graph-amplitude} as 
\begin{equation}
M_{\mathbf{b};\, a,\mathbf{i}}^{(j)}\left( h_{l} \right) = \sum_{\{ m_{vl}\}}^{}\left\lbrack \prod_{v = 0}^{5}\iota_{i_{v}}^{\, x_{v},\{ m_{vl}\}_{l \ni v}} \right\rbrack\left\lbrack \prod_{l = (v,w) \in L_{int}}^{}L_{m_{w}m_{v}}^{(l)} \right\rbrack
\label{eq:s3-src41}
\end{equation}
with $\mathbf{b}=\left( b_{1},\ldots,b_{5} \right)$ and $\mathbf{h}=\left( h_{l} \right)_{l\in E_{int}}$. Its explicit form is

\begin{equation}
\begin{aligned}
&M_{b_{1},\ldots,b_{5};\,\mu,\,i_{0},\ldots,i_{5}}^{(j_l)}
\left(h_l\right)
\\
&=
\sum_{\substack{
m_{01},m_{03},m_{05}\\
m_{10},m_{12},m_{15}\\
m_{21},m_{23},m_{24}\\
m_{30},m_{32},m_{34}\\
m_{42},m_{43},m_{45}\\
m_{50},m_{51},m_{54}
}}
\imath_{i_{0}}^{a m_{01},m_{03},m_{05}}
\imath_{i_{1}}^{b_{1}m_{10},m_{12},m_{15}}
\\
&\qquad\times
\imath_{i_{2}}^{b_{2}m_{21},m_{23},m_{24}}
\imath_{i_{3}}^{b_{3}m_{30},m_{32},m_{34}}
\\
&\qquad\times
\imath_{i_{4}}^{b_{4}m_{42},m_{43},m_{45}}
\imath_{i_{5}}^{b_{5}m_{50},m_{51},m_{54}}
\\
&\qquad\times
L_{m_{10}m_{01}}^{(01)}
L_{m_{30}m_{03}}^{(03)}
L_{m_{50}m_{05}}^{(05)}
\\
&\qquad\times
L_{m_{21}m_{12}}^{(12)}
L_{m_{51}m_{15}}^{(15)}
L_{m_{32}m_{23}}^{(23)}
\\
&\qquad\times
L_{m_{42}m_{24}}^{(24)}
L_{m_{43}m_{34}}^{(34)}
L_{m_{54}m_{45}}^{(45)} .
\end{aligned}
\label{eq:s3-src42}
\end{equation}
Thus, the effective cell tensor in Eq.~\eqref{eq:s3-src32} can be written as
\begin{equation}
T_{b_{1},\ldots,b_{5};\ \mu\ }^{(eff\ LQG)} = M_{b_{1},\ldots,b_{5};\ \mu,\ \ i_{0},\ldots,i_{5}}^{\left( j_{l} \right)}\mathcal{N}_{a\mu}^{i_{0},\ldots,i_{5}}
\label{eq:s3-src43}
\end{equation}
The mappings of the quantities above are
\begin{equation}
M_{\left( j_{l} \right)} :\ \mathbb{C}^{{d_{l}}^{7}}\  \rightarrow \ \mathbb{C}^{{d_{l}}^{5}}
\label{eq:s3-src44}
\end{equation}

\begin{equation}
\mathcal{N}\ :\ \mathbb{C}^{d_{l}}\  \rightarrow \ \mathbb{C}^{{d_{l}}^{7}}
\label{eq:s3-src45}
\end{equation}

\begin{equation}
T^{(eff\ LQG)} = M_{\left( j_{l} \right)}\mathcal{N} 
\label{eq:s3-src46}
\end{equation}
with the collective microscopic input index $(\mu,\mathbf{i}=i_{0},\ldots,i_{5})$ and the output index $(\mathbf{b}=b_{1},\ldots,b_{5})$. Thus, the exponent $7$ is not the number of effective open legs. One cell still has six effective legs $\left( \mu,b_{1},\ldots,b_{5} \right)$, with dimension $d^{7}$ corresponding to the dimension of the microscopic synthesis space before contraction with the tensor $\mathcal{N}$. Here there is a surjective mapping from the input index to the output index, showing that the matrices $M$ and $\mathcal{N}$ contain more internal parameters than the number of effective five-leg output amplitudes. The matrix product $M\mathcal \times {N}$ gives $T^{(eff\ LQG)}$ with $d^{6}$ components, as required for an effective six-leg tensor.

\subsection{Exact Synthesis of an Effective HaPPY-Type Perfect Tensor}

\subsubsection{Rank Condition and Right Inverse}

The determining microscopic condition is
\begin{equation}
\mathrm{Rank}\, M^{\left( j_{l} \right)} = d^{5}.
\label{eq:s3-src47}
\end{equation}
For fixed spin $j_{l}$, this condition is equivalent to the surjectivity of $M^{\left( j_{l} \right)}$ onto the entire effective five-leg output space. Since the matrix elements depend real-analytically on the nine internal holonomies, it is sufficient to find one nonzero $d^{5}\times d^{5}$ maximal minor to conclude that full row rank holds for Haar-almost every holonomy configuration. In~\hyperref[app:A]{Appendix A}, we prove that $\mathrm{Rank}\, M^{\left( j_{l} \right)}\left( \mathbf{h}_{l} \right) = {d_{l}}^{5}$. We formalize this proof as theorem 1, which establishes the microscopic six-tetrahedron graph used to realize one pentagonal cell of the tessellation $\{5,4\}$, the microscopic transfer map. The matrix $M^{\left( j_{l} \right)}$ is the microscopic LQG transfer map from the central magnetic/intertwiner data to the five effective pentagon legs. We then define the inverse matrix
\begin{equation}
R_{\left( j_{l} \right)} \equiv {M_{\left( j_{l} \right)}}^{\dagger}\left( M_{\left( j_{l} \right)}{M_{\left( j_{l} \right)}}^{\dagger} \right)^{- 1}
\label{eq:s3-src48}
\end{equation}
Since $M_{\left( j_{l} \right)}{M_{\left( j_{l} \right)}}^{\dagger}$ is positive definite and therefore invertible, it satisfies
\begin{equation}
M_{\left( j_{l} \right)}R_{\left( j_{l} \right)} = \mathbb{I}_{{d_{l}}^{5}}
\label{eq:s3-src49}
\end{equation}
This is the standard pseudoinverse construction for a full-row-rank matrix \cite{ref74,ref75}.

\subsubsection{Target Perfect Tensor and AME State}

We consider a perfect tensor $W_{\mathbf{b};\mu}$, which is a six-leg tensor in the HaPPY code having the isometry property for every partition containing up to three input indices:
\begin{equation}
\sum_{}^{}{\overline{W_{\mathbf{b}^{\mathbf{'}};\mu^{'}}}W_{\mathbf{b};\mu}} = |\chi|^{2}\delta_{\mu^{'}\mu}\delta_{\mathbf{b}^{\mathbf{'}}\mathbf{b}}
\label{eq:s3-src50}
\end{equation}
Here $\chi$ is a normalization constant ($\chi \neq 0$). The target we wish to obtain is a matrix representation $T^{(eff\ LQG)}$ that, locally on the pentagon, becomes an effective perfect tensor in the HaPPY code. Three indices are selected for the partition because we consider the tensor above with three input indices and three output indices: the $3\vert3$ splitting is used, from which the isometries for the $1\vert5$ and $2\vert4$ splittings follow. In addition, the tensor $W_{\mathbf{b};\mu}$ must be associated with an AME (absolutely maximally entangled) state, namely a state for which each set of three parties in the $3\vert3$ splitting considered here yields a maximally mixed reduced density matrix, as in \cite{ref4,ref5,ref10} for the HaPPY code:
\begin{equation}
\left. \ \left| \Psi_{W} \right.\  \right\rangle = \sum_{x_{0},\ldots,x_{5}}^{}W_{x_{0}x_{1}x_{2}x_{3}x_{4}x_{5}}\left. \ \left| x_{0}x_{1}x_{2}x_{3}x_{4}x_{5} \right.\  \right\rangle
\label{eq:s3-src51}
\end{equation}
and the reduced density matrix is
\begin{equation}
\rho_{r_{1}r_{2}r_{3}} = \left. \ \left| \Psi_{W} \right.\  \right\rangle\left\langle \left. \ \Psi_{W} \right| \right.\  = \frac{\mathbb{I}_{{d_{l}}^{3}}}{{d_{l}}^{3}}
\label{eq:s3-src52}
\end{equation}
with $\left\{r_{1},r_{2},r_{3}\right\}\subset\left\{0,1,2,3,4,5\right\}$. The state described by Eqs.~\eqref{eq:s3-src51} and \eqref{eq:s3-src52} is denoted AME$(6,d_{l})$. For $j=1/2$ ($d=2$), the standard six-qubit perfect tensor derived from the five-qubit code can be used directly \cite{ref10}. Because we require three-party bipartitions $\left\{r_{1},r_{2},r_{3}\right\}$ for the six-leg tensor, here we take the state $\left.\ \left|\Psi_{W}\right.\ \right\rangle$ as follows:
\begin{equation}
\left. \ \left| \Psi_{W} \right.\  \right\rangle = \frac{1}{{d_{l}}^{\frac{3}{2}}}\sum_{A,B,C}^{}\begin{matrix}
5 \\
\bigotimes_{}^{}1 \\
r = 0
\end{matrix}\left. \ \left| A_{1} + B_{2}t_{r} + C_{3}{t_{r}}^{2} \right.\  \right\rangle
\label{eq:s3-src53}
\end{equation}
With $t_{r}=\left\{0,1,2,3,4,5\right\}$ and $r=\left\{0,1,2,3,4,5\right\}$. The choice of polynomial basis state above is not unique, but it still describes $\left.\ \left|\Psi_{W}\right.\ \right\rangle$ as a basis state for a six-leg tensor satisfying AME$(6,d_{l})$. We then obtain the perfect-tensor representation of the state as
\begin{equation}
W_{x_{0}x_{1}x_{2}x_{3}x_{4}x_{5}} = \frac{1}{{d_{l}}^{\frac{3}{2}}}\sum_{A,B,C}^{}{\prod_{r = 0}^{5}\delta_{x_{r},A + Bt_{r} + C{t_{r}}^{2}}}
\label{eq:s3-src54}
\end{equation}
with $x_{0}=\mu$ and $x_{r}=b_{r}+j_{l}$, together with
\begin{equation}
\delta_{x_{r},A + Bt_{r} + C{t_{r}}^{2}} = \left\{ \begin{array}{r}
1,\ \ x_{r} = A + Bt_{r} + C{t_{r}}^{2} \  \\0,\ \ x_{r} \neq A + Bt_{r} + C{t_{r}}^{2}
\end{array} \right.\
\label{eq:s3-src55}
\end{equation}

For any three positions $r_{1},r_{2},r_{3}$, the three subsystem values are related to $(A,B,C)$ through the Vandermonde matrix
\begin{equation}
\begin{pmatrix}
x_{r_{1}} \\
x_{r_{2}} \\
x_{r_{3}}
\end{pmatrix} = \begin{pmatrix}
1 & t_{r_{1}} & t_{r_{1}}^{2} \\
1 & t_{r_{2}} & t_{r_{2}}^{2} \\
1 & t_{r_{3}} & t_{r_{3}}^{2}
\end{pmatrix}\begin{pmatrix}
A \\
B \\
C
\end{pmatrix}.
\label{eq:s3-src56}
\end{equation}
Its determinant is
\begin{equation}
\left( t_{r_{2}} - t_{r_{1}} \right)\left( t_{r_{3}} - t_{r_{1}} \right)\left( t_{r_{3}} - t_{r_{2}} \right) \neq 0,
\label{eq:s3-src57a}
\end{equation}
so that any three subsystem labels determine $(A,B,C)$ uniquely. Consequently, the reduced state on any three parties is $I_{{d_{l}}^{3}}/{d_{l}}^{3}$, so the construction is AME$\left(6,d_{l}\right)$. This is the Reed--Solomon/MDS mechanism underlying finite-field AME constructions \cite{ref71,ref72,ref73}.

\subsubsection{Isometric Microscopic Preimage}

Arrange the target perfect tensor as a matrix $T\in\mathbb{C}^{d^{5}\times d}$, with the logical index $\mu$ as the column label. Our target is to obtain the effective tensor $T^{(eff\ LQG)}$ in Eq.~\eqref{eq:s3-src32} equivalent to the perfect tensor $W_{\mathbf{b};\mu}$ in Eq.~\eqref{eq:s3-src54}; for this purpose, we propose a modification of the tensor $\mathcal{N}_{a\mu}^{i_{0},\ldots,i_{5}}$ in Eq.~\eqref{eq:s3-src43}. First, define
\begin{equation}
\mathcal{N}_{a\mu}^{i_{0},\ldots,i_{5}} = \chi{\mathcal{N}_{a\mu}^{i_{0},\ldots,i_{5}\ }}^{(0)} + \sum_{\nu = 0}^{d - 1}{\mathcal{N}_{a\nu}^{i_{0},\ldots,i_{5}\ }}^{(1)}\left( S^{\frac{1}{2}} \right)_{\mu\nu}
\label{eq:s3-src57b}
\end{equation}
The quantity ${\mathcal{N}_{a\mu}^{i_{0},\ldots,i_{5}\ }}^{(0)}$ in the first term of Eq.~\eqref{eq:s3-src57b} is written as the minimum-norm preimage defined by
\begin{equation}
{\mathcal{N}_{a\mu}^{i_{0},\ldots,i_{5}\ }}^{(0)} = \sum_{b_{1},\ldots,b_{5}}^{}{R_{{b_{1},\ldots,b_{5};\ a,\ \ i_{0},\ldots,i_{0}}_{\left( j_{l} \right)}}W_{b_{1},\ldots,b_{5};\mu}}
\label{eq:s3-src58a}
\end{equation}
Thus, ${\mathcal{N}_{a\mu}^{i_{0},\ldots,i_{5}\ }}^{(0)}\in \mathrm{rank}\,{M^{\dagger}}_{\left(j_{l}\right)}$ and
\begin{equation}
\Pi_{\mu^{'}\mu} = \sum_{i_{0},\ldots,i_{5}}^{}{\overline{{\mathcal{N}_{\mu^{'}}^{i_{0},\ldots,i_{5}\ }}^{(0)}}{\mathcal{N}_{\mu}^{i_{0},\ldots,i_{5}\ }}^{(0)}}
\label{eq:s3-src59}
\end{equation}
Here $\Pi\geq0$, and the eigenvalues of $\Pi$ are
\begin{equation}
0 \leq \lambda_{1}(\Pi) \leq \ldots \leq \lambda_{d_j}(\Pi) = \lambda_{\max}(\Pi)
\label{eq:s3-src60}
\end{equation}
Here we rescale $\mathcal{N}^{(0)}$ such that $\Pi$ does not exceed the identity, $\chi^{2}\Pi\leq\mathbb{I}_{\Pi}$, with

\begin{equation}
0 < \chi \leq \frac{1}{\sqrt{\lambda_{\max}(\Pi)}}
\label{eq:s3-src61}
\end{equation}

In general, Eqs.~\eqref{eq:s3-src58a} and \eqref{eq:s3-src59} do not yet define an isometric logical embedding. Therefore, the matrix ${\mathcal{N}_{a\nu}^{i_{0},\ldots,i_{5}\ }}^{(1)}$ in Eq.~\eqref{eq:s3-src57b} is defined by ${\mathcal{N}_{a\nu}^{i_{0},\ldots,i_{5}\ }}^{(1)} :\ \mathbb{C}^{d_{l}}\  \rightarrow \ \ker M_{\left(j_{l}\right)}$, so that
\begin{equation}
{\mathrm{rank}\, M^{\dagger}}_{\left( j_{l} \right)} = \left( \ker M_{\left( j_{l} \right)} \right)^{\dagger}
\label{eq:s3-src62}
\end{equation}
hence
\begin{equation}
{{\mathcal{N}_{a\nu}^{i_{0},\ldots,i_{5}\ }}^{(1)} }{\mathcal{N}_{a\mu}^{i_{0},\ldots,i_{5}\ }}^{(0)} = 0
\label{eq:s3-src63}
\end{equation}
Finally, the matrix $S$ is defined by
\begin{equation}
S = \mathbb{I}_{\Pi} - \chi^{2}\Pi \geq 0
\label{eq:s3-src64}
\end{equation}
Using Eqs.~\eqref{eq:s3-src57b}, \eqref{eq:s3-src58a}, \eqref{eq:s3-src59}, and \eqref{eq:s3-src63} above, we obtain
\begin{equation}
\begin{aligned}
\sum_{a,\mathbf{i}}
\overline{\mathcal{N}_{a\mu'}^{i_{0},\ldots,i_{5}}}
\mathcal{N}_{a\mu}^{i_{0},\ldots,i_{5}}
&=
\Bigg(
\chi\,
\overline{
\left(
\mathcal{N}_{a\mu'}^{i_{0},\ldots,i_{5}}
\right)^{(0)}
}
+
\sum_{\nu=0}^{d-1}
\overline{
{\mathcal{N}_{a\nu}^{i_{0},\ldots,i_{5}\ }}^{(1)}
}
\,
\overline{
\left(S^{1/2}\right)_{\mu'\nu}
}
\Bigg)
\\
&\qquad\times
\Bigg(
\chi\,
\left(
\mathcal{N}_{a\mu}^{i_{0},\ldots,i_{5}}
\right)^{(0)}
+
\sum_{\nu=0}^{d-1}
{\mathcal{N}_{a\nu}^{i_{0},\ldots,i_{5}\ }}^{(1)}
\left(S^{1/2}\right)_{\mu\nu}
\Bigg)
\\[2mm]
&=
\chi^{2}\Pi_{\mu'\mu}
+
S_{\mu'\mu}
\\[2mm]
&=
\delta_{\mu'\mu}.
\end{aligned}
\label{eq:s3-src65}
\end{equation}
Thus, the tensor $\mathcal{N}_{a\mu}^{i_{0},\ldots,i_{5}}$ in Eq.~\eqref{eq:s3-src43} is an isometric tensor. Returning to the effective tensor $T^{(eff\ LQG)}$ in Eq.~\eqref{eq:s3-src43}, we obtain
\begin{equation}
\begin{aligned}
T_{b_{1},\ldots,b_{5};\,\mu}^{(eff\,LQG)}
&=
\chi\,
M_{b_{1},\ldots,b_{5};\,\mu,\,i_{0},\ldots,i_{5}}^{(j_{l})}
\left(
\mathcal{N}_{a\mu}^{i_{0},\ldots,i_{5}}
\right)^{(0)}
\\
&\quad+
\sum_{\nu=0}^{d-1}
M_{b_{1},\ldots,b_{5};\,\mu,\,i_{0},\ldots,i_{5}}^{(j_{l})}
\left(
\mathcal{N}_{a\nu}^{i_{0},\ldots,i_{5}}
\right)^{(1)}
\left(
S^{1/2}
\right)_{\mu\nu}.
\end{aligned}
\label{eq:s3-src66}
\end{equation}
with
\begin{equation}
M_{b_{1},\ldots,b_{5};\ \mu,\ \ i_{0},\ldots,i_{5}}^{\left( j_{l} \right)}{\mathcal{N}_{a\mu}^{i_{0},\ldots,i_{5}\ }}^{(0)} = W_{b_{1},\ldots,b_{5};\ \mu}
\label{eq:s3-src67}
\end{equation}

\begin{equation}
M_{b_{1},\ldots,b_{5};\ \mu,\ \ i_{0},\ldots,i_{5}}^{\left( j_{l} \right)}\mathcal{N}_{a\nu}^{i_{0},\ldots,i_{5}\ } = 0
\label{eq:s3-src68}
\end{equation}
therefore
\begin{equation}
T_{b_{1},\ldots,b_{5};\ \mu\ }^{(eff\ LQG)} = \chi W_{b_{1},\ldots,b_{5};\ \mu}
\label{eq:s3-src69}
\end{equation}
We obtain the effective tensor $T^{(eff\ LQG)}$ as a perfect tensor that can be constructed as the tensor for the HaPPY-code tessellation $\{5,4\}$. Here, the orthonormality of the logical embedding $\mathcal{N}^{(0)}$ and the perfectness of the effective tensor $T^{(eff\ LQG)}$ can be satisfied simultaneously. Thus, the logical embedding can be made isometric while the effective cell is proportional to the prescribed perfect tensor.

Recent no-go theorem results constrain the existence of AME or perfect holographic tensors under certain assumptions of local continuous-symmetry covariance \cite{ref68}. The construction above does not contradict those results because $W$ is a coarse-grained effective tensor obtained after contracting several intertwiners and holonomy or Bell kernels; the logical index $\mu$ is introduced through $\mathcal{N}$ and is not required to be an additional bare local $SU(2)$-covariant leg.

\subsection{Global Bulk-to-Boundary Encoding on a Finite Patch of the Tessellation $\{5,4\}$}

Let $\Gamma$ be a finite patch of the tessellation $\{5,4\}$ and $F_{\Gamma}$ the set of its pentagonal cells. Using Eq.~\eqref{eq:s3-src32}, which acts locally on one pentagon, we construct an encoding operator for the entire tessellation $\{5,4\}$, referred to as the global $\{5,4\}$ bulk-to-boundary operator. For a pentagon $f$, if side $r$ of cell $f$ is adjacent to side $s$ of cell $g$, we connect them with normalized holonomy-dressed Bell states (see Eq.~\eqref{eq:s3-src36}) together with $T^{(eff\ LQG)}$ in Eq.~\eqref{eq:s3-src32}, yielding the raw global encoding operator
\begin{equation}
\left(
V_{{\Gamma=\{5,4\}}_{LQG}}^{\mathrm{Raw}}
\right)_{\mathbf{L};\boldsymbol{\mu}}
=
\sum_{\{x\}_{\mathrm{int}}}
\left[
\prod_{f\in F_{\Gamma}}
\left(
T_{f}^{(eff\,LQG)}
\right)_{\{x_{fr}\};\mu_{f}}
\right]
\left[
\prod_{
l=((f,r),(g,s))\in E_{\Gamma}^{\mathrm{int}}
}
L_{x_{gs}x_{fr}}^{(l)}
\right].
\label{eq:global-raw-encoding}
\end{equation}
Its explicit form is as follows:

\begin{equation}
\begin{aligned}
\left(
V_{{\Gamma=\{5,4\}}_{LQG}}^{\mathrm{Raw}}
\right)_{\mathbf{L};\mu}
&=
\sum_{x_{1},\ldots,x_{5}}
\sum_{y_{1},\ldots,y_{5}}
\left(T_{0}\right)_
{x_{1}x_{2}x_{3}x_{4}x_{5};\mu_{0}}
\\
&\quad\times
\left(T_{1}\right)_
{y_{1}b_{12}b_{13}b_{14}b_{15};\mu_{1}}
\\
&\quad\times
\left(T_{2}\right)_
{y_{2}b_{22}b_{23}b_{24}b_{25};\mu_{2}}
\\
&\quad\times
\left(T_{3}\right)_
{y_{3}b_{32}b_{33}b_{34}b_{35};\mu_{3}}
\\
&\quad\times
\left(T_{4}\right)_
{y_{4}b_{42}b_{43}b_{44}b_{45};\mu_{4}}
\\
&\quad\times
\left(T_{5}\right)_
{y_{5}b_{52}b_{53}b_{54}b_{55};\mu_{5}}
\\
&\quad\times
L_{\{y_{1}\}x_{1}}^{(01)}
L_{\{y_{2}\}x_{2}}^{(02)}
L_{\{y_{3}\}x_{3}}^{(03)}
L_{\{y_{4}\}x_{4}}^{(04)}
L_{\{y_{5}\}x_{5}}^{(05)} .
\end{aligned}
\label{eq:S3-src92}
\end{equation}
The index $\mathbf{x}=\left(x_{1},x_{2},x_{3},x_{4},x_{5}\right)$ is the input index for the five effective network indices associated with the five sides of the pentagon, each carrying spin $j=1/2$, or $x_{r}\in\left\{-\frac{1}{2},\frac{1}{2}\right\},\ r=1,\ldots,5$. The mapping $x_{r}\rightarrow y_{r}$ applies to the neighboring pentagon, so that the index $y$ denotes the output index; $\mathbf{\mu}=\left(\mu_{0},\mu_{1},\mu_{2},\mu_{3},\mu_{4},\mu_{5}\right)$ is the bulk logical index, and $\mathbf{b}=\left\{b_{12}b_{13}b_{14}b_{15};\ldots;b_{52}b_{53}b_{54}b_{55}\right\}$ is the collective boundary index. The normalized holonomy-dressed Bell state is written as
\begin{equation}
L_{y_{r}x_{r}}^{(0r)} = \frac{( - 1)^{j_{l} + x_{r}}}{\sqrt{2j_{l} + 1}}D_{y_{r} - x_{r}}^{\left( j_{l} \right)}\left( h_{0r} \right)\ ,\ \ \ \ \ \ \ r = 1,\ldots,\ 5
\label{eq:s3-src93}
\end{equation}

Equations ~\eqref{eq:s3-src32} and ~\eqref{eq:S3-src92} involve two uses of holonomy, but they have different roles. In Eq.~\eqref{eq:s3-src32}, the holonomy can modify the interference among microscopic paths and therefore can change the singular spectrum of $T^{(eff\ LQG)}$. By contrast, in Eq.~\eqref{eq:S3-src92}, the holonomy-dressed Bell spin-network links connecting pentagons do not spoil the global isometry because they produce a scalar normalization but do not generate spectral anisotropy, as shown by the normalization conditions (see Eqs.~\eqref{eq:s3-src37}--\eqref{eq:s3-src38}). Consequently, Eq.~\eqref{eq:S3-src92} carries only the parallel-transport geometry between cells and does not generate anisotropy within $T^{(eff\ LQG)}$.

Global isometry requires an additional condition beyond local perfectness: the finite patch must possess an acyclic radial orientation such that each pentagonal cell is used with no more than two incoming inter-cell legs. Together with the logical leg, the number of inputs of each cell is at most three, which remains within the isometric range of a six-leg perfect tensor. This is the standard radial condition in the HaPPY pentagon code \cite{ref10}.

To examine global isometry for the entire tessellation tiling $\{5,4\}$, consider the graph $\Gamma\subset\{5,4\}$. Let $P_{\Gamma}=\left\{f_{1},\ldots,f_{P}\right\}$ be the set of the first through the $P$th pentagons in the graph $\Gamma$. The bulk logical-qubit Hilbert space is
\begin{equation}
\mathcal{H}_{bulk} = \bigotimes_{f \in P_{\Gamma}}\mathcal{H}_{\mu_{f}}
\label{eq:s3-src94}
\end{equation}
and the boundary Hilbert space is

\begin{equation}
\mathcal{H}_{\partial\Gamma} = \bigotimes_{l = 1}^{P_{\partial}}\mathcal{H}_{L_{l}}
\label{eq:s3-src95}
\end{equation}
Here $\mathbf{\mu}=\left\{\mu_{f_{1}},\ldots,\mu_{f_{p}}\right\}$ and $\mathbf{L}=\left\{L_{1},\ldots,L_{P_{\partial}}\right\}$ are, respectively, the collective bulk index and boundary index. Next, define radial layers for each patch in the graph:
\begin{equation}
\mathcal{L}_{1},\mathcal{L}_{2},\ldots,\ \mathcal{L}_{n}
\label{eq:s3-src96}
\end{equation}
where $\mathcal{L}_{1},\ldots,\mathcal{L}_{n}$ are the radial shells and the frontier after layer $n$ is $\sum_{n}$; the collective frontier index is
\begin{equation}
\mathbf{x}^{(n)} = \left( x_{1}^{(n)},\ldots.,x_{n}^{(n)} \right)
\label{eq:s3-src97}
\end{equation}
and the logical indices on shell $n$ are
\begin{equation}
\mathbf{\mu}^{(n)} = \left\{ \mu_{f_{1}}|f \in \mathcal{L}_{n} \right\}
\label{eq:s3-src98}
\end{equation}

Define the normalized shell map as
\begin{equation}
{\mathcal{K}}_{(n)_{\mathbf{L};\ \mathbf{x}^{(n - 1)},\mathbf{\mu}^{(n)}}} = \sum_{\mathbf{L}}^{}{\prod_{f \in \mathcal{L}_{n}}^{}{\widetilde{T}}_{f\left\{ \mathbf{L} \right\};\mu_{f}}} \times \prod_{l \subset \mathcal{L}_{n}}^{}U_{m_{w}m_{v}}^{(l)}\prod_{l \in \left( \sum_{n - 1}^{}{,\mathcal{L}_{n}} \right)}^{}U_{m_{w}m_{v}}^{(l)}
\label{eq:s3-src99}
\end{equation}
Each shell defines an isometry with the mapping
\begin{equation}
\mathcal{K}_{n}:\mathcal{H}_{\mathbf{x}^{(n - 1)}} \otimes \mathcal{H}_{\mathbf{\mu}^{(n)}} \rightarrow \mathcal{H}_{\mathbf{x}^{(n)}}
\label{eq:s3-src100}
\end{equation}
where $T_{f\left\{\mathbf{x}^{(n)}\right\};\mu_{f}}$ and $U_{m_{w}m_{v}}^{(l)}$ in the normalized equation use
\begin{equation}
{\widetilde{T}}_{f\left\{ \mathbf{x}^{(n)} \right\};\mu_{f}} = \frac{1}{\sqrt{c_{f,\mu_{f}}}}T_{f\left\{ \mathbf{x}^{(n)} \right\};\mu_{f}}
\label{eq:s3-src101}
\end{equation}

\begin{equation}
U_{m_{w}m_{v}}^{(l)} = \frac{1}{\sqrt{2j_{l} + 1}}L_{m_{w}m_{v}}^{(l)}
\label{eq:s3-src102}
\end{equation}
where $c_{f,\mu_{f}}$ is a scalar. Then, using Eqs.~\eqref{eq:s3-src37} and \eqref{eq:s3-src38}, the normalized shell map satisfies

\begin{equation}
\sum_{\mathbf{L}}^{}{\overline{{\mathcal{K}}_{n_{\mathbf{L};{\mathbf{x}^{\mathbf{'}}}^{(n - 1)},{\mathbf{\mu}^{\mathbf{'}}}^{(n)}}}}{\mathcal{K}}_{n_{\mathbf{L};\mathbf{x}^{(n - 1)},\mathbf{\mu}^{(n)}}}} = \delta_{{\mathbf{x}^{\mathbf{'}}}^{(n - 1)},\mathbf{x}^{(n - 1)}}\delta_{{\mathbf{\mu}^{\mathbf{'}}}^{(n)},\mathbf{\mu}^{(n)}}
\label{eq:s3-src103}
\end{equation}

We may then write Eq.~\eqref{eq:S3-src92} for the entire layer $(n)$ as
\begin{equation}
\begin{aligned}
\left(
V_{{\Gamma=\{5,4\}}_{LQG}}
\right)_{\mathbf{L};\mu}
&=
\frac{1}{\kappa_{\Gamma}}
V_{{\Gamma=\{5,4\}}_{LQG}}^{\mathrm{Raw}}
\\[1mm]
&=
\frac{1}{\kappa_{\Gamma}}
\sum_{\mathbf{x}^{(1)},\ldots,\mathbf{x}^{(n-1)}}
\mathcal{K}_{(L)\mathbf;
\mathbf{x}^{(n-1)},\mu^{(n)}}
\\
&\quad\times
\mathcal{K}_{(L-1)\mathbf{x}^{(n-1)};
\mathbf{x}^{(n-2)},\mu^{(n-1)}}
\\
&\quad\times \cdots
\\
&\quad\times
\mathcal{K}_{(2)\mathbf{x}^{(2)};
\mathbf{x}^{(1)},\mu^{(2)}}
\\
&\quad\times
\mathcal{K}_{(1)\mathbf{x}^{(1)};
\mu^{(1)}} .
\end{aligned}
\label{eq:S3-src104}
\end{equation}
with $\kappa_{\Gamma}=\left\lbrack\prod_{f}^{}\sqrt{c_{f,\mu_{f}}}\right\rbrack\left\lbrack\prod_{l\in\left(\sum_{n-1}^{}{,\mathcal{L}_{n}}\right)}^{}\sqrt{2j_{l}+1}\right\rbrack$; and we obtain
\begin{equation}
\left( {V}_{{\Gamma = \{ 5,4\}}_{LQG}}^{\dagger}{V}_{{\Gamma = \{ 5,4\}}_{LQG}} \right)_{\mathbf{\mu}^{\mathbf{'}\left( \mathbf{1} \right)}\mathbf{\ldots.\ }\mathbf{\mu}^{\mathbf{'}\left( \mathbf{n} \right)}\mathbf{;}\mathbf{\mu}^{\left( \mathbf{1} \right)}\mathbf{\ldots.}\mathbf{\mu}^{\left( \mathbf{n} \right)}} = \prod_{n - 1}^{l}\delta_{\mathbf{\mu}^{\mathbf{'}\left( \mathbf{n} \right)},\mathbf{\mu}^{\left( \mathbf{n} \right)}}
\label{eq:s3-src105}
\end{equation}
The result above proves the global isometry property of the global encoding operator constructed from the entanglement of quantum tetrahedra by holonomy-dressed spin-network states. In addition, Eq.~\eqref{eq:s3-src105} shows how a finite Bell--LQG tessellation exactly defines an exact HaPPY-like bulk-to-boundary isometry, subject to the rank, synthesis, and radial-admissibility conditions stated explicitly above.

Equation ~\eqref{eq:S3-src104} is the operator we propose for encoding from the bulk to the boundary on the holographic tessellation $\{5,4\}$, based on the holonomy-dressed Bell spin-network-state formulation constructed previously. The tensor $V_{\Gamma}$ for the entire hyperbolic tessellation $\{5,4\}$ can be interpreted, first, as a coarse-grained bulk-to-boundary mapping for the entire hyperbolic patch:
\begin{equation}
{V}_{\Gamma_{LQG}}\ :\ \bigotimes_{f}\left( \mathcal{H}_{\mu} \right)_{f\ } \rightarrow \ \bigotimes_{l \in \partial\Gamma}V_{j_{l}}
\label{eq:s3-src106}
\end{equation}
Second, the presence of holonomy-dressed spin-network states in ${V}_{\Gamma_{LQG}}$ provides a mechanism for quantum gluing between adjacent effective pentagons. Third, the presence of $T_{eff\ LQG}$ in $V_{\Gamma_{LQG}}$ means that ${V}_{\Gamma_{LQG}}$ automatically stores the quantum geometry within one effective pentagon: the intertwiner states, the entanglement among the six tetrahedra, and the microscopic holonomies. Our entire construction for obtaining an effective perfect tensor refers to the required condition $rank\ M^{\left(j_{l}\right)}\left(\mathbf{h}_{l}\right)={d_{l}}^{5}$ in Eq.~\eqref{eq:s3-src47}, which is proved in Appendix A.

\subsection{Discrete Quantum RT and Generalized-Entropy Relation}

Consider a boundary subregion $A\subset\partial\Gamma$ and its complement $\overline{A}$. Choose a discrete cut $\gamma_{c}$ separating $A$ and $\overline{A}$. Suppose that the cut intersects $L_{\gamma_{c}}$ internal links carrying holonomy-dressed Bell states. For each cut edge $e$, define $x_{e_{A}}$ (the half-edge on the $A$ side) and $y_{e_{\overline{A}}}$ (the half-edge on the $\overline{A}$ side), together with the collective cut indices
\begin{equation}
\mathbf{x} = \left( x_{e_{A}} \right)_{e_{A} \in \gamma_{c}}\ ,\ \ \ \ \ \ \ \ \ \mathbf{y =}\left( y_{e_{\overline{A}}} \right)_{e_{A} \in \gamma_{c}}
\label{eq:s3-src107}
\end{equation}
For one link $(l\in e)$, the holonomy-dressed spin-network state is
\begin{equation}
L_{y_{l_{\overline{A}}}\ x_{l_{A}}}^{(l)} = \frac{1}{\sqrt{2j_{l} + 1}}U_{y_{l_{\overline{A}}}\ x_{l_{A}}}^{(l)}
\label{eq:s3-src108}
\end{equation}
For the entire cut, we have
\begin{equation}
L_{\gamma_{c}\ \mathbf{y;x}}^{(l)} = \prod_{l \in \gamma_{c}}^{}{\frac{1}{\sqrt{2j_{l} + 1}}U_{y_{l_{\overline{A}}}\ x_{l_{A}}}^{(l)}}
\label{eq:s3-src109}
\end{equation}
The reduced density matrix for subregion $A$ is
\begin{equation}
\left( \rho_{\gamma_{c},A} \right)_{\mathbf{x;}\mathbf{x}^{\mathbf{'}}} = \prod_{l \in \gamma_{c}}^{}\lbrack \frac{1}{{(2j_{l} + 1)}}\sum_{y_{e_{\overline{A}}}}^{}{U_{y_{l_{\overline{A}}}\ x_{l_{A}}}^{(l)}\overline{U_{y_{l_{\overline{A}}}\ x_{l_{A}}}^{(l)}}} = \prod_{l \in \gamma_{c}}^{}\frac{\delta_{\mathbf{x;}\mathbf{x}^{\mathbf{'}}}}{{(2j_{l} + 1)}}
\label{eq:s3-src110}
\end{equation}

The entanglement entropy of the cut region can therefore be calculated using the von Neumann entropy formula:
\begin{equation}
\begin{aligned}
S_{\mathrm{cut}}\!\left(\gamma_{c}\right)
&=
-\operatorname{Tr}
\left(
\rho_{\gamma_{c},A}
\log \rho_{\gamma_{c},A}
\right)
\\[2mm]
&=
-
\prod_{l\in\gamma_{c}}
{(2j_{l}+1)}
\,
\prod_{l\in\gamma_{c}}
\frac{\mathbb{I}_{\gamma_{c}}}
{{(2j_{l}+1)}}
\,
\log
\left[
\prod_{l\in\gamma_{c}}
\frac{\mathbb{I}_{\gamma_{c}}}
{{2j_{l}+1}}
\right]
\\[2mm]
&=
\sum_{l\in\gamma_{c}}
\log{(2j_{l}+1)}
\\[2mm]
&=
\left|\gamma_{c}\right|
\log{(2j_{l}+1)}.
\end{aligned}
\label{eq:s3-src111}
\end{equation}
References \cite{ref31,ref32,ref34,ref35,ref38,ref45,ref46,ref47,ref48} obtain the same structure: for $m$ spin-$j$ links, the boundary entropy acquires the additional term $m\log\sqrt{2j_{l}+1}$ in addition to the bulk entropy. For $j_{l}=\frac{1}{2}$, the entropy becomes
\begin{equation}
S_{cut}\left( \gamma_{c} \right) = \left| \gamma_{c} \right|\log 2
\label{eq:s3-src112}
\end{equation}
which is exactly the same as the discrete entropy formula for the pentagon \cite{ref4,ref6,ref10}.

Next, we consider the bulk regions on both sides of the cut. The cut $\gamma_{c}$ divides the bulk logical indices into two subregions. We define the bulk density matrix as follows:
\begin{equation}
\left( \rho_{bulk} \right)_{\mathbf{\mu}_{a},\mathbf{\mu}_{\overline{a}};\mathbf{\nu}_{a}\mathbf{\nu}_{\overline{a}}}
\label{eq:s3-src113}
\end{equation}
The indices $\mathbf{\mu}=\left(\mathbf{\mu}_{a},\mathbf{\mu}_{\overline{a}}\right)$ and $\mathbf{\nu}=\left(\mathbf{\nu}_{a},\mathbf{\nu}_{\overline{a}}\right)$ are pairs of collective logical indices, while the indices $a$ and $\overline{a}$ are, respectively, the bulk logical indices in subregions $A$ and $\overline{A}$. The reduced density matrix for subregion $A$ is therefore
\begin{equation}
\left( \rho_{a} \right)_{\mathbf{\mu}_{a},;\mathbf{\nu}_{a}} = \sum_{\mathbf{\mu}_{\overline{a}},\ \mathbf{\nu}_{\overline{a}}}^{}\left( \rho_{bulk} \right)_{\mathbf{\mu}_{a},\mathbf{\mu}_{\overline{a}};\mathbf{\nu}_{a}\mathbf{\nu}_{\overline{a}}}
\label{eq:s3-src114}
\end{equation}
The global encoding operator (see Eq.~\eqref{eq:S3-src104}) maps the bulk density matrix to the boundary density matrix:
\begin{equation}
\rho_{\partial} = {V}_{{\Gamma = \{ 5,4\}}_{LQG}}\rho_{bulk}{V}_{{\Gamma = \{ 5,4\}}_{LQG}}^{\dagger}
\label{eq:s3-src115}
\end{equation}
The bulk and boundary Hilbert spaces are, respectively,
\begin{equation}
\mathcal{H}_{bulk} = \mathcal{H}_{a} \otimes \mathcal{H}_{\overline{a}}\ \ \ \ ,\ \ \ \ \ \mathcal{H}_{\partial} = \mathcal{H}_{A} \otimes \mathcal{H}_{\overline{A}}
\label{eq:s3-src116}
\end{equation}
Using the basis $\left|\left.\ \mu_{a}\right\rangle\right.\in\mathcal{H}_{a}$ and $\left|\left.\ \mu_{\overline{a}}\right\rangle\right.\in\mathcal{H}_{\overline{a}}$, the total bulk basis is $\left|\left.\ \mu_{a},\mu_{\overline{a}}\right\rangle\right.=\left|\left.\ \mu_{a}\right\rangle\right.\otimes\left|\left.\ \mu_{\overline{a}}\right\rangle\right.$, and similarly for the indices $\nu_{a}$ and $\nu_{\overline{a}}$. We assume that the tensor networks on both sides of $\gamma$ can be reduced to isometries:
\begin{equation}
U_{A}:\mathcal{H}_{a_{\gamma}} \otimes \mathcal{H}_{\gamma,A} \rightarrow \mathcal{H}_{A},\quad\quad U_{\bar{A}}:\mathcal{H}_{{\bar{a}}_{\gamma}} \otimes \mathcal{H}_{\gamma,\bar{A}} \rightarrow \mathcal{H}_{\bar{A}}.
\label{eq:s3-src117}
\end{equation}
We write the bulk density matrix as
\begin{equation}
\rho_{bulk} = \sum_{\mu_{a},\mu_{\overline{a}};\nu_{a}\nu_{\overline{a}}}^{}\left( \rho_{bulk} \right)_{\mu_{a},\mu_{\overline{a}};\nu_{a}\nu_{\overline{a}}}\left| \left. \ \mu_{a},\mu_{\overline{a}} \right\rangle \right.\ \left\langle \left. \ \nu_{a},\nu_{\overline{a}} \right| \right.\
\label{eq:s3-src118}
\end{equation}
For each edge $l\in\gamma_{c}$, using Eq.~\eqref{eq:S3-src104} and Eq.~\eqref{eq:s3-src115}, we obtain
\begin{equation}
\begin{aligned}
\rho_{\partial}
&=
\sum_{\substack{
\mu_{a},\mu_{\overline{a}};
\nu_{a}\nu_{\overline{a}}\\
\mathbf{x},\mathbf{y},
\mathbf{x}',\mathbf{y}'
}}
\Bigg[
\left(
\widetilde{T}_{A}
\right)_{\mathbf{A};\mu_{a};\mathbf{x}}
\left(
\widetilde{T}_{\overline{A}}
\right)_{\overline{\mathbf{A}};
\mu_{\overline{a}};\mathbf{x}}
\left(
U_{\gamma_{c}}
\right)_{\mathbf{y};\mathbf{x}}
\Bigg]
\\
&\qquad\times
\rho_{bulk}
\Bigg[
\overline{
\left(
U_{\gamma_{c}}
\right)_{\mathbf{y}';\mathbf{x}'}
}
\;
\overline{
\left(
\widetilde{T}_{\overline{A}}
\right)_{\overline{\mathbf{A}'};
\nu_{\overline{a}};\mathbf{x}'}
}
\\
&\hspace{40mm}\times
\overline{
\left(
\widetilde{T}_{A}
\right)_{\mathbf{A}';\nu_{a};\mathbf{x}'}
}
\Bigg]
\\[3mm]
&=
\sum_{\substack{
\mu_{a};\nu_{a}\\
\mathbf{x},\mathbf{x}'
}}
\left(
\widetilde{T}_{A}
\right)_{\mathbf{A};\mu_{a};\mathbf{x}}
\Bigg[
\left(
\rho_{a}
\right)_{\boldsymbol{\mu}_{a};
\boldsymbol{\nu}_{a}}
\otimes
\prod_{l\in\gamma_{c}}
\frac{
\delta_{\mathbf{x}\mathbf{x}'}
}{{(2j_{l}+1)}
}
\Bigg]
\\
&\qquad\times
\overline{
\left(
\widetilde{T}_{A}
\right)_{\mathbf{A}';\nu_{a};\mathbf{x}'}
}.
\end{aligned}
\label{eq:s3-src119}
\end{equation}

An isometry preserves the nonzero spectrum, while the entropy of a tensor product is additive. Therefore, using the von Neumann entanglement-entropy formula, we obtain
\begin{equation}
S(A) = S_{bulk}\left( \rho_{a} \right) + \sum_{l \in \gamma_{c}}^{}{\log{(2j_{l} + 1)}}
\label{eq:s3-src120}
\end{equation}
The equation above is the exact discrete quantum Ryu--Takayanagi formula for a cut-isometric $\gamma_{c}$. Next, define the generalized entropy \cite{ref10,ref11}:
\begin{equation}
S_{gen}\left( \gamma_{c} \right) = S_{bulk}\left( a_{\gamma_{c}} \right) + S_{cut}\left( \gamma_{c} \right)
\label{eq:s3-src121}
\end{equation}
Comparing the cut entropy equation with $S(A)$ (two-sided cut-isometric cut analogies), the relation can be written as
\begin{equation}
S_{gen}\left( \gamma_{c} \right) = S_{bulk}\left( a_{\gamma_{c}} \right) + \left| \gamma_{c} \right|\log{(2j_{l} + 1)}
\label{eq:s3-src123}
\end{equation}
If there exists a minimal cut $(\gamma_{c})$ homologous to subregion $A$, separating the bulk into $a_{\gamma_{c}}$ and its complement, allowing a two-sided isometric reduction, and minimizing the generalized entropy among admissible cuts, then
\begin{equation}
S(A) = \min_{\gamma_{c}\sim A}\left\lbrack S_{gen}\left( \gamma_{c} \right) \right\rbrack
\label{eq:s3-src124}
\end{equation}
and we obtain
\begin{equation}
S(A) = \min_{\gamma_{c}\sim A}\left\lbrack S_{bulk}\left( a_{\gamma_{c}} \right) + \left| \gamma_{c} \right|\log{(2j_{l} + 1)} \right\rbrack
\label{eq:s3-src125}
\end{equation}
This is the relation between the entanglement entropy of subregion $A$ and the minimal generalized entropy of the model of quantum-tetrahedron entanglement structures in LQG. This is the appropriate form of the discrete quantum-RT/generalized-entropy relation. The structure parallels the minimal-cut mechanism in HaPPY \cite{ref10,ref11} and the bulk-entropy correction in quantum RT and random tensor networks \cite{ref12,ref14,ref45,ref46,ref47,ref48,ref49,ref50}. The formula must not be claimed for an arbitrary cut without the hypotheses of two-sided isometry and minimality.

\subsubsection{Relation to the LQG Area Spectrum}\label{Relation to the LQG Area Spectrum}

The LQG area-operator formula carried by a cut is
\begin{equation}
\widehat{A}h_{\left( j_{l} \right)} = 8\pi l_{Pl}^{2}\gamma_{BI}\sqrt{j_{l}\left( j_{l} + 1 \right)}h_{\left( j_{l} \right)}
\label{eq:s3-src126}
\end{equation}
where $\widehat{A}=\sqrt{{\widehat{F}}_{S}^{i}{\widehat{F}}_{S}^{i}}$ is the area operator, ${\widehat{F}}_{S}^{i}$ is the $SU(2)$ flux operator, $l_{Pl}$ is the Planck length, and $\gamma_{BI}\in\mathbb{R}_{+}$ is the Barbero--Immirzi parameter \cite{ref19,ref21,ref61}. In the large-spin or semiclassical limit $(j_{l}\gg1)$, the Ryu--Takayanagi entropy equation is obtained $(c=\hslash=G=1)$:
\begin{equation}
S(A) = S_{bulk}\left( \rho_{a} \right) + \frac{A}{4l_{Pl}^{2}A_{j_{l}}}
\label{eq:s3-src127}
\end{equation}
This is the quantum RT formula. If $j_{l}=1/2$ is taken, then $\frac{A}{4l_{Pl}^{2}A_{j_{l}=1/2}}\rightarrow\frac{A}{4l_{Pl}^{2}}$, with $\gamma_{BI}=\frac{\log2}{\sqrt{3\pi}}$ in this single-spin counting convention. This particular matching must not be presented as a universal derivation of the Bekenstein--Hawking coefficient. The most robust result of this tensor-network construction is the discrete bond term $\sum_{l}^{}\log\left(2j_{l}+1\right)$.

Here we have shown that our formulation succeeds in constructing the HaPPY code from the quantum-tetrahedron entanglement structure, both for the effective tensor structure constructed to become a perfect tensor (see Eq.~\eqref{eq:s3-src69}) for the case of one pentagon and for the global encoding operator over the entire surface of the tessellation $\{5,4\}$, while also obtaining the generalized entropy and discrete quantum RT entropy formula. We next generalize this formulation to a family of holographic tensor networks on tessellations $\{p,q\}$ with specified $p$ and $q$.

\subsection{Generalization to Hyperbolic Tessellations $\{p,q\}\in \mathcal{T}_{\mathrm{arch}}$}

In this section, we generalize the previous microscopic holonomy dressed Bell spin-network-LQG construction on the pentagonal $\{5,4\}$ cell to a class of specific hyperbolic tessellations $\{p,q\}$ within the family of HaPPY-code tessellations. The purpose of this generalization is not merely to extend the tiling shape, but to explicitly isolate the four structural layers required for the construction to function as a holographic quantum code: (i) the geometric hyperbolic structure conditions of the tessellation and the combinatorics of the microscopic four-valent quantum-tetrahedron graph; and (ii) the admissible microscopic transfer map conditions relating to the algebraic structure so that the microscopic transfer map can synthesize effective perfect tensors, and the global structure such that the effective tensors can be oriented as compositions of isometries within a finite patch. This separation is crucial because hyperbolicity itself does not imply perfectness, full-row rank, or a global bulk-to-boundary isometry.

The microscopic foundation relies on Bell-network states and the tensor-network representation of spin networks. The Bell-network construction provides inter-node entanglement supporting gluing/vector-geometric correlations between quantum polyhedra ~\cite{ref31,ref32}, while the explicit representation of $\mathrm{SU}(2)$ holonomy, invariant metric, and four-valent intertwiner as tensor-network ingredients follows the formulation in Refs.~\cite{ref36,ref38}. Perfect tensors and the global encoding mechanism subsequently refer to the structure of the HaPPY code~\cite{ref10}.

\subsubsection{Hyperbolic geometry conditions}
A regular Schläfli tessellation $\{p,q\}$ consists of regular $p$-gons with exactly $q$ polygons meeting at each vertex. Such a tessellation corresponds to hyperbolic space if the inequality in Eq.~\eqref{eq:hyperbolic-condition} holds. Since $p > 2$, dividing by $p-2$ preserves the direction of the inequality, yielding
\begin{equation}
q > \frac{2p}{p-2}
\end{equation}
This condition only establishes that the tiling resides on a hyperbolic geometry. However, we find that this condition does not yet guarantee that the local tensor can be used as an isometry, nor does it guarantee the existence of a radial orientation that makes the entire network a bulk-to-boundary isometry, as shown in the previous section.

To generalize the six-tetrahedron cell at $p=5$, one effective $p$-gonal cell is constructed from $P_{\mathrm{tet}} = p+1$ four-valent quantum tetrahedra. For each tetrahedron, one magnetic leg is left open, and the other three magnetic legs are used for internal holonomy-dressed Bell contractions. Since each tetrahedron is 4-valent, the total number of magnetic half-links is
\begin{equation}
P_{\mathrm{half}} = 4(p+1)
\label{eq 4.1}
\end{equation}
The number of preserved open half-links is
\begin{equation}
P_{\mathrm{open}} = p+1
\label{eq 4.2}
\end{equation}
Therefore, the number of internal half-links is
\begin{equation}
P_{\mathrm{half,int}} = P_{\mathrm{half}} - P_{\mathrm{open}} = 4(p+1) - (p+1) = 3(p+1)
\label{eq 4.3}
\end{equation}
One internal link utilizes two half-links. Consequently, the number of internal links must satisfy
\begin{equation}
L_{\mathrm{int}} = \frac{3(p+1)}{2}
\label{eq 4.4}
\end{equation}
For $L_{\mathrm{int}}$ to be an integer, $p+1$ must be even. In other words, $p = 2k-1$. In the family considered here, we take $k \ge 3$, so that $p \ge 5$. With this parameterization, the condition $k \ge 3$ represents the choice of family relevant to generalizing the pentagonal HaPPY construction, rather than an absolute logical consequence of four-valency. The case $p=3$ can be treated separately, but is not included in the family used in the formulation here.

The most natural choice of a 3-regular graph on $2k$ vertices is the $k$-prism graph, where vertex labels are written as
\begin{equation}
V_{\mathrm{Prism}_k} = \{u_0, \dots, u_{k-1}, v_0, \dots, v_{k-1}\}
\label{eq 4.5}
\end{equation}
with the index $r$ understood modulo $k$. The internal link set can be written as
\begin{equation}
L_{\mathrm{Prism}_n} = \{(u_r, u_{r+1}), (v_r, v_{r+1}), (u_r, v_r)\}_{r=0}^{k-1}
\label{eq 4.6}
\end{equation}
There are $k$ links on the upper cycle, $k$ links on the lower cycle, and $k$ rungs, so that
\begin{equation}
|L_{\mathrm{Prism}_k}| = 3k = L_{\mathrm{int}}
\label{eq 4.7}
\end{equation}
where each vertex has three internal incident links. After one open magnetic leg is added, the total degree of each node becomes four, exactly like the four-valent quantum tetrahedron in LQG. For $k=3$, we obtain $p=5$, six tetrahedra, and nine internal links, thereby recovering the $\{5,4\}$ case. We distinguish between the connectivity graph used in the proof and the illustrative embedding consisting of five peripheral tetrahedra and one tetrahedron at the center of the pentagon.

By combining the parity and hyperbolicity conditions, we define the architectural family for the hyperbolic tessellation structure in the constructed model, namely:
\begin{equation}
\mathcal{T}_{\mathrm{arch}} = \left\{ \{p,q\} \;\middle|\; p = 2k-1, \; k \ge 3, \; q \in \mathbb{Z}_{\ge 3}, \; (p-2)(q-2) > 4 \right\}
\label{eq 4.8}
\end{equation}
Membership in $\mathcal{T}_{\mathrm{arch}}$ means that the hyperbolic tiling geometry and the microscopic cell half-link counting are compatible. However, this condition is not yet sufficient to conclude that a global holographic code has been obtained.

\subsubsection{Hilbert spaces, intertwiner labels, and holonomy-dressed Bell links}
As in the discussion of the $\{5,4\}$ tiling tessellation for the HaPPY code, we review here the Hilbert space formulation for intertwiners and holonomy-dressed Bell links. Take a homogeneous spin on all microscopic links,
\begin{equation}
j = j_l, \qquad d \equiv d_{j_l} = 2j_l + 1
\label{eq 4.9}
\end{equation}
At each four-valent equal-spin node, the intertwiner space is
\begin{equation}
\mathcal{I}_{j_l} = \mathrm{Inv}_{\mathrm{SU}(2)}\!\left( V_{j_l}^{\otimes 4} \right)
\label{eq 4.10}
\end{equation}
In the standard recoupling basis, the intermediate spin $i$ takes values $i = 0, 1, \dots, 2j_l$, so that
\begin{equation}
\dim \mathcal{I}_{j_l} = 2j_l + 1 = d_{j_l}
\label{eq 4.11}
\end{equation}
An intertwiner basis tensor $(\iota_i)_{m_1 m_2 m_3 m_4}$ is written as in Eq.~\eqref{eq:s3-src35}. For each oriented internal link $l = (s_l, t_l)$, define the invariant metric $g_{mn}^{(j)} = (-1)^{j-m}\delta_{m,-n}$ and the spin-$j$ Wigner matrix $D^{(j_l)}(h_l)$, with $h_l \in \mathrm{SU}(2)$. We use the notation
\begin{equation}
U_l^{(j)} \equiv D^{(j_l)}(h_l) g^{(j_l)}
\label{eq 4.12}
\end{equation}
and the normalized holonomy-dressed Bell spin-network:
\begin{equation}
L_l(m_t, m_s) = \frac{1}{\sqrt{d_{j_l}}} U_l^{(j_l)}(m_t, m_s)
\label{eq 4.13}
\end{equation}
Since $D^{(j_l)}(h_l)$ is unitary and $(g^{(j_l)})^\dagger g^{(j_l)} = \mathbb{I}_{d_{j_l}}$, the normalization condition in Eq.~\eqref{eq:s3-src37} is satisfied. In open indices,
\begin{equation}
\sum_{m_t} \left( U_l^{(j_l)} \right)^\dagger(m_s', m_t) U_l^{(j_l)}(m_t, m_s) = \delta_{m_s', m_s}
\label{eq 4.14}
\end{equation}
For normalized $L_l$,
\begin{equation}
\sum_{m_t} L_l^\dagger(m_s', m_t) L_l(m_t, m_s) = \frac{1}{d_{j_l}} \delta_{m_s', m_s}
\label{eq 4.15}
\end{equation}
Therefore, one side of a holonomy-dressed Bell pair has a reduced state $\mathbb{I}_{d_{j_l}} / d_{j_l}$, and the von Neumann entropy of one cut link is (see Eq.~\eqref{eq:s3-src111}:
\begin{equation}
S_{\mathrm{cut}, l}(j_l) = \log d_{j_l} = \log(2j_l + 1) 
\label{eq:144}
\end{equation}
The holonomy $D^{(j_l)}(h_l)$ does not change the Schmidt spectrum; it merely performs a unitary parallel transport on one of the endpoint Hilbert spaces. This structure represents a synthesis of Bell gluing ~\cite{ref31,ref32} and explicit holonomy/intertwiner tensor-network representations ~\cite{ref38}.

\subsubsection{Microscopic transfer map with open indices}
Consider a vertex and choose one distinguished vertex, e.g., $u_0$, and write its open magnetic index as $a$. The remaining $p = 2n-1$ open magnetic legs are labeled $b = (b_1, \dots, b_p)$. The intertwiner labels of all $2n = p+1$ tetrahedra are collected as $i = (i_{u_0}, \dots, i_{u_{n-1}}, i_{v_0}, \dots, i_{v_{n-1}})$. For each vertex $v$, write the three internal magnetic indices as $m_v^{(e_{1v})}, m_v^{(e_{2v})}, m_v^{(e_{3v})}$. We obtain the microscopic transfer tensor defined by the contraction:
\begin{equation}
M_{p,j_l}(b_1 \dots b_p; a, i) = \sum_{\{m_v^{(l)}\}} \prod_{v \in V_{\mathrm{Prism}_n}} (\iota_{i_v})_{x_v, m_v^{(e_{1v})}, m_v^{(e_{2v})}, m_v^{(e_{3v})}} \prod_{l = (v,w) \in L_{\mathrm{Prism}_n}} L_l(m_w^{(l)}, m_v^{(l)})
\label{eq:145}
\end{equation}
with $x_{u_0} = a$ and all other $x_v$ being $b_1, \dots, b_p$ according to the node labeling. The collective microscopic input index is $A = (a, i)$, where there is one distinguished magnetic index plus $p+1$ intertwiner indices, so the number of elementary indices in the domain is $p+2$. Since each index has dimension $d$,
\begin{equation}
\dim \mathrm{Dom}(M_{p,j_l}) = d^{p+2} \label{eq:146}
\end{equation}
The output has $p$ magnetic indices,
\begin{equation}
\dim \mathrm{Codom}(M_{p,j_l}) = d^p \label{eq:147}
\end{equation}
Thus,
\begin{equation}
M_{p,j_l} : \mathbb{C}^{d^{p+2}} \to \mathbb{C}^{d^p} 
\label{eq:148}
\end{equation}
The exponent $p+2$ does not count the number of effective open legs. The effective cell after logical embedding still has $p+1$ legs $(\mu, b_1, \dots, b_p)$. The domain $\mathbb{C}^{d^{p+2}}$ is the microscopic synthesis space consisting of one open magnetic channel $a$ and $p+1$ intertwiner degrees of freedom.

\phantomsection
\subsubsection{Theorem 2}
\label{theo:2}
Here we show a theorem regarding the generic surjectivity property of the prism-cell Bell-LQG transfer map.

\bigskip
\noindent\textbf{Theorem 2.} \textit{For $p=2n-1$, $n \ge 3$, homogeneous spin $j > 0$, and microscopic connectivity $V_{\mathrm{Prism}_n} = \{u_0, \dots, u_{n-1}, v_0, \dots, v_{n-1}\}$, there exists an open dense subset with full Haar measure:}
\begin{equation}
\mathcal{G}_{p,j} \subset \mathrm{SU}(2)^{3n} 
\label{eq:149}
\end{equation}
\textit{such that for every internal holonomy configuration $h \in \mathcal{G}_{p,j}$,}
\begin{equation}
\mathrm{Rank}\, M_{p,j_l}(h) = d^p 
\label{eq:150}
\end{equation}
\textit{Thus, $M_{p,j_l}$ is generically surjective onto its $d^p$-dimensional effective output space.} 

In Theorem 2, we provide an explicit microscopic construction relating the holonomy-dressed Bell states—previously developed within the framework of spin network states in LQG to the effective five-legged tensor representing a perfect tensor for the $\{p,q\} \in \mathcal{T}_{\mathrm{arch}}$ HaPPY code family. Specifically, this theorem establishes the necessary prerequisites to precisely construct a perfect tensor from such a microscopic structure. The proof of Theorem 2 can be found in \hyperref[app:B]{Appendix B}.

\subsubsection{From generic surjectivity to effective tensor synthesis}
Full row rank has a direct consequence:
\begin{equation}
\mathrm{Rank}\, M_{p,j_l} = d^p \label{eq:151}
\end{equation}
Define the right inverse
\begin{equation}
R_{p,j_l} = M_{p,j_l}^\dagger \left( M_{p,j_l} M_{p,j_l}^\dagger \right)^{-1}
\label{eq:152}
\end{equation}
Since $M_{p,j_l}$ is full row rank, $M_{p,j_l} M_{p,j_l}^\dagger$ is positive definite and invertible, hence
\begin{equation}
M_{p,j_l} R_{p,j_l} = \mathbb{I}_{d^p}
\label{eq:153}
\end{equation}
Take an arbitrary target tensor $W_{b;\mu} \in \mathbb{C}^{d^p \times d}$ with $b = (b_1, \dots, b_p)$ and logical index $\mu = 0, \dots, d-1$. Next, define the preliminary microscopic coefficient tensor (see Eq.~\eqref{eq:s3-src58a}:
\begin{equation}
\begin{split}
\mathcal{N}^{(0)}_{A;\mu} &= \sum_b (R_{p,j_l})_{A;b} W_{b;\mu}
\end{split}
\label{eq:154b}
\end{equation}
Then we can define the effective tensor by (see Eq.~\eqref{eq:s3-src46}):
\begin{equation}
T_{a;\mu}^{(N)} = \sum_A (M_{p,j_l})_{a;A} \mathcal{N}_{A;\mu}
\label{eq:154c}
\end{equation}
For $\mathcal{N} = \mathcal{N}^{(0)}$, we obtain
\begin{equation}
T_{a;\mu}^{(N^{(0)})} = \sum_{A,b} (M_{p,j_l})_{a;A} (R_{p,j_l})_{A;b} W_{b;\mu} = \sum_b (M_{p,j_l} R_{p,j_l})_{a;b} W_{b;\mu} = W_{a;\mu}
\label{eq:154d}
\end{equation}
Thus, we show here that $T = W$. This result is an exact algebraic consequence of surjectivity and the explicit right-inverse construction \cite{ref74,ref75}.

\subsubsection{Explicit $2n$-leg perfect target from finite-field MDS construction}
To realize a HaPPY-like cell, the target $W$ must be perfect. Since one effective $p$-gonal cell has $p+1 = 2n$ legs, the required target is a $2n$-leg perfect tensor. A perfect tensor with $2n$ indices is proportional to an isometry for every bipartition with at most $n$ input indices; equivalently, the pure state whose amplitudes are given by the tensor is an absolutely maximally entangled state $\mathrm{AME}(2n,d)$ ~\cite{ref10,ref71,ref72}.

To provide an explicit infinite family, use the local dimension $d = d_{j_l}$, with $d_{j_l}$ being a prime power. The symbol $d_{j_l}$ is distinguished from the Schläfli parameter $q$. In the following simple Vandermonde construction, we require $d_{j_l} \ge 2n = p+1$ so that $2n$ distinct field elements are available. Choose $t_0, t_1, \dots, t_{2n-1} \in \mathbb{F}_Q$ all distinct. For a coefficient vector $c = (c_0, c_1, \dots, c_{n-1}) \in \mathbb{F}_Q^n$, define a polynomial of degree at most $n-1$,
\begin{equation}
P_c(t) = \sum_{s=0}^{n-1} c_s t^s
\label{eq 4.18}
\end{equation}
Define the normalized tensor
\begin{equation}
W_{x_0 x_1 \dots x_{2n-1}} = \frac{1}{d_{j_l}^{n/2}} \sum_{c \in \mathbb{F}_Q^n} \prod_{r=0}^{2n-1} \delta_{x_r, P_c(t_r)}
\label{eq 4.19}
\end{equation}
This is the quantum-state version of the Reed-Solomon/MDS evaluation construction ~\cite{ref71,ref72}. The finite field here serves as an algebraic device to construct the target code tensor; it does not replace $\mathrm{SU}(2)$ as the microscopic gauge structure.

\bigskip
\noindent\textbf{Proposition 1.} \textit{The target $W$ is perfect.}

\smallskip
\noindent\textit{Proof.} Take an arbitrary subset $S = \{r_1, \dots, r_n\}$ of exactly $n$ tensor positions. The evaluation values satisfy
\begin{equation}
\begin{pmatrix} x_{r_1} \\ x_{r_2} \\ \vdots \\ x_{r_n} \end{pmatrix} = V_S \begin{pmatrix} c_0 \\ c_1 \\ \vdots \\ c_{n-1} \end{pmatrix}
\label{eq 4.20}
\end{equation}
with the Vandermonde matrix
\begin{equation}
V_S = \begin{pmatrix}
1 & t_{r_1} & \dots & t_{r_1}^{n-1} \\
1 & t_{r_2} & \dots & t_{r_2}^{n-1} \\
\vdots & \vdots & \ddots & \vdots \\
1 & t_{r_n} & \dots & t_{r_n}^{n-1}
\end{pmatrix}
\label{eq 4.21}
\end{equation}
Its determinant is
\begin{equation}
\det V_S = \prod_{1 \le a < b \le n} (t_{r_b} - t_{r_a})
\label{eq 4.22}
\end{equation}
Since all $t_{r_a}$ are distinct, every factor on the right-hand side is nonzero in the field $\mathbb{F}_Q$. Since the product of nonzero field elements is nonzero, $\det V_S \neq 0$. Thus, the evaluation map $E_S : \mathbb{F}_Q^n \to \mathbb{F}_Q^n$ is bijective. The same applies to the complement $\bar{S}$, which also consists of $n$ distinct evaluation points. We define the normalized pure state associated with $W$ as
\begin{equation}
|W\rangle = \frac{1}{d_{j_l}^{n/2}} \sum_{c \in \mathbb{F}_Q^n} |E_S c\rangle_S \otimes |E_{\bar{S}} c\rangle_{\bar{S}}
\label{eq 4.23}
\end{equation}
The reduced density matrix on $S$ is
\begin{equation}
\rho_S = \mathrm{Tr}_{\bar{S}} |W\rangle \langle W| = \frac{1}{d_{j_l}^n} \sum_{c, c'} \langle E_{\bar{S}} c' | E_{\bar{S}} c \rangle |E_S c\rangle \langle E_S c'|
\label{eq 4.24}
\end{equation}
Orthogonality of the computational basis gives
\begin{equation}
\langle E_{\bar{S}} c' | E_{\bar{S}} c \rangle = \delta_{E_{\bar{S}} c', E_{\bar{S}} c}
\label{eq 4.25}
\end{equation}
Since $E_{\bar{S}}$ is bijective,
\begin{equation}
E_{\bar{S}} c' = E_{\bar{S}} c \implies c' = c
\label{eq 4.26}
\end{equation}
Hence,
\begin{equation}
\rho_S = \frac{1}{d_{j_l}^n} \sum_{c} |E_S c\rangle \langle E_S c|
\label{eq 4.27}
\end{equation}
Since $E_S$ is also bijective, as $c$ ranges over $\mathbb{F}_{d_{j_l}}^n$, $E_S c$ traverses each basis vector of $\mathbb{F}_{d_{j_l}}^n$ exactly once. Therefore,
\begin{equation}
\rho_S = \frac{\mathbb{I}_{d_{j_l}^n}}{d_{j_l}^n}
\label{eq 4.27b}
\end{equation}
The subset $S$ was chosen arbitrarily. Thus, every balanced $n|\bar{n}$ bipartition is maximally entangled. By further tracing out, all subsystem reduced states with $k \le n$ are also maximally mixed. Therefore, $W$ is $\mathrm{AME}(2n,Q)$ and hence a $2n$-leg perfect tensor, proving the proposition. $\blacksquare$

To match the local Hilbert-space dimension of the perfect target with the spin-$j$ magnetic Hilbert space, choose
\begin{equation}
d_{j_l} = 2j_l + 1 \implies j_l = \frac{d_{j_l} - 1}{2}
\label{eq 4.28}
\end{equation}
Since there exist arbitrarily large primes, there exists an infinite sequence $d_{j_l} \to \infty$ and consequently
\begin{equation}
j_l = \frac{d_{j_l}-1}{2} \to \infty
\label{eq 4.29}
\end{equation}
Thus, exact target perfect tensors are available at arbitrarily large spin values, even though this simple finite-field construction does not necessarily cover every half-integer spin.

\subsubsection{Isometric microscopic preimage via kernel completion}
The previous preliminary solution $\mathcal{N}^{(0)} = R_{p,j_l} T$ guarantees $M_{p,j_l} \mathcal{N}^{(0)} = T$, but does not automatically guarantee that the logical embedding $\mathcal{N}^{(0)}$ itself is isometric. Next, define the matrix:
\begin{equation}
\Pi = (\mathcal{N}^{(0)})^\dagger \mathcal{N}^{(0)} \succeq 0
\label{eq 4.30}
\end{equation}
We choose a real scalar $0 < \chi \le \lambda_{\max}(\Pi)^{-1/2}$. Since all eigenvalues $\lambda_k(\Pi)$ satisfy $0 \le \lambda_k \le \lambda_{\max}$, we have $1 - \chi^2 \lambda_k(\Pi) \ge 0$, and the residual matrix
\begin{equation}
S = \mathbb{I}_{d_{j_l}} - \chi^2 \Pi \succeq 0
\label{eq 4.31}
\end{equation}
has a unique positive-semidefinite square root $S^{1/2}$. Then, rank-nullity gives
\begin{equation}
\dim \ker M_{p,j_l} = d^{p+2} - d^p = d^p(d^2 - 1)
\label{eq 4.32}
\end{equation}
For $d_{j_l} \ge 2$, we have $\dim \ker M_{p,j_l} \ge d_{j_l}$. Therefore, we can choose an isometry satisfying
\begin{equation}
\mathcal{N}^{(1)} : \mathbb{C}^d \to \ker M_{p,j_l}
\label{eq 4.33}
\end{equation}
with
\begin{equation}
(\mathcal{N}^{(1)})^\dagger \mathcal{N}^{(1)} = \mathbb{I}_{d_{j_l}}, \qquad M_{p,j_l} \mathcal{N}^{(1)} = 0
\label{eq 4.34}
\end{equation}

We define the final microscopic logical embedding
\begin{equation}
\mathcal{N} = \chi \mathcal{N}^{(0)} + \mathcal{N}^{(1)} S^{1/2}
\label{eq 4.35}
\end{equation}
Since $\mathcal{N}^{(0)} = M_{p,j_l}^\dagger (M_{p,j_l} M_{p,j_l}^\dagger)^{-1} T$, every column of $\mathcal{N}^{(0)}$ lies in $\mathrm{Ran}(M_{p,j_l}^\dagger)$. The standard finite-dimensional decomposition gives $\mathrm{Ran}(M_{p,j_l}^\dagger) = (\ker M_{p,j_l})^\perp$ (see Eq.~\eqref{eq:s3-src62}, so
\begin{equation}
(\mathcal{N}^{(0)})^\dagger \mathcal{N}^{(1)} = 0 = (\mathcal{N}^{(1)})^\dagger \mathcal{N}^{(0)}
\label{eq 4.36}
\end{equation}
Now compute
\begin{equation}
\mathcal{N}^\dagger \mathcal{N} = \chi^2 (\mathcal{N}^{(0)})^\dagger \mathcal{N}^{(0)} + \chi (\mathcal{N}^{(0)})^\dagger \mathcal{N}^{(1)} S^{1/2} + \chi S^{1/2} (\mathcal{N}^{(1)})^\dagger \mathcal{N}^{(0)} + S^{1/2} (\mathcal{N}^{(1)})^\dagger \mathcal{N}^{(1)} S^{1/2}
\label{eq 4.37}
\end{equation}
Since the cross terms are zero and $(\mathcal{N}^{(1)})^\dagger \mathcal{N}^{(1)} = \mathbb{I}_{d_{j_l}}$,
\begin{equation}
\mathcal{N}^\dagger \mathcal{N} = \chi^2 \Pi + S = \chi^2 \Pi + \mathbb{I}_{d_{j_l}} - \chi^2 \Pi = \mathbb{I}_{d_{j_l}}
\label{eq 4.38}
\end{equation}
Thus, $\mathcal{N}$ is an exact isometric logical embedding. The resulting effective tensor is
\begin{equation}
T^{(N)} = M_{p,j_l} N = \chi M_{p,j_l} N^{(0)} + M_{p,j_l} N^{(1)} S^{1/2} = \chi W
\label{eq 4.39}
\end{equation}
This construction explicitly shows that $T$ is not assumed to be equal to a perfect tensor. What is proven is the existence of a microscopic isometric preimage $N$ that maps, via the microscopic Bell-LQG transfer map, to a tensor proportional to the prescribed perfect target.

\subsubsection{HaPPY-admissible finite patches}
The results up to this point are local. Furthermore, for the local perfect tensors to yield a global bulk-to-boundary isometry, the finite patch here must possess an orientation compatible with the isometric directions of each local tensor. We call this condition the HaPPY-admissible finite patches requirement. First, we define:

\bigskip
\noindent\textbf{Definition} (HaPPY-admissible radial orientation). \textit{A finite patch $\Gamma \subset \{p,q\}$ is called HaPPY-admissible if the internal cell adjacency graph can be assigned an acyclic orientation from the bulk toward the boundary such that for each cell $f$, the number of incoming inter-cell legs $r_f$ satisfies}
\begin{equation}
r_f \le n-1
\label{eq 4.40}
\end{equation}
With the logical leg $\mu_f$ as an additional input, the total number of local input legs is $1 + r_f \le n$, so that perfectness guarantees that the tensor can be used as a local isometry.

This condition is an additional combinatorial requirement; it does not follow solely from $(p-2)(q-2) > 4$. For the original $\{5,4\}$ HaPPY tiling, radial layering ensures that each pentagonal tensor has at most two inward network legs, so that together with one logical leg the number of inputs does not exceed three~\cite{ref10}. For general $\{p,q\}$, radial admissibility must be proven or verified for the patch and layering being used.

\subsubsection{Global bulk-to-boundary encoding operator}
For each cell $f$, write the effective tensor $T_f(x_{f1} \dots x_{fp}; \mu_f)$. If side $r$ of cell $f$ is adjacent to side $s$ of cell $g$, the inter-cell bond is
\begin{equation}
L_{x_{gs} x_{fr}}^{\langle f,g \rangle} = \frac{D^{(j_l)}(H_{fg}) g^{(j_l)}(x_{gs}, x_{fr})}{\sqrt{d_{j_l}}}
\label{eq 4.41}
\end{equation}
For a finite patch, the collective bulk index is $\mu = \{\mu_f\}_{f \in F_\Gamma}$ and the collective physical boundary index is $B$. The raw global encoding matrix is
\begin{equation}
V_\Gamma^{\mathrm{raw}}(B; \mu) = \sum_{\{x\}_{\mathrm{int}}} \prod_{f \in F_\Gamma} T_f(x_{f1} \dots x_{fp}; \mu_f) \prod_{\langle f,r, g,s \rangle \in E_\Gamma^{\mathrm{int}}} L_{x_{gs} x_{fr}}^{\langle f,g \rangle}
\label{eq 4.42}
\end{equation}
Since each $T_f$ is proportional to a perfect tensor and each $L_{fg}$ is proportional to a unitary matrix, $V_\Gamma^{\mathrm{raw}}$ differs from a composition of exact local isometries only by a product of nonzero scalar factors. To prove isometry without carrying scalars, define the normalized local maps $\tilde{T}_f$ and unitary bond maps $U_{fg} = \sqrt{d_{j_l}} L_{fg}$.

Group cells according to radial layers $\mathcal{L}_1, \dots, \mathcal{L}_L$. Suppose the normalized map at layer $n$ is (see. Eq.~\eqref{eq:s3-src99})
\begin{equation}
\widetilde{K}^{(n)}(\mathcal{L}_n; x_{n-1}, \mu_n) = \prod_{f \in \mathcal{L}_n} \tilde{T}_f(x_f; \mu_f) \times \prod_{l \subset \mathcal{L}_n} U_l \prod_{l \in (\mathcal{L}_{n-1}, \mathcal{L}_n)} U_l
\label{eq 4.44}
\end{equation}
where effective tensor $\tilde{T}$ use Eq.~\eqref{eq:s3-src101}, $x_{n-1}$ are the frontier indices coming from the previous layer and $\mathcal{L}_n$ are the outward frontier indices. Since each constituent local map is isometric and all interposed bond maps are unitary, in open indices,
\begin{equation}
\sum_{\mathcal{L}_n} (\widetilde{K}^{(n)})^\dagger(\mathcal{L}_n; x_{n-1}', \mu_n') \widetilde{K}^{(n)}(\mathcal{L}_n; x_{n-1}, \mu_n) = \delta_{x_{n-1}', x_{n-1}} \delta_{\mu_n', \mu_n}
\label{eq 4.45}
\end{equation}
In other words, $(\widetilde{K}^{(n)})^\dagger \widetilde{K}^{(n)} = \mathbb{I}$.

The normalized global encoder is
\begin{equation}
\begin{split}
V_\Gamma(\mathcal{L}_L; \mu_1 \dots \mu_L) = \sum_{x_1, \dots, x_{L-1}} \widetilde{K}^{(L)}(\mathcal{L}_L; x_{L-1}, \mu_L) \widetilde{K}^{(L-1)}(\mathcal{L}_{L-1}; x_{L-2}, \mu_{L-1}) \dots \widetilde{K}^{(1)}(\mathcal{L}_1; \mu_1)
\label{eq 4.45}
\end{split}
\end{equation}

\phantomsection
\subsubsection{Theorem 3}
\label{theo:3}
Based on the Eq.~\eqref{eq 4.44}, ~\eqref{eq 4.45} and ~\eqref{eq 4.45}), we show the exact global isometry theorem for the constructed global encoding operator.

\bigskip
\noindent\textbf{Theorem 3.} \textit{If each local cell is realized as $T_f = \chi_f W_f$ with $W_f$ being perfect and the finite patch $\Gamma$ has a HaPPY-admissible radial orientation, then the scalar-normalized global encoder satisfies}
\begin{equation}
V_\Gamma^\dagger V_\Gamma = \mathbb{I}_{\mathcal{H}_{\mathrm{bulk}}} \label{eq:thm2}
\end{equation}
Theorem 3 is a local-to-global composition theorem, rather than establishing the formation of a perfect tensor. Specifically, we demonstrate that once every local cell has been reduced to a tensor proportional to the HaPPY perfect tensor (see Eq.~\ref{eq 4.39}), contracting all cells within the patch up to the boundary generates an exact isometric global encoding operator, provided the patch's radial orientation geometry aligns with the $\{p,q\}$ tessellation family structure as in the HaPPY code $\{5,4\}$  ($\{p,q\} \in \mathcal{T}_{\mathrm{arch}}$). The proof of Theorem 3 can be found in \hyperref[app:C]{Appendix C}.

\subsubsection{Cut decomposition and quantum RT formula}
Take a boundary region $A \subset \partial\Gamma$ and its complement $\bar{A}$. Suppose a cut $\gamma$ divides the network into sides $A$ and $\bar{A}$ and cuts links $l \in \gamma$. The cut divides the bulk logical Hilbert space into
\begin{equation}
\mathcal{H}_{\mathrm{bulk}} = \mathcal{H}_{a_\gamma} \otimes \mathcal{H}_{\bar{a}_\gamma}
\label{eq 4.46}
\end{equation}
For a single cut link $l$, the normalized state has the amplitude
\begin{equation}
L_l(y_l, x_l) = \frac{1}{\sqrt{d_{j_l}}} U_l(y_l, x_l)
\label{eq 4.47}
\end{equation}
For the entire cut,
\begin{equation}
L_\gamma(y; x) = \prod_{l \in \gamma} \frac{1}{\sqrt{d_{j_l}}} U_l(y_l, x_l)
\label{eq 4.48}
\end{equation}
Define
\begin{equation}
D_\gamma = \prod_{l \in \gamma} d_{j_l}
\label{eq 4.49}
\end{equation}
Tracing out the Bell partners on side $\bar{A}$ yields
\begin{equation}
\tau_{\gamma, A}(x; x') = \sum_y L_\gamma(y; x) L_\gamma^\dagger(y; x') = \prod_{l \in \gamma} \frac{1}{d_{j_l}} \sum_{y_l} U_l(y_l, x_l) U_l^\dagger(y_l, x_l') = \frac{1}{D_\gamma} \delta_{x, x'}
\label{eq 4.50}
\end{equation}
Therefore,
\begin{equation}
\tau_{\gamma, A} = \frac{\mathbb{I}_{D_\gamma}}{D_\gamma}
\label{eq 4.51}
\end{equation}
and
\begin{equation}
S_{\mathrm{cut}}(\gamma) = \log D_\gamma = \sum_{l \in \gamma} \log d_{j_l} = \sum_{l \in \gamma} \log(2j_l + 1)
\label{eq 4.52}
\end{equation}
Now assume that the networks on both sides of the cut can be reduced to isometries
\begin{equation}
U_A : \mathcal{H}_{a_\gamma} \otimes \mathcal{H}_{\gamma, A} \to \mathcal{H}_A, \qquad U_{\bar{A}} : \mathcal{H}_{\bar{a}_\gamma} \otimes \mathcal{H}_{\gamma, \bar{A}} \to \mathcal{H}_{\bar{A}}
\label{eq 4.53}
\end{equation}
Use collective bulk basis indices $\mu_a, \bar{\mu}_a$ and bra indices $\nu_a, \bar{\nu}_a$. The cut decomposition of the global encoding matrix is
\begin{equation}
V_\Gamma(A, \bar{A}; \mu_a, \bar{\mu}_a) = \sum_{x,y} \widetilde{T}_A(A; \mu_a, x) \widetilde{T}_{\bar{A}}(\bar{A}; \bar{\mu}_a, y) U_\gamma(y; x)
\label{eq 4.54}
\end{equation}
For an arbitrary bulk density matrix,
\begin{equation}
\rho_\partial = V_\Gamma \rho_{\mathrm{bulk}} V_\Gamma^\dagger
\label{eq 4.55}
\end{equation}
In fully open indices,
\begin{equation}
\begin{split}
\rho_\partial(A, \bar{A}; A', \bar{A}') = \sum_{\substack{\mu_a, \bar{\mu}_a, \nu_a, \bar{\nu}_a \\ x, y, x', y'}} 
& \widetilde{T}_A(A; \mu_a, x) \widetilde{T}_{\bar{A}}(\bar{A}; \bar{\mu}_a, y) U_\gamma(y; x) \\
& \times \rho_{\mathrm{bulk}}(\mu_a, \bar{\mu}_a; \nu_a, \bar{\nu}_a) U_\gamma^\dagger(y'; x') \widetilde{T}_A^\dagger(A'; \nu_a, x') \widetilde{T}_{\bar{A}}^\dagger(\bar{A}'; \bar{\nu}_a, y')
\label{eq 4.56}
\end{split}
\end{equation}
Tracing over $\bar{A}$ means identifying $\bar{A}' = \bar{A}$ and summing. Isometry of $U_{\bar{A}}$ gives
\begin{equation}
\sum_{\bar{A}} \widetilde{T}_{\bar{A}}(\bar{A}; \bar{\mu}_a, y) \widetilde{T}_{\bar{A}}^\dagger(\bar{A}; \bar{\nu}_a, y') = \delta_{\bar{\mu}_a, \bar{\nu}_a} \delta_{y, y'}
\label{eq 4.57}
\end{equation}
The first delta yields the reduced bulk density matrix
\begin{equation}
\rho_a(\mu_a; \nu_a) = \sum_{\bar{\mu}_a} \rho_{\mathrm{bulk}}(\mu_a, \bar{\mu}_a; \nu_a, \bar{\mu}_a)
\label{eq 4.58}
\end{equation}
while the second delta enables the Bell contraction
\begin{equation}
\sum_y U_\gamma(y; x) U_\gamma^\dagger(y; x') = \frac{1}{D_\gamma} \delta_{x, x'}
\label{eq 4.59}
\end{equation}
Then,
\begin{equation}
\rho_A = T_A \left( \rho_{a_\gamma} \otimes \frac{\mathbb{I}_{D_\gamma}}{D_\gamma} \right) T_A^\dagger
\label{eq 4.60}
\end{equation}
Isometry preserves all nonzero eigenvalues, and using the von Neumann entanglement entropy on a tensor product,
\begin{equation}
S(A) = S(\rho_A) = S\left( \rho_{a_\gamma} \otimes \frac{\mathbb{I}_{D_\gamma}}{D_\gamma} \right) = S_{\mathrm{bulk}}(a_\gamma) + \log D_\gamma
\label{eq 4.61}
\end{equation}
Thus, an exact discrete quantum RT relation is obtained for every two-sided cut-isometric cut,
\begin{equation}
S(A) = S_{\mathrm{bulk}}(a_\gamma) + \sum_{l \in \gamma} \log(2j_l + 1). \label{eq:RT_discrete}
\end{equation}
The second term has a clear microscopic origin: each cut holonomy-dressed Bell link contributes a maximally mixed factor $\mathbb{I}_{d_{j_l}}/d_{j_l}$. The first term is the entropy of the bulk logical degrees of freedom located on side $A$ of the cut. This area-plus-bulk structure parallels quantum corrections to RT \cite{ref11, ref12, ref14}, while the Bell-network literature provides an independent area-law motivation for spin-network entanglement \cite{ref32}.

\subsubsection{Generalized minimal-cut prescription}
Define the discrete generalized entropy
\begin{equation}
S_{\mathrm{gen}}(\gamma) = S_{\mathrm{bulk}}(a_\gamma) + \sum_{l \in \gamma} \log(2j_l + 1)
\label{eq 4.62}
\end{equation}
The preceding equation proves $S(A) = S_{\mathrm{gen}}(\gamma)$ for a cut $\gamma$ that is two-sided cut-isometric. To incorporate minimization, an additional hypothesis is required: that there exists an admissible cut $\gamma_A$ which is simultaneously two-sided cut-isometric and generalized-minimal among candidate cuts homologous to $A$. Only under this hypothesis can we write
\begin{equation}
S(A) = \min_{\gamma \sim A} \left[ S_{\mathrm{bulk}}(a_\gamma) + \sum_{l \in \gamma} \log(2j_l + 1) \right]
\label{eq 4.63}
\end{equation}
For homogeneous spin $j_l$, this yields
\begin{equation}
S(A) = \min_{\gamma \sim A} \left[ S_{\mathrm{bulk}}(a_\gamma) + |\gamma| \log(2j_l + 1) \right]
64
\end{equation}
If the bulk state is a product state with respect to $a_\gamma | \bar{a}_\gamma$, then $S_{\mathrm{bulk}}(a_\gamma) = 0$ and the standard discrete HaPPY form is obtained, namely:
\begin{equation}
S(A) = |\gamma_A| \log d_{j_l}
\label{eq 4.65}
\end{equation}
Minimization cannot be claimed solely from Bell maximal entanglement. Bell links determine the cut entropy; the perfect/cut-isometric structure transfers that spectrum to the boundary; the generalized-minimal-cut property is an additional global network statement~\cite{ref10,ref11,ref12, ref14,ref45,ref46,ref47,ref48,ref49,ref50}.

\subsubsection{Theorem 4}

Building upon these previous results, we formulate the theorem of Bell-LQG holographic realizability on admissible hyperbolic tessellations. This theorem relates the holonomy-dressed Bell-LQG framework which constructs $p$-gonals from entangled quantum tetrahedra to the realization of an effective perfect tensor ($T^{(\mathcal{N})}$) via a surjective microscopic transfer map, thereby achieving a globally isometric encoding (${V}_{\Gamma}$). 

\bigskip
\noindent\textbf{Theorem 4.} \textit{Let $\{p,q\} \in \mathcal{T}_{\mathrm{arch}} = \left\{ \{p,q\} \;\middle|\; p = 2k-1, \; k \ge 3, \; q \in \mathbb{Z}_{\ge 3}, \; (p-2)(q-2) > 4 \right\}$ and in each $p$-gonal cell place $2n = p+1$ four-valent quantum tetrahedra with internal connectivity $V_{\mathrm{Prism}_n} = \{u_0, \dots, u_{k-1}, v_0, \dots, v_{k-1}\}$. Each internal link carries a holonomy-dressed Bell kernel $L_l = d_{j_l}^{-1/2} D^{(j_l)}(h_l) g^{(j_l)}$. Then:}
\begin{enumerate}
\item \textit{For every fixed spin $j > 0$, the microscopic transfer map $M_{p,j} : \mathbb{C}^{d^{p+2}} \to \mathbb{C}^{d^p}$ has full row rank for Haar-almost every internal holonomy configuration.}
\item \textit{If $d_{j_l} = 2j_l + 1$ is a prime power with $d_{j_l} \ge p+1$, there exists an explicit $2n$-leg perfect target tensor $W$ from the finite-field MDS/Reed-Solomon construction and a microscopic logical embedding $\mathcal{N}$ satisfying $\mathcal{N}^\dagger \mathcal{N} = \mathbb{I}_{d_{j_l}}$, such that $T^{(\mathcal{N})} = \chi W$ for some $\chi \neq 0$. Consequently, all local perfectness defects on partitions with at most $n$ input legs vanish exactly.}
\item \textit{If a finite patch $\Gamma \subset \{p,q\}\in \mathcal{T}_{\mathrm{arch}}$ has a HaPPY-admissible acyclic radial orientation with $r_f \le n-1$ incoming network legs on each cell, the scalar-normalized global encoder satisfies $V_\Gamma^\dagger V_\Gamma = \mathbb{I}_{\mathcal{H}_{\mathrm{bulk}}}$.}
\item \textit{For every boundary region $A$ possessing a common two-sided cut-isometric generalized-minimal cut $\gamma_A$, the boundary entropy satisfies}
\begin{equation}
S(A) = \min_{\gamma \sim A} \left[ S_{\mathrm{bulk}}(a_\gamma) + \sum_{l \in \gamma} \log(2j_l + 1) \right]
\label{eq 4.63}
\end{equation}
\end{enumerate}

\smallskip
\noindent\textit{Proof.} Statement (1) is \hyperref[theo:2]{Theorem 2}. Statement (2) follows from the finite-field perfectness proposition, right-inverse synthesis, and kernel completion: surjectivity yields $\mathcal{N}^{(0)} = R_{p,j_l} W$ with $M_{p,j} \mathcal{N}^{(0)} = W$; orthogonal kernel completion produces a normalized embedding $N$ without altering the image up to an overall scalar, resulting in $T^{(\mathcal{N})} = \chi W$. Statement (3) is \hyperref[theo:3]{Theorem 3}: radial admissibility ensures each local tensor is applied to a partition with at most $n$ inputs, so shell maps are isometries and telescopic contraction yields $V_\Gamma^\dagger V_\Gamma = \mathbb{I}$. Statement (4) follows from the cut-density-matrix calculation: cut Bell pairs yield $\mathbb{I}_{D_\gamma}/D_\gamma$, two-sided isometries preserve nonzero spectra, and generalized minimality selects the minimizing admissible cut. $\blacksquare$

\section{Applications to Black Hole Interiors}

In this section, we examine the application of the method presented in Section 3 to black hole interiors. The model developed here uses the hyperbolic tessellation \{5,4\} as a discrete spacetime model for a static black hole in AdS geometry~\cite{ref16}. Other tessellation models for studying black holes have been considered in~\cite{ref17}. We employ the \{5,4\} tessellation model of~\cite{ref16} to compare the results of our model with those of ~\cite{ref15}, focusing on two aspects: (i) the non-isometric nature of the encoding operator in the black hole setting and (ii) black hole entropy, described by the generalized entropy and following the Page curve.

\subsection{The \{5,4\} Tessellation Model with Entanglement between Quantum Tetrahedra as a Semiclassical Black Hole Model}

In this section we adopt the black hole model in Ref.~\cite{ref16}, in which the continuous spacetime structure is replaced by holographic tensor networks. Using qubits, these networks describe bulk locality constructed from the quantum informational properties of the boundary. The geometry provided by such tensor network constructions makes available tools such as Ryu--Takayanagi (RT)-like formulas for studying QES and operator reconstruction within entanglement wedges. The authors use the \{5,4\} tessellation as the framework for the black hole structure. This choice is based on the bulk--boundary mapping properties exhibited by this tessellation. An early study of the implementation of holographic tensor networks for black holes was undertaken by Akers, Engelhardt, Harlow, Penington, and Vardhan in Ref.~\cite{ref15}, referred to as the PHEVA code. The difference between Ref.~\cite{ref16} and the PHEVA code is that the former introduce an additional degree of freedom, namely locality, into the PHEVA model, allowing the interior and exterior structures of a black hole and their dynamics to be examined within a more physical framework. This provides an explanation of black hole horizon dynamics and a resolution of the unitarity issue for infalling matter using quantum extremal surfaces (QES). The model in Ref.~\cite{ref16}. is thus known as a generalization of PHEVA, in which the non-isometric properties of the PHEVA code for the black hole interior are formulated with locality in the \{5,4\} tessellation.

We begin by examining the black hole model in the PHEVA code. The non-isometric properties arising in the study of black holes originate from the horizon, which separates the interior and exterior regions. Two descriptions of the black hole are given. First, the effective description gives the perspective of an observer falling into a black hole in a semiclassical bulk. Interior degrees of freedom behind the horizon are divided into left-moving (radially ingoing) modes $\ell$ and right-moving (radially outgoing) modes $r$. The left movers ($\ell$) keep track of what fell into the black hole from the exterior, while the right movers ($r$) represent interior Hawking partners that are maximally entangled with outgoing radiation. Second, the fundamental description provides the account of an exterior (or boundary) observer watching the infalling observer from a safe distance. This observer describes the black hole using fundamental quantum gravitational degrees of freedom $B$; the number of $B$ qubits is determined by the area of the event horizon. Both descriptions share the same exterior, since the accounts of both observers should agree outside the horizon. Any exterior degrees of freedom are removed to a reservoir $R$, which is divided into two parts: infalling modes \(R_{in}\) that will eventually cross the horizon to become $\ell$ modes in the effective description, and outgoing Hawking partners \(R_{out}\) that are maximally entangled with the $r$ modes in the effective description. The resulting Hilbert spaces for both descriptions are:
\begin{align}
\mathcal{H}_{\mathrm{eff}} &= \mathcal{H}_{\ell} \otimes \mathcal{H}_{r} \otimes \mathcal{H}_{R_{\mathrm{in}}} \otimes \mathcal{H}_{R_{\mathrm{out}}} \nonumber \\
&\quad \in \mathcal{H}_{\mathrm{intertwiner}} = \operatorname{Inv}_{SU(2)}\!\left[ \left( V_{j_{\ell} = \frac{1}{2}} \right)^{\otimes 4} \right] = 2, \\
\mathcal{H}_{\mathrm{fun}} &= \mathcal{H}_{B} \otimes \mathcal{H}_{R_{\mathrm{in}}} \otimes \mathcal{H}_{R_{\mathrm{out}}} \nonumber \\
&\quad \in \mathcal{H}_{\mathrm{intertwiner}} = \operatorname{Inv}_{SU(2)}\!\left[ \left( V_{j_{\ell} = \frac{1}{2}} \right)^{\otimes 4} \right] = 2.
\end{align}
For qubits, \(V_{j_{l}}\) is given by \(V_{j_{l}} = V_{1/2} = \operatorname{span}\ \left\{ \left. \ \left| \frac{1}{2}, + \frac{1}{2} \right.\  \right\rangle,\left. \ \left| \frac{1}{2}, - \frac{1}{2} \right.\  \right\rangle \right\} \simeq \left\{ \left. \ \left| 0 \right.\  \right\rangle,\left. \ 1 \right\rangle \right\}\). Unlike in Ref.~\cite{ref16}, we use qubits characterized by four-valent intertwiner nodes rather than qudits, in accordance with the tetrahedral structure, each face of which carries spin-up and spin-down states as a consequence of satisfying the Gauss constraint (see Fig.~\ref{fig:a four-valent node}), so that each factor in the Hilbert space is compressed into a $1/2$-dimensional qubit. The Hilbert spaces are invariant subspaces (singlets) of four spin-$1/2$ Hilbert spaces.

The semiclassical black hole description proposed in Ref.~\cite{ref15,ref17} using non-isometric holographic codes within a \{5,4\} tessellation structure representing locality~\cite{ref16}, we propose that this tessellation structure can also be realized through entanglement between tetrahedra in LQG. The discrete space described by hyperbolic tessellations can thus emerge from the entanglement structure between tetrahedra and provide a microscopic description of semiclassical black holes based on non-isometric holographic codes.

\begin{figure}[t]
    \centering
\includegraphics[width=1\columnwidth]{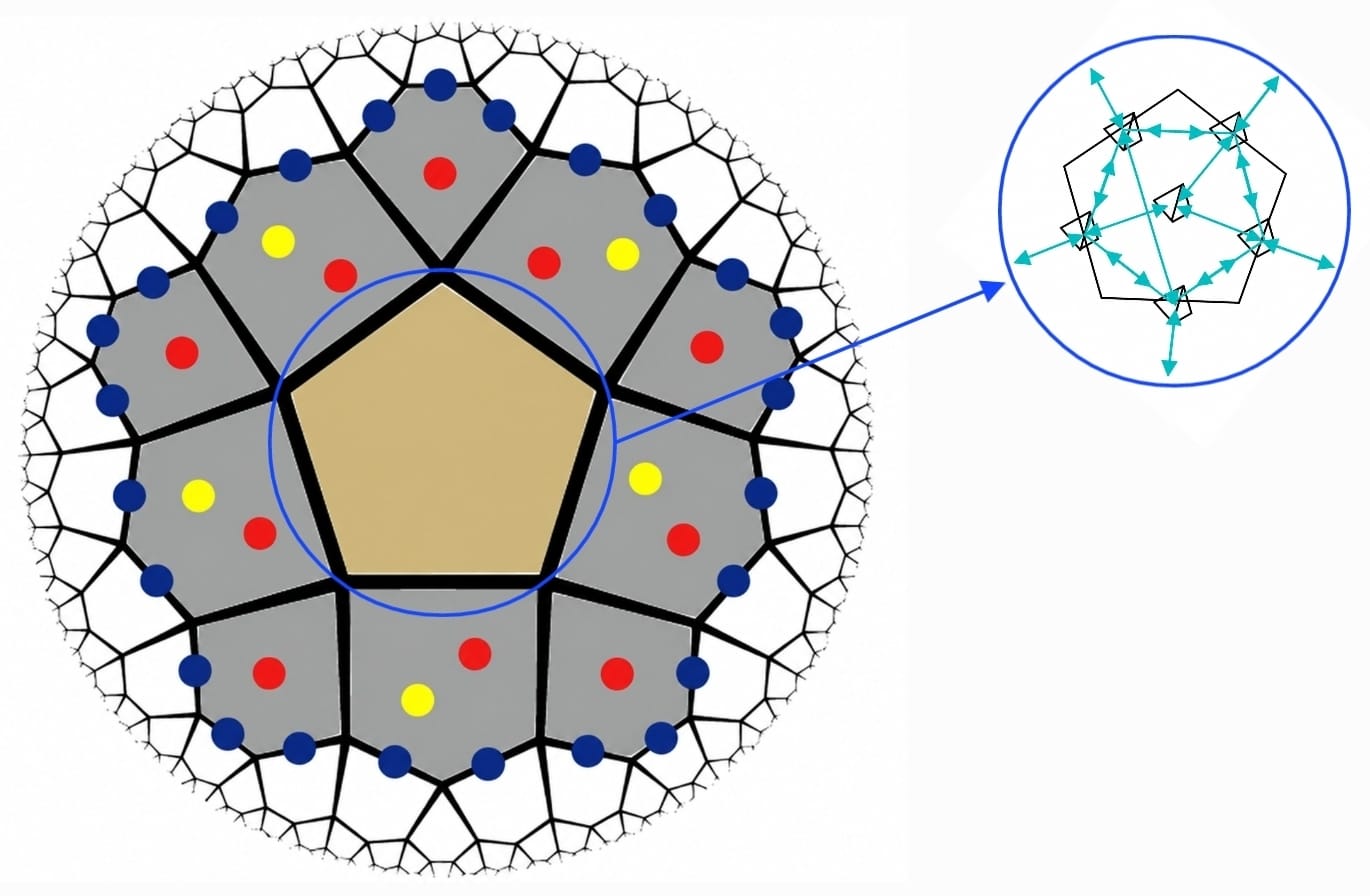}
    \caption{\small A \{5,4\} hyperbolic tessellation constructed by placing four pentagons around each vertex. Shaded cells describe the black hole interior and are colored according to their row: row 0 is gold and row 1 is gray. Qubits are represented by dots and colored by type according to Fig. 1. The $r$ (yellow) and $\ell$ (red) qudits are associated with cells, while the $B$ (dark blue) qubits are associated with edges along the horizon $\Lambda_{\mathrm{BH}}$.}
    \label{fig:bhtessellation}
\end{figure}

In the PHEVA code, the holographic map \(\mathcal{U}\) maps the interior degrees of freedom in the effective description to the fundamental black hole degrees of freedom:
\begin{equation}
{\mathcal{U\ :\ }\mathcal{H}_{l}\ \bigotimes\ \mathcal{H}_{r} \rightarrow \mathcal{H}_{B}}
\end{equation}
For an evaporating black hole, the dimension of \(\mathcal{H}_{B}\) continually decreases as more Hawking radiation leaves into \(\mathcal{H}_{R_{out}}\) than enters from \(\mathcal{H}_{R_{in}}\), and the horizon area shrinks. From the perspective of the effective description, however, the dimension of \(\mathcal{H}_{l}\ \bigotimes\ \mathcal{H}_{r}\) continually increases as more Hawking pairs $rR_{\mathrm{out}}$ are generated. Therefore, there will be a time (close to the Page time) when the effective interior degrees of freedom $\ell r$ out number the fundamental black hole degrees of freedom $B$, and $V$ must become non-isometric. In this model, non-isometry gives rise to many important features of black hole information physics. We consider the following \{5,4\} hyperbolic tessellation of a spatial slice of AdS$_3$ in Fig.~\ref{fig:bhtessellation}.

\subsection{Non-Isometric Structure of the Holographic Inner Code}

The holographic map in the equation (\(\mathcal{U\ :\ }\mathcal{H}_{l}\ \bigotimes\ \mathcal{H}_{r} \rightarrow \mathcal{H}_{B}\)) maps states in the effective description, obtained from the tensor product of the Hilbert spaces associated with the reservoir $R$, to the fundamental description. During black hole evaporation, we have
\begin{equation}
{\dim(lr) > \dim B}
\end{equation}
This shows that \(\mathcal{U}\) is a non-isometric rather than an isometric map. Consequently, for any linear map \(\mathcal{U\ :\ }\mathcal{H}_{l}\ \bigotimes\ \mathcal{H}_{r} \rightarrow \mathcal{H}_{B}\), there must necessarily be a sizeable subspace of ``null states'' annihilated by \(\mathcal{U}\)~\cite{ref15}. As stated in Ref.~\cite{ref15}, the effective field theory Hilbert space inside a black hole contains a large set of ``null states,'' each annihilated by the holographic map to the fundamental degrees of freedom.

\begin{figure}[t]
    \centering
\includegraphics[width=1\columnwidth]{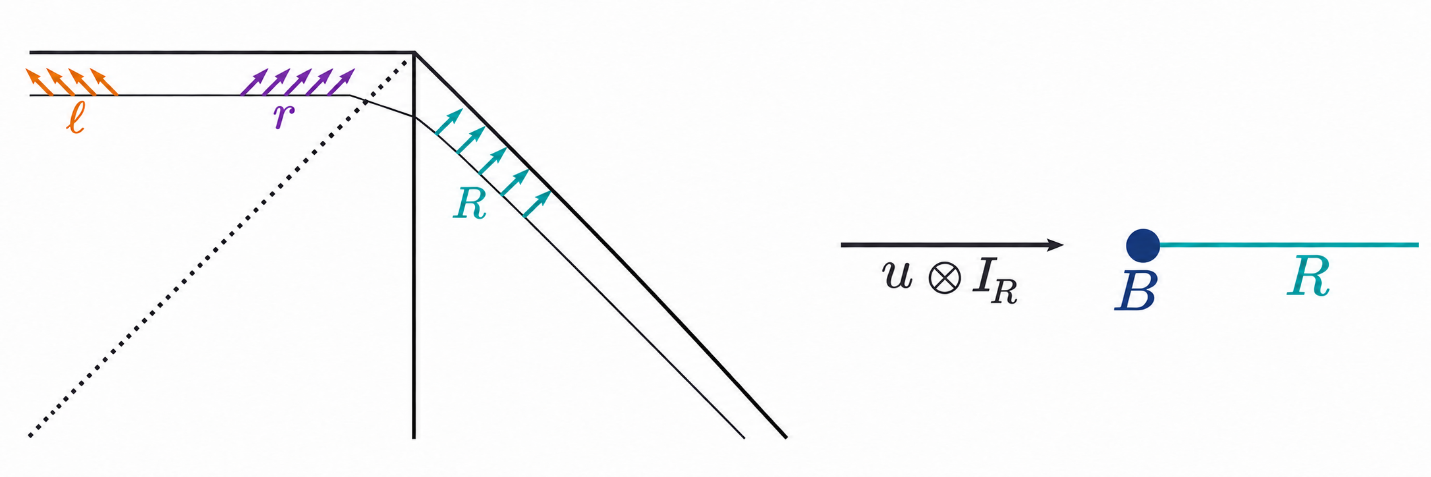}
    \caption{\small Holography for a black hole $B$ (dark blue), representing the fundamental black hole degrees of freedom, radiating into a reservoir $R$ (turquoise) that contains radiation and is consistent on both sides of the mapping of \(l\) and $r$. In the effective description, the left-moving modes $\ell$ (orange) are left-moving interior modes, whereas the right-moving modes $r$ (purple) are the interior partners of Hawking radiation. The holographic map \(\mathcal{U\ :\ }\mathcal{H}_{l}\ \bigotimes\ \mathcal{H}_{r} \rightarrow \mathcal{H}_{B}\) describes how these interior modes are encoded in the fundamental degrees of freedom. We always take \(\mathcal{U}\) to be a linear map.}
    \label{fig:structurehorizon}
\end{figure} 

Here we use the effective tensor formulation for a single pentagon in the tessellation constructed previously (\(T_{\left\{ 5,4 \right\}}^{(eff\ LQG)}\)) as an operator representing the holographic map. This differs from Ref.~\cite{ref15} and~\cite{ref16}, where the holographic map acting inside the black hole is non-isometric, whereas the exterior is described by an operator originating from an isometric holographic tensor network. In our model, \(T_{\left\{ 5,4 \right\}}^{(eff\ LQG)}\) is used to formulate both the inner and outer codes, which are holographic tensor networks describing the interior and exterior of the black hole, respectively. We modify the initially isometric inner code constructed from \(T_{\left\{ 5,4 \right\}}^{(eff\ LQG)}\) into a non-isometric code using the postselection formulation~\cite{ref15,ref16}, in which an isometric tensor network is subjected to a postselection operation. To this end, we examine the constituent parts of \(T_{\left\{ 5,4 \right\}}^{(eff\ LQG)} = \mathcal{N}M_{\left( j_{l} \right)}\), with \(\mathcal{N}\) and \(M_{\left( j_{l} \right)}\) as given in Eqs. () and (), and employ the formulation in Ref.~\cite{ref16}.

We have shown that \(T_{\left\{ 5,4 \right\}}^{(eff\ LQG)}\) is an isometric holographic tensor network. When this tensor is promoted to an encoding operator, the inner and outer codes it describes complete the holographic map \(lr \rightarrow B \rightarrow CFT\), entirely within the \{5,4\} tessellation (~\ref{fig:structurehorizon}), thereby preserving all information in the $B$ qubits as it is encoded in the CFT degrees of freedom.

Our construction is inspired by Bueller's encoding operator based on a \{5,4\} hyperbolic tessellation model for the black hole interior, with degrees of freedom $lr$ associated with the tessellation cells and fundamental degrees of freedom $B$ along the edges of $\Lambda_{\mathrm{BH}}$. In this description, the inner holographic code is constructed as a tensor network mapping $\ell r\to B$ (see Fig.~\ref{fig:structurehorizon}). We first consider \(T_{\left\{ 5,4 \right\}}^{(eff\ LQG)}\) as follows:
\begin{equation}
{T_{\left\{ 5,4 \right\}}^{(eff\ LQG)}\mathcal{= N}M_{\left( j_{l} \right)} = \chi W}
\end{equation}
Here \(W\) is a six-leg tensor constructed from an AME perfect tensor and satisfies
\begin{equation}
{W^{\dagger}W = D^{1 - k}\mathbb{I}_{k}}
\end{equation}
Here \(D\) is the dimension of the local Hilbert space on each leg of the effective tensor $W$. A pentagonal cell has \(\left( c_{x} \right)\) legs directed into the cell along its sides. Together with the leg representing the logical qubit \(\left( s_{x} \right)\), the total number of legs in each cell \(\left( k_{x} \right)\) of the \{5,4\} tessellation is therefore
\begin{equation}
{k_{x} = c_{x} + s_{x}}
\end{equation}
In our model, the number of legs is restricted to \(c_{x} \leq 2\) and \(s_{x} = 1\) for each cell in the tessellation.

We define the isometric tensor \(H_{x}\) representing a tessellation cell as
\begin{equation}
{H_{x} = D^{c_{x}/2}W_{x} = \frac{D^{c_{x}/2}}{\chi_{x}}\left( T_{\left\{ 5,4 \right\}}^{(eff\ LQG)} \right)_{x}}
\label{eq1}
\end{equation}
Furthermore, according to Fig.~\ref{fig:bhtessellation}, each pentagon $x$ in the \{5,4\} tessellation has five legs, and \(5 - c_{x}\) denotes the number of legs connected to the subsequent tensor \(H_{x}\) toward the boundary. Specifically, we denote the numbers of incoming and outgoing legs of the pentagon represented by the tensor \(H_{x}\) by \(\mathcal{J}_{in}\) and \(\mathcal{J}_{out}\), respectively:
\begin{equation}
{\mathcal{J}_{in} \equiv k_{x} = c_{x} + s_{x}}
\label{eq2}
\end{equation}
\begin{equation}
{\mathcal{J}_{out} \equiv 5 - c_{x}}
\label{eq3}
\end{equation}
The difference between \(\mathcal{J}_{in}\) and \(\mathcal{J}_{out}\) implies that
\begin{equation}
\Delta_{x} = \mathcal{J}_{\mathrm{out},x} - \mathcal{J}_{\mathrm{in},x}
\left\{
\begin{array}{@{}l @{\,:\,} l@{}}
> 0 & \text{yields an isometric tensor } H_x, \\
= 0 & \text{yields a unitary tensor } H_x, \\
< 0 & \text{yields a non-isometric tensor through postselection on } H_x.
\end{array}
\right.
\end{equation}
The difference in the number of legs \(\mathrm{\Delta}_{x}\) of the tensor \(H_{x}\) provides the mechanism by which isometry becomes non-isometry. In our model, we use the definition of \(H_{x}\) in Eq.~\ref{eq1}, with \(H_{x}\) mapping
\begin{equation}
{H_{x}:\mathbb{C}^{D^{\mathcal{J}_{in,x}}} \rightarrow \mathbb{C}^{D^{\mathcal{J}_{out,x}}}}
\end{equation}
The tensor \(H_{x}\) can also be written as
\begin{equation}
{H_{x} = D^{\left( k_{x} - 1 \right)/2}W_{x}}
\end{equation}
We identify the number of legs \(k_{x}\) of the pentagon as follows. First, for \(k_{x} \leq 3\), we have
\begin{equation}
{{H_{x}}^{\dagger}H_{x} = \mathbb{I}_{D^{k_{x}}}}
\label{eq5}
\end{equation}
Second, for \(k_{x} > 3\), we have
\begin{equation}
{H_{x}{H_{x}}^{\dagger} = D^{2\left( k_{x} - 3 \right)}\mathbb{I}_{D^{6 - k_{x}}}}
\end{equation}
In the second case, the number of outgoing legs of the pentagon satisfies \(6 - k_{x} \leq 3\). Using the definitions of \(\mathcal{J}_{in}\) and \(\mathcal{J}_{out}\) in Eq.~\ref{eq2} and~\ref{eq3}, we obtain
\begin{equation}
{{H_{x}}^{\dagger}H_{x} = D^{- \left( \mathcal{J}_{out,x} - \mathcal{J}_{in,x} \right)}\mathbb{I}_{D^{6 - k_{x}}} = D^{- \mathrm{\Delta}_{x}}\mathbb{I}_{D^{6 - k_{x}}} = \left| P_{x} \right|\mathbb{I}_{D^{6 - k_{x}}}}
\label{eq4}
\end{equation}
Here \(\left| P_{x} \right| = D^{ \mathrm{\Delta}_{x}}\) is the postselection variable~\cite{ref16}, which serves as a prefactor included to preserve normalization and determines the non-isometric nature of \(H_{x}\). This prefactor appears when the number of legs of the pentagon satisfies \(k_{x} > 3\).

These two results show that the holographic inner code \(H_{x}\) constructed from \(T_{\left\{ 5,4 \right\}}^{(eff\ LQG)}\) can naturally be either isometric or non-isometric, depending on the number of legs in each pentagon of the \{5,4\} tessellation. In particular, for \(k_{x} \leq 3\), the input dimension is smaller than the output dimension:
\begin{equation}
{D^{\mathcal{J}_{in,x}} < D^{\mathcal{J}_{out,x}}\ \  \rightarrow \ \ \ D^{k_{x}} < D^{6 - k_{x}}}
\end{equation}
We then introduce an ancilla \(A_{x}\) with dimension
\begin{equation}
{\left| A_{x} \right| = D^{\mathrm{\Delta}_{x}} = D^{- 2\left( k_{x} - 3 \right)}}
\end{equation}
The input space is enlarged to match the dimension of the output space when \(k_{x} = 3\), namely when
\begin{equation}
{D^{\mathcal{J}_{in,x}}\left| A_{x} \right| = D^{k_{x}} = D^{- 2\left( k_{x} - 3 \right)} = D^{6 - k_{x}} = D^{\mathcal{J}_{out,x}}}
\end{equation}
This agrees with Eq.~\ref{eq5}. We obtain a representation of the isometric tensor \(H_{x}\) by inserting an ancilla $A$ and applying a unitary \(\mathcal{U}_{x}\):
\begin{equation}
{H_{x} = \mathcal{U}_{x}\left( \mathbb{I}_{in,x}\ \bigotimes\ \left.\left| 0 \right. \right\rangle_{A_{x}} \right)\ ,\ \ \ \ \ \ \mathcal{J}_{out,x} > \mathcal{J}_{in,x}}
\label{eq6}
\end{equation}
The unitary operator \(\mathcal{U}_{x}\) has the same dimension as \(\mathcal{J}_{out,x}\) and maps
\begin{equation}
{\mathcal{U}_{x}:\mathcal{H}_{\mathcal{J}_{in,x}}\bigotimes\ \mathcal{H}_{A_{x}} \rightarrow \mathcal{H}_{\mathcal{J}_{out,x}}\bigotimes\ \mathcal{H}_{P_{x}}}
\end{equation}
Moreover, the encoding operator represented by the tensor \(H_{x}\) acts as a linear map from \(\mathcal{J}_{in}\) incoming legs to \(\mathcal{J}_{out}\) outgoing legs. The state \(\left. \left| 0 \right. \right\rangle_{A}\) is defined as an ancilla state storing information from the black hole interior, namely information about the $l$ and $r$ modes, within the dimensional space of the ancilla state.

For \(k_{x} > 3\), however, the input dimension is larger than the output dimension:
\begin{equation}
{D^{\mathcal{J}_{in,x}} > D^{\mathcal{J}_{out,x}}\ \  \rightarrow \ \ \ D^{k_{x}} > D^{6 - k_{x}}}
\end{equation}
The previously introduced prefactor \(P_{x}\) is defined to have dimension
\begin{equation}
{\left| P_{x} \right| = D^{- \mathrm{\Delta}_{x}} = D^{2\left( k_{x} - 3 \right)}}
\end{equation}
This yields Eq.~\ref{eq4}. As in the case \(k_{x} \leq 3\), matching the dimensions of the input and output spaces requires
\begin{equation}
{D^{\mathcal{J}_{out,x}}\left| P_{x} \right| = D^{6 - k_{x}}D^{2\left( k_{x} - 3 \right)} = D^{k_{x}}}
\end{equation}
The tensor \(H_{x}\) is obtained by applying a unitary $U_x$ to all incoming legs and then postselecting:
\begin{equation}
{H_{x} = \sqrt{\left| P_{x} \right|}\left( \mathbb{I}_{out,x}\ \bigotimes\ \left.  \left\langle 0 \right. \right|_{P_{x}} \right)\mathcal{U}_{x}\ ,\ \ \ \ \ \ \mathcal{J}_{out,x} < \mathcal{J}_{in,x}}
\label{eq7}
\end{equation}
The unitary operator \(\mathcal{U}_{x}\) has the same dimension as \(\mathcal{J}_{in,x}\), effectively characterizes the black hole horizon, and maps
\begin{equation}
{\mathcal{U}_{x}:\mathcal{H}_{\mathcal{J}_{in,x}} \rightarrow \mathcal{H}_{\mathcal{J}_{out,x}}\bigotimes\ \mathcal{H}_{P_{x}}}
\end{equation}
The state \(\left. \left\langle 0 \right. \right|_{P}\) is defined as the state implementing postselection on a subsystem of dimension \(\left| P_{x} \right|\), yielding the non-isometric tensor \(H_{x}\). In addition to its role as a postselection variable, the prefactor \(\sqrt{\left| P_{x} \right|}\) also serves as a normalization-preserving coefficient.

Combining Eqs.~\ref{eq6} and~\ref{eq7}, we obtain the full representation of \(H_{x}\):
\begin{equation}
{H_{x} = \sqrt{\left| P_{x} \right|}\left( \mathbb{I}_{out,x}\ \bigotimes\ \left. \left\langle 0 \right. \right|_{P_{x}} \right)\mathcal{U}_{x}\left( \mathbb{I}_{in,x}\ \bigotimes\ \left.\left| 0 \right.  \right\rangle_{A_{x}} \right)}
\end{equation}
This is the holographic inner-code tensor-network representation providing the linear map \(H_{x}:\mathbb{C}^{D^{\mathcal{J}_{in,x}}} \rightarrow \mathbb{C}^{D^{\mathcal{J}_{out,x}}}\) on a single pentagonal cell in the \{5,4\} holographic tessellation, describing a black hole with event horizon $\Lambda_{\mathrm{BH}}$. This equation has a form similar to the constructions in Ref.~\cite{ref15} and Ref.~\cite{ref16}, applied to a single pentagonal cell in the \{5,4\} tessellation.

We next construct a global non-isometric representation of the tensor \(H_{x}\), generalizing the PHEVA holographic map formulation in Ref.~\cite{ref15} and Ref.~\cite{ref16} to include a notion of locality through hyperbolic tensor networks originating from the microscopic constituents of an entanglement network of LQG quantum tetrahedra. We also examine the non-isometric properties of this global representation of \(H_{x}\) using the conditions of the PHEVA code and their relation to quantum extremal surfaces (QES) and generalized entropy.

\subsection{Global Non-Isometric Encoding Operator from the Holographic Inner Code}\label{subsec:Global Non-Isometric Encoding Operator from the Holographic Inner Code}

We construct a global non-isometric encoding operator from the tensor \(H_{x}\), previously built from \(T_{\left\{ 5,4 \right\}}^{(eff\ LQG)}\), an effective tensor equivalent to the six-leg AME perfect tensor \(W\) and constructed from the tensor map \(\mathcal{N}\) and the holonomy-dressed spin-network state amplitude \(M_{\left( j_{l} \right)}\). We first define $v$ (vertices), \(l_{in}\) (links representing legs between tetrahedra in the bulk), and \(l_{out}\) (links representing boundary legs), which satisfy the following relation in the \{5,4\} tessellation:
\begin{equation}
{5v = 2l_{in} + l_{out}}
\label{eq8}
\end{equation}
Relating these quantities to the notation for the number of legs in each cell gives \(\sum_{}^{}s_{x} = l_{in}\), and hence
\begin{equation}
{\prod_{x}^{}D^{\left( k_{x} - 1 \right)/2} = D^{\sum_{}^{}s_{x}/2} = D^{l_{in}/2}}
\end{equation}

In our treatment of the non-isometric inner code for the black hole considered here, \(H_{x}\) is identified as a unitary drawn from the Haar measure acting on the entire Hilbert space \(\mathcal{H}_{l}\ \bigotimes\ \mathcal{H}_{A}\ \bigotimes\ \mathcal{H}_{r}\). Accordingly, we write the global non-isometric inner code as
\begin{equation}
{V_{\Gamma = \left\{ 5,4 \right\}} = \prod_{x}^{}H_{x} = \prod_{x}^{}\left\lbrack \sqrt{\left| P_{x} \right|}\left( \mathbb{I}_{out,x}\ \bigotimes\ \left.\left\langle 0 \right. \right|_{P_{x}} \right)\mathcal{U}_{x}\left( \mathbb{I}_{in,x}\ \bigotimes\ \left. \left| 0 \right. \right\rangle_{A_{x}} \right) \right\rbrack}
\end{equation}
Here \(\sqrt{|P|} = \prod_{x}^{}\sqrt{\left| P_{x} \right|}\), \(\left.\left| 0 \right.  \right\rangle_{A} = \bigotimes\ \left. \left| 0 \right.\right\rangle_{A_{x}}\), \(\left. \left| 0 \right. \right\rangle_{P} = \bigotimes\ \left. \left| 0 \right. \right\rangle_{P_{x}}\), and \(\mathcal{U}_{\Gamma = \left\{ 5,4 \right\}} = \prod_{x}^{}\mathcal{U}_{x}\). Thus,
\begin{equation}
{V_{\Gamma = \left\{ 5,4 \right\}} = \sqrt{|P|}\left( \mathbb{I}_{out}\ \bigotimes\ \left. \left\langle 0 \right.\right|_{P} \right)\mathcal{U}_{\Gamma = \left\{ 5,4 \right\}}\left( \mathbb{I}_{in}\ \bigotimes\ \left. \left| 0 \right. \right\rangle_{A} \right)}
\label{eq10}
\end{equation}

The Hilbert spaces of the ancilla and postselection states satisfy \(\mathcal{H}_{A} = \bigotimes\mathcal{H}_{A_{x}}\) and \(\mathcal{H}_{P} = \bigotimes\mathcal{H}_{P_{x}}\), respectively. The global non-isometric operator \(V_{\Gamma = \left\{ 5,4 \right\}}\) provides the generic holographic map \(\mathcal{U\ :\ }lr \rightarrow B\), with
\begin{equation}
{V_{\Gamma = \left\{ 5,4 \right\}}\ :\ \mathcal{H}_{l}\ \bigotimes\ \mathcal{H}_{A}\ \bigotimes\ \mathcal{H}_{r} \rightarrow \mathcal{H}_{B}\bigotimes\ \mathcal{H}_{P}}
\label{eq9}
\end{equation}
Here \(V_{\Gamma = \left\{ 5,4 \right\}}\) is an encoding operator capable of annihilating states in the Hilbert space \(\mathcal{H}_{l}\ \bigotimes\ \mathcal{H}_{A}\ \bigotimes\ \mathcal{H}_{r}\), thereby satisfying the linear map \(\mathcal{H}_{l}\ \bigotimes\ \mathcal{H}_{A}\ \bigotimes\ \mathcal{H}_{r} \rightarrow \mathcal{H}_{B}\bigotimes\ \mathcal{H}_{P}\). This annihilation results from the non-isometric nature of the encoding operator. 

Consider
\begin{equation}
{\log_{D}\left| A_{x} \right| - \log_{D}\left| P_{x} \right| = 2\left( 3 - k_{x} \right) - 2\left( k_{x} - 3 \right) = 4\left( 3 - k_{x} \right)}
\end{equation}
and
\begin{equation}
{\log_{D}|A| - \log_{D}|P| = \sum_{x}^{}{4\left( 3 - k_{x} \right)} = 4\left( 2v - l_{in} \right)}
\end{equation}
Using the relation Eq.~\ref{eq8}, we obtain
\begin{equation}
{\log_{D}|A| - \log_{D}|P| = 2\left( l_{out} - v \right)}
\end{equation}
and hence
\begin{equation}
{D^{2v}|A| = D^{2l_{out}}|P|}
\end{equation}
Identifying this with the mapping implemented by the global non-isometric operator (see Eq.~\ref{eq9}), we obtain
\begin{equation}
{\dim\left( \mathcal{H}_{l}\bigotimes\ \mathcal{H}_{r}\bigotimes\ \mathcal{H}_{A} \right) = \dim\left( \mathcal{H}_{B}\bigotimes\ \mathcal{H}_{P} \right) \leftrightarrow D^{2v}|A| = D^{2l_{out}}|P|}
\end{equation}
With \(\dim\left( \mathcal{H}_{l}\bigotimes\ \mathcal{H}_{r} \right) = D^{2v} = |l||r|\), \(\dim\left( \mathcal{H}_{A} \right) = |A|\), \(\dim\left( \mathcal{H}_{B} \right) = D^{2l_{out}} = |B|\), and \(\dim\left( \mathcal{H}_{P} \right) = |P|\). We have therefore shown that our representation of the global non-isometric operator \(V_{\Gamma = \left\{ 5,4 \right\}}\), whose legs represent entanglement between tetrahedra, consistently interprets the input and output dimensions of the unitary operator \(\mathcal{U}_{\Gamma = \left\{ 5,4 \right\}}\).

Next, we demonstrate the non-isometric nature of \(V_{\Gamma = \left\{ 5,4 \right\}}\) in Eq.~\ref{eq10} by following the same procedure as in the PHEVA code, namely Haar integration. We first consider an ensemble of unitary operators \(\mathcal{U}_{\Gamma = \left\{ 5,4 \right\}}\) as exact unitaries in the group $U(N)$:
\begin{equation}
{\mathcal{U}_{\Gamma = \left\{ 5,4 \right\}}\sim \operatorname{Haar}\ U(N)}
\end{equation}
where $N$ denotes the Hilbert-space dimension:
\begin{equation}
{N = D^{2v}|A| = D^{2l_{out}}|P|}
\end{equation}
We assume that \(\mathcal{U}_{\Gamma = \left\{ 5,4 \right\}}\sim \operatorname{Haar}\ U(N)\) describes a random variable in the matrix representation of $U(N)$ with a specified probability distribution. In our model, links between tetrahedra carry holonomies belonging to $SU(2)$, while the Gauss constraint restricts states to an $SU(2)$-invariant code space; the unitary operator \(\mathcal{U}_{\Gamma = \left\{ 5,4 \right\}}\) is assumed to be Haar-distributed on $U(N)$. This Haar assumption is an additional statistical assumption, rather than a direct consequence of the Gauss constraint. With appropriate input--output factorization and dimensions, the global encoding operator \(\mathcal{U}_{\Gamma = \left\{ 5,4 \right\}}\) can retain a form analogous to the PHEVA code.

We next examine the isometric properties of the global encoding operator at the local scale, namely the action of \(V_{\Gamma = \left\{ 5,4 \right\}}\) on states inside the black hole. We first consider
\begin{equation}
\begin{split}
V_{\Gamma=\{5,4\}}^{\dagger} V_{\Gamma=\{5,4\}} 
&= \sqrt{|P'||P|} \bigl( (\mathbb{I}_{\mathrm{in}} \otimes \langle 0|_{A'}) \mathcal{U}_{\Gamma=\{5,4\}}^{\dagger} (\mathbb{I}_{\mathrm{out}} \otimes |0\rangle_{P'}) \\
&\quad (\mathbb{I}_{\mathrm{out}} \otimes \langle 0|_{P}) \mathcal{U}_{\Gamma=\{5,4\}} (\mathbb{I}_{\mathrm{in}} \otimes |0\rangle_{A}) \bigr)
\end{split}
\end{equation}
Using the identity
\begin{equation}
{\mathcal{U}_{\Gamma = \left\{ 5,4 \right\}}^{\dagger}X\mathcal{U}_{\Gamma = \left\{ 5,4 \right\}} = \frac{\mathbb{I}_{N}}{N}\operatorname{Tr}(X)}
\end{equation}
with \(X = \left( \mathbb{I}_{out}\ \bigotimes\ \left.  \left| 0 \right.  \right\rangle_{P^{'}} \right)\left( \mathbb{I}_{out}\ \bigotimes\ \left.  \left\langle 0 \right.  \right|_{P} \right)\) and \(\operatorname{Tr}(X) = D^{2l_{out}}\), and identifying the indices \(A^{'} = A\) and \(P^{'} = P\), we obtain
\begin{equation}
{V_{\Gamma = \left\{ 5,4 \right\}}^{\dagger}V_{\Gamma = \left\{ 5,4 \right\}} = |P|\left( \left( \mathbb{I}_{in}\ \bigotimes\ \left. \left\langle 0 \right.  \right|_{A} \right)\frac{\mathbb{I}_{N}}{|P|}\left( \mathbb{I}_{in}\ \bigotimes\ \left.  \left| 0 \right. \right\rangle_{A} \right) \right) = \mathbb{I}_{in,N}}
\label{eq11}
\end{equation}
For states \(\left. \ \left| \psi_{1} \right.\  \right\rangle,\left. \ \left| \psi_{2} \right.\  \right\rangle \in \mathcal{H}_{l}\ \bigotimes\ \mathcal{H}_{r}\), Eq.~\ref{eq11} implies
\begin{equation}
{\left\langle \psi_{2} \middle| V_{\Gamma = \left\{ 5,4 \right\}}^{\dagger}V_{\Gamma = \left\{ 5,4 \right\}} \middle| \psi_{1} \right\rangle = \left\langle \psi_{2} \middle| \psi_{1} \right\rangle}
\label{eq24}
\end{equation}
The Eq.~\ref{eq24} shows that the inner product between \(\left. \ \left| \psi_{1} \right.  \right\rangle\) and \(\left. \ \left| \psi_{2} \right. \right\rangle\) is preserved locally under the action of the global encoding operator \(V_{\Gamma = \left\{ 5,4 \right\}}\). This demonstrates that \(V_{\Gamma = \left\{ 5,4 \right\}}\) exhibits isometric behavior at the local scale within the black hole interior. Thus, the discrete holographic tensor-network structure in the interior remains isometric, as it does outside the black hole horizon.

We next consider the global scale, namely the entire tiling describing the black hole, from the horizon $\Lambda_{\mathrm{BH}}$ to its center (see Fig.~\ref{fig:bhtessellation}). We introduce a reference system $L$ with $|L|=|\ell|$ and a reservoir $R$ with $|R|=|r|$, together with a second Haar moment over $U(N)$:
\begin{equation}
{X\bigotimes X = \alpha\mathbb{I}_{N^{2}} + \beta F_{N}}
\label{eq13}
\end{equation}
With \(\alpha\) and \(\beta\) are constants to be determined and depend on $N$, \(\alpha = \alpha(N);\ \beta = \beta(N)\), while \(F_{N}\) is the operator exchanging two states, in this case \(\left. \ \left| \psi_{1} \right.  \right\rangle,\left. \left| \psi_{2} \right.  \right\rangle \in \mathcal{H}_{l}\ \bigotimes\ \mathcal{H}_{r}\). Taking the trace of \(X\bigotimes X\) gives
\begin{equation}
{\operatorname{Tr}(X\bigotimes X) = \alpha\operatorname{Tr}\left( \mathbb{I}_{N^{2}} \right) + \beta\operatorname{Tr}\left( F_{N} \right)}
\label{eq12}
\end{equation}
The left-hand side of Eq.~\ref{eq12} becomes
\begin{equation}
\begin{split}
\operatorname{Tr}(X \otimes X) &= \operatorname{Tr}(X)\operatorname{Tr}(X) \\
&= \operatorname{Tr}\bigl( (\mathbb{I}_{\mathrm{out}} \otimes |0\rangle_{P'}) (\mathbb{I}_{\mathrm{out}} \otimes \langle 0|_{P}) \bigr) \\
&\quad \times \operatorname{Tr}\bigl( (\mathbb{I}_{\mathrm{out}} \otimes |0\rangle_{P'}) (\mathbb{I}_{\mathrm{out}} \otimes \langle 0|_{P}) \bigr) \\
&= \dim(\mathcal{H}_{B}) \times \dim(\mathcal{H}_{B}) = |B|^{2}.
\end{split}
\label{eq15}
\end{equation}
The right-hand side of Eq.~\ref{eq12} becomes:
\begin{equation}
{{\alpha\operatorname{Tr}\left( \mathbb{I}_{N^{2}} \right) + \beta\operatorname{Tr}\left( F_{N} \right) = \alpha N^{2} + \beta N}}
\label{eq16}
\end{equation}
From Eqs.~\ref{eq15} and~\ref{eq16} yielding
\begin{equation}
{|B|^{2} = \alpha N^{2} + \beta N}
\label{eq17}
\end{equation}
Next, multiplying the Eq.~\ref{eq13} and taking the trace gives
\begin{equation}
{\operatorname{Tr}\left( F_{N}(X\bigotimes X) \right) = \alpha\operatorname{Tr}\left( F_{N}\mathbb{I}_{N^{2}} \right) + \beta\operatorname{Tr}\left( {F_{N}}^{2} \right)}
\label{eq14}
\end{equation}
The left-hand side of Eq.~\ref{eq14} becomes 
\begin{equation}
{\operatorname{Tr}\left( F_{N}(X\bigotimes X) \right) = \operatorname{Tr}\left( X^{2} \right) = \operatorname{Tr}(X) = |B|}
\end{equation}
whereas the right-hand side of Eq.~\ref{eq14} becomes
\begin{equation}
{\alpha\operatorname{Tr}\left( F_{N}\mathbb{I}_{N^{2}} \right) + \beta\operatorname{Tr}\left( {F_{N}}^{2} \right) = \alpha\operatorname{Tr}\left( F_{N} \right) + \beta\operatorname{Tr}\left( F_{N^{2}} \right) = \alpha N + \beta N^{2}}
\end{equation}
and Eq.~\ref{eq14} yielding 
\begin{equation}
{|B| = \alpha N + \beta N^{2}}
\label{eq18}
\end{equation}
From Eqs.~\ref{eq17} and~\ref{eq18}, we obtain
\begin{equation}
{\alpha = \frac{|B|^{2} - |B|/N}{N^{2} - 1}\ ,\ \ \ \ \ \beta = \frac{|B| - |B|^{2}/N}{N^{2} - 1}}
\label{eq20}
\end{equation}
Next, defining \(G = V_{\Gamma = \left\{ 5,4 \right\}}^{\dagger}V_{\Gamma = \left\{ 5,4 \right\}}\), we have
\begin{equation}
\begin{split}
G \otimes G &= \left( V_{\Gamma=\{5,4\}}^{\dagger} V_{\Gamma=\{5,4\}} \right) \otimes \left( V_{\Gamma=\{5,4\}}^{\dagger} V_{\Gamma=\{5,4\}} \right) \\
&= |P|^{2} \left[ \alpha \mathbb{I}_{N^{2}} + \beta \left( Y_{0}^{\dagger} \otimes Y_{0}^{\dagger} \right) F_{N} \left( Y_{0} \otimes Y_{0} \right) \right].
\end{split}
\label{eq19}
\end{equation}
Here \(Y_{0} = \mathbb{I}_{in}\bigotimes\ \left. \ \left| 0 \right. \right\rangle_{A} = \mathbb{I}_{|l||r|}\bigotimes\ \left. \ \left| 0 \right.\right\rangle_{A}\) and \(Y_{0}^{\dagger} = \mathbb{I}_{in}\bigotimes\left. \ \left\langle 0 \right.\  \right|_{A} = \mathbb{I}_{|l||r|}\bigotimes\left. \ \left\langle 0 \right.\  \right|_{A}\), with \(\mathbb{I}_{in} = \mathbb{I}_{|l||r|}\) and \(\mathbb{I}_{out} = \mathbb{I}_{|B|}\). The second term of Eq.~\ref{eq19} gives
\begin{equation}
{{\left( Y_{0}^{\dagger} \otimes Y_{0}^{\dagger} \right)F_{N}\left( Y_{0}\bigotimes Y_{0} \right) = F_{N}}}
\label{eq21}
\end{equation}
Using Eqs.~\ref{eq20} and~\ref{eq21}, we obtain
\begin{equation}
{G\bigotimes G = \left( \frac{|P|^{2}|B|^{2} - |P|}{|P|^{2}|B|^{2} - 1} \right)\mathbb{I}_{N^{2}} + \left( \frac{|P||B|\left( |P| - 1 \right)}{|P|^{2}|B|^{2} - 1} \right)F_{N}}
\label{eq22}
\end{equation}
Applying Eq.~\ref{eq22} to \(s = \left\langle \psi_{1} \middle| \psi_{2} \right\rangle\) and \(z = \left\langle \psi_{2} \middle| G\bigotimes\mathbb{I}_{|B|} \middle| \psi_{1} \right\rangle\) gives
\begin{equation}
\begin{split}
(G \otimes G)|z|^{2} &= \biggl[ \left( \frac{|P|^{2}|B|^{2} - |P|}{|P|^{2}|B|^{2} - 1} \right) \mathbb{I}_{N^{2}} \\
&\quad + \left( \frac{|P||B|(|P| - 1)}{|P|^{2}|B|^{2} - 1} \right) F_{N} \biggr] \left| \left\langle \psi_{2} \middle| G \otimes \mathbb{I}_{|B|} \middle| \psi_{1} \right\rangle \right|^{2} \\
&= \left( \frac{|P|^{2}|B|^{2} - |P|}{|P|^{2}|B|^{2} - 1} \right) |z|^{2} + \left( \frac{|P||B|(|P| - 1)}{|P|^{2}|B|^{2} - 1} \right) \operatorname{Tr}(\rho_{1}, \rho_{2}).
\end{split}
\end{equation}
Here \(\rho_{i} = \left. \ \left| \psi_{i} \right.\  \right\rangle\left\langle \left. \ \psi_{i} \right| \right.\ \) and \(i = 1,2\). We then obtain
\begin{equation}
\begin{split}
|(G \otimes G)z - s|^{2} &= (G \otimes G)|z|^{2} - s^{*}(G \otimes G)z - s(G \otimes G)z^{*} + |s|^{2} \\
&= (G \otimes G)|z|^{2} - |s|^{2} \\
&= \left( \frac{|P| - 1}{|P|^{2}|B|^{2} - 1} \right) \left[ |P||B|\operatorname{Tr}(\rho_{1}, \rho_{2}) - |s|^{2} \right].
\end{split}
\label{eq23}
\end{equation}
The quantity \(\left| (G\bigotimes G)z - s \right|^{2}\) indicates whether the output overlaps of the global encoding operator $V$ have an isometric or non-isometric distribution. Eq.~\ref{eq23}  exhibits non-isometric behavior because \(\left| (G\bigotimes G)z - s \right|^{2} > 1\) in the global average of the encoding operator, except when \(\left| \widetilde{P} \right| = 1\), as required because in this case there is no postselection and \(\mathcal{U}_{\Gamma_{\left\{ 5,4 \right\} - BH}}\) is an isometry.

For \(|P|,|B| > 1\), we obtain
\begin{equation}
{\left| (G\bigotimes G)z - s \right|^{2} \leq \frac{1}{|B|}}
\label{eq25}
\end{equation}
The Eq.~\ref{eq25} shows that our model, like the PHEVA code, exhibits non-isometry within the stated bound. Since $|B|$ is exponential in the black hole entropy, although \(\mathcal{U}_{\Gamma_{\left\{ 5,4 \right\} - BH}}\) is not an isometry, deviations of the overlap between $(V_{\Gamma=\{5,4\}} \otimes \mathbb{I}_{|l||r|}) |\psi_{1}\rangle$ and $(V_{\Gamma=\{5,4\}} \otimes \mathbb{I}_{|l||r|}) |\psi_{2}\rangle$ from that between $|\psi_{1}\rangle$ and $|\psi_{2}\rangle$ are exponentially small in the entropy. Thus, $V_{\Gamma=\{5,4\}}$ almost always preserves the inner product.

\subsection{Generalized Entropy, Quantum Extremal Surfaces (QES), and the Page Curve}

The preceding discussion in Sec.~\ref{subsec:Global Non-Isometric Encoding Operator from the Holographic Inner Code} demonstrated that the encoding operator \(V_{\Gamma = \left\{ 5,4 \right\}}\) is non-isometric. One indication of non-isometry in the black hole model based on the \{5,4\} tessellation is the presence of quantum extremal surfaces (QES). A QES is a surface that makes the generalized entropy stationary under variations of its position or shape. The generalized entropy includes both a geometric area contribution and a quantum entanglement contribution from bulk fields. The QES position is therefore determined by both the geometry and the quantum state~\cite{ref75}. The generalized entropy of a black hole (\(S_{gen}(A)\)) can be written as in Eq. (2). The QES is selected when \(A_{QESs} = \min{S_{gen}(A)}\).

In the quantum extremal surface paradigm~\cite{ref14,ref16,ref75}, black hole evaporation begins with an early phase (before the Page time), during which the entropy continually increases as outgoing radiation in modes $R$ becomes entangled with infalling modes $r$, forming Hawking pairs. This process is described by the Hawking state~\cite{ref15}:
\begin{equation}
{\left. \left| \Psi_{H} \right.\right\rangle = \left.\left| \psi^{in} \right.  \right\rangle_{Ll}\bigotimes\left.\left| \psi^{out} \right.\right\rangle_{rR}}
\label{eq27}
\end{equation}
Here the region $L$ purifies the state of the infalling matter in mode $l$.

Following Ref.~\cite{ref15}, the global encoding operator \(V_{\Gamma = \left\{ 5,4 \right\}}\) gives rise to a large number of null states, producing a generalized entropy that follows the Page curve. Based on \(V_{\Gamma = \left\{ 5,4 \right\}}\) obtained above, we find the same result. In this subsection, we relate the entropy of a system $A$ in the fundamental and effective descriptions. We begin by expressing the Hawking state in Eq.~\ref{eq27} to its Schmidt decomposition:
\begin{equation}
{\left.\left| \psi^{out} \right.  \right\rangle_{rR} = \sum_{m}^{}\sqrt{d_{m}}\left.\left| m \right.\  \right\rangle_{r}\left.\left| m \right.\  \right\rangle_{R}\ ;\ \ \ \ \ \sum_{m}^{}{d_{m} = 1}}
\end{equation}
\begin{equation}
{\left.\left| \psi^{in} \right.  \right\rangle_{Ll} = \sum_{i}^{}\sqrt{c_{i}}\left.\left| i \right.  \right\rangle_{L}\left.\left| i \right.  \right\rangle_{l}\ ;\ \ \ \ \ \sum_{i}^{}{c_{i} = 1}}
\end{equation}
Thus,
\begin{equation}
{\left. \left| \Psi_{H} \right.\right\rangle = \sum_{i,m}^{}\sqrt{c_{i}d_{m}}\left.\left| i \right.\right\rangle_{L}\left.\left| i \right.  \right\rangle_{l}\left.\left| m \right.  \right\rangle_{r}\left.\left| m \right. \right\rangle_{R}}
\end{equation}
The density matrices are \(\rho_{R}^{out} = \sum_{m}^{}d_{m}\left. \left| m \right. \right\rangle\left\langle \left. \ m \right| \right.\ \) and \(\rho_{L}^{in} = \sum_{i}^{}c_{i}\left. \left| i \right. \right\rangle\left\langle \left. i \right| \right. \). We define
\begin{equation}
{{u_{n} = \sum_{i}^{}{c_{i}}^{n}}\ ,\ \ \ \ \ v_{n} = \sum_{m}^{}{d_{m}}^{n}}
\label{eq28}
\end{equation}
To calculate the von Neumann entropy, we employ the replica trick based on the R\'enyi entropy:
\begin{equation}
{S_{n} = - \left( \frac{1}{n - 1} \right)\log\left(\operatorname{Tr}\rho^{n} \right)}
\label{eq29}
\end{equation}
Here $n$ denotes the entropy order, which determines how strongly the entropy emphasizes dominant states relative to less probable states in the spectrum of the density matrix (\(\rho\)). Using the definitions from Eqs.~\ref{eq28} and the definition of the R\'enyi entropy~\ref{eq29}, we obtain
\begin{equation}
{u_{n} = \operatorname{Tr}\left( \rho_{L}^{in} \right)^{n}}\ ,\ \ \ \ \ {v_{n} = \operatorname{Tr}\left( \rho_{R}^{out} \right)^{n}}
\end{equation}
We define the Hawking state acted upon by the global encoding operator (\(V_{\Gamma = \left\{ 5,4 \right\}}\)) as
\begin{equation}
\begin{split}
|\widetilde{\Psi}_{H}\rangle &= \left( \mathbb{I}_{L} \otimes V_{\Gamma=\{5,4\}} \otimes \mathbb{I}_{R} \right) |\Psi_{H}\rangle \\
&= \sum_{i,m,\beta} \sqrt{c_{i} d_{m}} \left( V_{\Gamma=\{5,4\}} \right)_{\beta;im} |i\rangle_{L} |\beta\rangle_{|B|} |m\rangle_{R}.
\end{split}
\label{eq31}
\end{equation}
The trace of the reduced density matrix of the state~\ref{eq31} is
\begin{equation}
{\operatorname{Tr}{\widetilde{\rho}}_{R} = \operatorname{Tr}_{L}\left( \left. \left| {\widetilde{\Psi}}_{H} \right.  \right\rangle\left\langle \left.  {\widetilde{\Psi}}_{H} \right| \right.  \right) = \operatorname{Tr}\rho_{R}^{out} = \sum_{m}^{}d_{m}}
\end{equation}
Similarly, for \({\widetilde{\rho}}_{L}\), we have \(\operatorname{Tr}{\widetilde{\rho}}_{L} = \operatorname{Tr}\rho_{L}^{in} = \sum_{i}^{}c_{i}\), and at second order we obtain
\begin{equation}
\begin{split}
\operatorname{Tr}(\widetilde{\rho}_{R})^{2} &= \sum_{m,m'} (\widetilde{\rho}_{R})_{mm'} (\widetilde{\rho}_{R})_{m'm} \\
&= \sum_{i,i',m,m',\beta,\beta'} c_{i} c_{i'} d_{m} d_{m'} \left( V_{\Gamma=\{5,4\}} \right)_{\beta;im} \left( V_{\Gamma=\{5,4\}}^{\dagger} \right)_{\beta;im'} \\
&\quad \times \left( V_{\Gamma=\{5,4\}} \right)_{\beta';i'm} \left( V_{\Gamma=\{5,4\}}^{\dagger} \right)_{\beta';i'm}.
\end{split}
\label{eq34}
\end{equation}
Since \(V_{\Gamma = \left\{ 5,4 \right\}} = V_{\Gamma = \left\{ 5,4 \right\}}\left( \mathcal{U} \right)\) and \(\mathcal{U}_{\Gamma}\sim \operatorname{Haar}\ \left( U(N) \right)\), we use the following identity for the group $U(N)$:
\begin{equation}
\begin{split}
\mathcal{U}_{i_{1}j_{1}}\mathcal{U}_{i_{2}j_{2}}\mathcal{U}_{j_{1}'i_{1}'}^{\dagger}\mathcal{U}_{j_{2}'i_{2}'}^{\dagger} 
&= \frac{1}{N^{2} - 1} \biggl[ \delta_{i_{1}i_{1}'}\delta_{i_{2}i_{2}'}\delta_{j_{1}j_{1}'}\delta_{j_{2}j_{2}'} + \delta_{i_{1}i_{1}'}\delta_{i_{2}i_{2}'}\delta_{j_{1}j_{1}'}\delta_{j_{2}j_{2}'} \\
&\quad - \frac{1}{N}\left( \delta_{i_{1}i_{1}'}\delta_{i_{2}i_{2}'}\delta_{j_{1}j_{2}'}\delta_{j_{2}j_{1}'} + \delta_{i_{1}i_{2}'}\delta_{i_{2}i_{1}'}\delta_{j_{1}j_{1}'}\delta_{j_{2}j_{2}'} \right) \biggr]
\end{split}
\end{equation}
Eq.~\ref{eq34} then becomes
\begin{equation}
{\operatorname{Tr}\left( {\widetilde{\rho}}_{R} \right)^{2} = \frac{|P|^{2}|B|^{2}}{|P|^{2}|B|^{2} - 1}\left\lbrack \left( 1 - \frac{1}{|P||B|^2} \right)v_{2} + \left( 1 - \frac{1}{|P|} \right)\frac{u_{2}}{|B|} \right\rbrack}
\label{eq35}
\end{equation}
For \(|P|,|B| \gg 1\), Eq.~\ref{eq35} gives
\begin{equation}
{\operatorname{Tr}\left( {\widetilde{\rho}}_{R} \right)^{2} \approx v_{2} + \frac{u_{2}}{|B|}}
\label{eq36}
\end{equation}
Using the R\'enyi entropy formula (see Eq.~\ref{eq29} for $n=2$ and Eq.~\ref{eq36}), we obtain
\begin{equation}
{S_{2}\left( {\widetilde{\rho}}_{R} \right) \approx \min\left\{ S_{2}\left( \rho_{R}^{out} \right),\log|B| + S_{2}\left( \rho_{L}^{in} \right) \right\}}
\label{eq37}
\end{equation}

For general integer replica order ($n$th-order), we define the reduced matrix elements in the Schmidt basis are
\begin{equation}
 (\widetilde\rho_R)_{mm'}
 =p_{\mathrm{ps}}\sqrt{d_m d_{m'}}
 \sum_{i,\beta}c_i\,
 (U_{\Gamma})_{\beta0;im0}
 (U_{\Gamma})^*_{\beta0;im'0},
 \label{eq:replica_matrix_elements}
\end{equation}
where $\beta=1,\ldots,b$. Taking the trace of the $n$th power produces
$n$ matrix elements of $U_{\Gamma}$ and $n$ conjugate matrix elements.
Their Haar average is a sum over two permutations in the symmetric
group $S_n$.

Let $C(\pi)$ denote the number of cycles of $\pi\in S_n$, and define
\begin{equation}
 u_{\pi}=\prod_{c\in\mathrm{cyc}(\pi)}u_{|c|},
 \qquad
 v_{\pi}=\prod_{c\in\mathrm{cyc}(\pi)}v_{|c|},
 \qquad
 \gamma_n=(1\,2\,\cdots\,n),
 \label{eq:replica_cycle_notation}
\end{equation}
where $|c|$ is the length of a cycle. With permutation products acting
from right to left, the unitary Weingarten formula gives
\begin{equation}
\begin{split}
\mathcal{M}_n &\equiv \operatorname{Tr}(\widetilde{\rho}_{R})^{n} \\
&= p_{\mathrm{ps}}^{n} \sum_{\sigma,\tau \in S_n} b^{C(\sigma)}\, \operatorname{Wg}_N(\sigma^{-1}\tau)\, u_{\tau} v_{\gamma_n\tau}.
\end{split}
\label{eq:replica_weingarten_exact}
\end{equation}
Here $\operatorname{Wg}_N$ is the unitary Weingarten function, and we use its standard expression for $N\geq n$. The factor $b^{C(\sigma)}$ counts the closed $B$-index loops, while $u_{\tau}$ and $v_{\gamma_n\tau}$ arise from the incoming and outgoing Schmidt
indices, respectively. Eq.~\eqref{eq:replica_weingarten_exact}
is exact for each such integer $n$ and reproduces $\mathcal M_1=1$ and Eq.~\eqref{eq35} for $n=2$.

For fixed $n$, the large-$N$ expansion of the Weingarten function selects the diagonal contractions $\sigma=\tau$ at leading order,
provided the remaining dimension factors do not compensate the suppression of off-diagonal terms. In the large-postselection regime
used in Ref.~\cite{ref15}, this gives
\begin{equation}
 \mathcal M_n\simeq
 \sum_{\tau\in S_n}b^{C(\tau)-n}u_{\tau}v_{\gamma_n\tau}.
 \label{eq:replica_diagonal_moments}
\end{equation}
This expression is an expansion at fixed integer replica order; its uniformity near $n=1$ must be addressed separately. The identity permutation contributes $v_n$, whereas
$\tau=\gamma_n^{-1}$ contributes $b^{1-n}u_n$. 

\begin{figure}[t]
    \centering
\includegraphics[width=1\columnwidth]{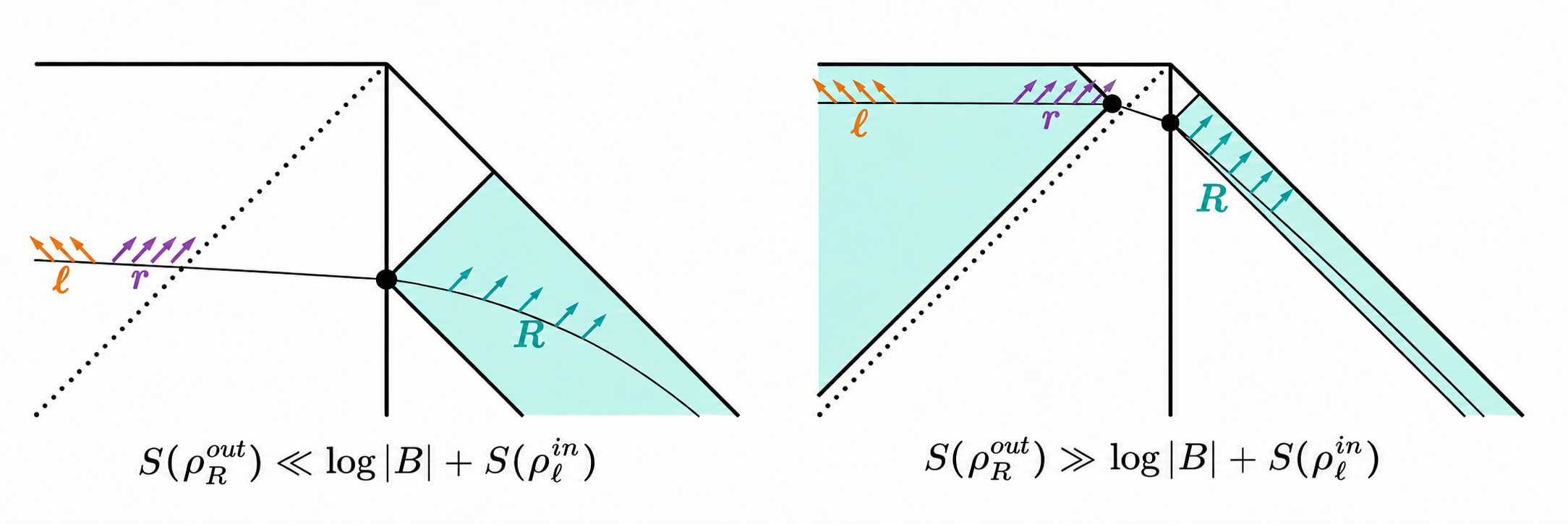}
    \caption{\small Determination of the radiation entropy through quantum extremal surfaces (QES). At the onset of evaporation, entanglement between radiation and matter in mode $r$ encompasses only the domain of dependence of region $(R)$, indicated by blue shading in the left panel. At later times, this region expands to include an island in the black hole interior. The island boundary, indicated by the second black point in the right panel, is the new QES.}
    \label{fig:structurehorizon2}
\end{figure}

Using the Eqs.~\ref{eq36} and ~\ref{eq:replica_diagonal_moments} takes the following form for the $n$th-order R\'enyi entropy:
\begin{equation}
{S_{n}\left( {\widetilde{\rho}}_{R} \right) \approx \min\left\{ S_{n}\left( \rho_{R}^{\left\{ out \right\}} \right),\log|B| + S_{n}\left( \rho_{L}^{\left\{ in \right\}} \right) \right\}}
\end{equation}
In the limit \(n \rightarrow 1\), the R\'enyi entropy yields the entanglement entropy:
\begin{equation}
{S\left( {\widetilde{\Psi}}_{H} \right) = \lim_{n \rightarrow 1}{S_{n}\left( {\widetilde{\rho}}_{R} \right)}}
\end{equation}
We therefore obtain the entanglement entropy
\begin{equation}
{S\left( {\widetilde{\Psi}}_{H} \right) \approx \min\left\{ S\left( \rho_{R}^{\left\{ out \right\}} \right),\ \log|B| + S\left( \rho_{L}^{\left\{ in \right\}} \right) \right\}}
\label{eq40}
\end{equation}
According to~\cite{ref15}, the minimum generalized entropy is proportional to the  entanglement entropy, namely
\begin{equation}
{S\left( {\widetilde{\Psi}}_{H} \right) = \min{S_{gen,\Gamma}\left( \rho_{R} \right)} = \min\left\{ S_{gen,\Gamma}(\varnothing,R),\ S_{gen,\Gamma}(I,R) \right\}}
\label{eq38}
\end{equation}
Here \(S_{gen,\Gamma}(\varnothing,R)\) is the entropy arising from Hawking radiation, \(\left( S_{gen,\Gamma}(\varnothing,R) = S_{Hawking} \right)\), whereas \(S_{gen,\Gamma}(I,R)\) in Eq.~\ref{eq38} can be decomposed as
\begin{equation}
{S_{gen,\Gamma}(I,R) = S_{A_{\Gamma}}(\partial I) + S_{bulk}(I \cup R)}
\label{eq39}
\end{equation}
From Eqs.~\ref{eq38} and~\ref{eq39} provide a mechanism for the formation of a QES, signaled by the appearance of an island in our model. Eq.~\ref{eq38} shows that the total black hole entropy follows the Page curve from the beginning to the end of evaporation. The process described here is illustrated in Fig.~\ref{fig:structurehorizon2}.

We first consider the onset of black hole evaporation, when the black hole entropy is dominated by the Hawking radiation entropy and no island has yet formed (\(I = \varnothing\)). Eq.~\ref{eq40} and~\ref{eq38} then give
\begin{equation}
{S_{gen,\Gamma}(\varnothing,R) = S\left( \rho_{R}^{\left\{ out \right\}} \right) \ll \log|B| + S\left( \rho_{L}^{\left\{ in \right\}} \right)}
\end{equation}
The black hole entropy in this phase is proportional to the Hawking radiation entropy:
\begin{equation}
{S\left( \rho_{R}^{\left\{ out \right\}} \right) = S_{Hawking}(\rho)}
\end{equation}
As time progresses, evaporation continues until a critical point, or the Page time, is reached. At this time the entropy attains its maximum, marked by the formation of a QES bounding an island $(I)$ behind the black hole horizon, where radiation in modes $r$ and $l$ accumulates. In this phase, we obtain
\begin{equation}
{S_{gen,\Gamma}(I,R) = \log|B| + S\left( \rho_{L}^{\left\{ in \right\}} \right) \ll S\left( \rho_{R}^{\left\{ out \right\}} \right)}
\end{equation}

With \(S_{A_{\Gamma}}(\partial I) = \log|B|\) and \(S_{bulk}(I \cup R) = S\left( \rho_{L}^{\left\{ in \right\}} \right)\). The presence of an island can reduce the second term in \(S_{gen,\Gamma}(I,R)\) because some entanglement partners of the radiation are now included in \((I \cup R)\). In this phase, non-isometry develops at the black hole horizon (see Fig.~\ref{fig:structurehorizon2}), as represented by the encoding operator (see the encoding-operator equation). Finally, in the late evaporation phase (after the Page time), the entropy decreases as the entangled radiation decreases in the presence of the island. In this phase, the island causes the entropy to be calculated from both \(\rho_{R}^{\left\{ out \right\}}\) and the $r$ and $l$ modes. At the end of evaporation, the information contained in the black hole is returned to the exterior, providing an alternative resolution of the black hole information paradox.

For specified time-dependent states and dimensions, a change of dominant branch can occur when
\begin{equation}
 S(\rho_R^{\left\{ out \right\}}(t_{\mathrm P}))
 =\log|B| +S(\rho_L^{\left\{ in \right\}}(t_{\mathrm P})).
 \label{eq:replica_page_crossing}
\end{equation}
This is the leading crossover condition. A Page-curve maximum requires the appropriate growth and decrease of the competing branches; neither that behavior nor a unique crossing follows from the replica calculation alone. Moreover, determining a quantum extremal surface requires varying the generalized entropy over admissible surfaces, or specifying and analyzing the corresponding discrete cut prescription \cite{ref13}. The crossover in Eq.~\eqref{eq:replica_page_crossing}
does not determine its location or motion. Finally, parameters that specify only the incoming or outgoing entanglement entropy do not fix the full sequences $u_n$ and $v_n$. A replica calculation for such parameter families must also specify the underlying Schmidt spectra.

Finally, in our model, the number of boundary legs in the fundamental description, represented by \(|B|\), is related to entanglement between quantum tetrahedra carrying holonomies and can be represented as an area operator in LQG~\cite{ref19,ref20,ref38}:
\begin{equation}
{\log|B| = \sum_{l}^{}{\log\left( 2j_{l} + 1 \right)}}
\end{equation}
Therefore, as discussed in the Sec.~\ref{Relation to the LQG Area Spectrum}, we interpret \(\log|B|\) as a quantum Ryu - Takayanagi formula using the LQG area operator (see Eq.~\ref{eq:s3-src127}):
\begin{equation}
{S_{A_{\Gamma}}(\partial I) = \log|B| = \frac{A}{4l_{Pl}^{2}}}
\end{equation}
Thus, in addition to reproducing precisely the results described by the PHEVA code and other studies of black hole evaporation, our model provides a microscopic physical interpretation of the origin of space as a network formed by entanglement between quantum tetrahedra, constituting the discrete structure of space at the Planck scale.

\section{Conclusion}
\label{sec:conclusions}

We have developed a constructive connection between quantum tetrahedra in loop quantum gravity (LQG) and holographic tensor networks on hyperbolic tessellations. The construction starts from four-valent intertwiner spaces and contractions of holonomy-dressed Bell states. These structures provide the microscopic data required to construct effective tensors with distinct network legs and logical indices. Within this framework, holonomies implement the parallel transport of internal frames along the links, while intertwiners enforce gauge invariance at the nodes. The connection between the spin-network description and the holographic tensor network is therefore expressed through an explicit linear map.

The main result of the local construction is the ability to synthesize an effective tensor proportional to a perfect tensor through an appropriate choice of microscopic embedding. When the transfer map has full row rank, its right inverse determines the component of the embedding that reproduces the target tensor, while the freedom in the kernel allows the embedding to satisfy the isometry condition. This mechanism is implemented for a pentagonal cell constructed from six tetrahedra and extended to the family $p=2n-1$, provided that the local dimension supports the required absolutely maximally entangled state. These results establish the existence of microscopic data that exactly realize the holographic code structure. However, this existence result does not establish that perfect tensors arise generically from arbitrary LQG states; the choice of embedding remains an essential part of the construction.

At the network level, local isometry yields an isometric bulk-to-boundary map when the contraction pattern satisfies the radial-orientation conditions and the bound on the number of input legs at each cell. These requirements delimit the class of networks covered by the construction: hyperbolicity of the tessellation alone does not guarantee global isometry. For a cut that induces isometric maps on both sides, the entropy of a boundary region takes the form
\begin{equation}
S(A) = S_{bulk}\left( \rho_{a} \right) + \sum_{l \in \gamma_{c}}^{}{\log{(2j_{l} + 1)}}
\label{eq:s3-src120}
\end{equation}
This expression shows how microscopic link contributions and bulk entanglement combine into the structure of generalized entropy. Its interpretation as a quantum Ryu--Takayanagi formula with a minimum-cut prescription additionally requires that the cut belong to the relevant class of candidates homologous to $A$ and minimize the generalized entropy.

The relation between the link contribution and geometric area also requires a careful distinction. The quantity $\log(2j_{\ell}+1)$ measures the entanglement capacity of a link, whereas the LQG area spectrum depends on $\sqrt{j_{\ell}(j_{\ell}+1)}$. Since these quantities have different spin dependence, identifying the link contribution with $A/(4\ell_{\mathrm P}^{2})$ requires an explicitly specified choice of spin sector, parameters, or coarse-graining prescription. The network entropy therefore does not, by itself, constitute a universal derivation of the Bekenstein--Hawking coefficient for arbitrary spin distributions. This distinction separates the exact algebraic results of the network construction from their interpretation in terms of semiclassical gravitational geometry.

The application to black-hole interiors extends the construction through non-isometric encoding based on postselection. When the dimension of the effective interior Hilbert space exceeds that of the fundamental black-hole Hilbert space, the encoding map cannot preserve all inner products and necessarily has a nontrivial kernel. An additional Haar-ensemble assumption then permits a statistical analysis of overlaps and density-matrix moments. Preservation of inner products on average must be distinguished from exact isometry for each realization of the encoding. Moreover, the Haar distribution on the effective unitary space is an additional assumption that has not been derived from the SU(2) holonomies, the Gauss constraint, or the microscopic dynamics of the tetrahedra.

Within the two-branch approximation, the radiation entropy discussed
here has the structure
\begin{equation}
S(R)
\approx\min\!\left\{
S\!\left(\rho_R^{\left\{ out \right\}}\right),\log|B|+
S\!\left(\rho_L^{{\left\{ in \right\}}}\right)\right\}.
\label{eq:conclusion_radiation_entropy}
\end{equation}
A change in the dominant branch provides a mechanism compatible with the Page transition when the individual contributions evolve under suitable conditions. However, the crossing of two entropy branches does not determine the location or motion of a quantum extremal surface. Such a determination requires the dependence of the generalized entropy on candidate surfaces or network cuts. Likewise, the radiation entropy need not return to zero if the initial or infalling matter is entangled with an external purifying system that is excluded from the radiation subsystem. A vanishing entropy at the end of evaporation requires appropriate purity conditions and a corresponding choice of subsystem; it does not follow solely from the shrinking dimension of the black-hole Hilbert space.

Further work should connect the mathematical existence results with the stability and practical implementation of the construction. Numerical calculations of the singular-value spectrum of the transfer map, its sensitivity to holonomies, and the normalization factor of the embedding can determine whether perfect-tensor synthesis remains stable under perturbations of the microscopic configuration. Extensions to inhomogeneous spins and approximately perfect tensors are also needed to assess the robustness of the isometry conditions and entropy formula under departures from the ideal construction. In parallel, the effective encoding ensemble should be derived from a specified dynamics or local quantum circuit, allowing the statistical assumptions used in the black-hole application to be tested directly.

Furthermore, a complementary direction is the implementation of the tessellated tensor-network construction on quantum processors. In the $j=1/2$ sector, the two-dimensional four-valent intertwiner space provides a natural encoding of a logical qubit associated with a quantum tetrahedron. Building on quantum simulations of intertwiners and holonomy-based tensor networks~\cite{ref36,ref38}, quantum circuits could be developed to prepare the constituent states, implement holonomy-dependent link operations, and realize the effective embedding for finite tessellation patches. Such implementations would allow the isometry conditions, boundary entanglement, and sensitivity to microscopic perturbations to be tested directly. For non-isometric encodings, ancilla-assisted postselection would also permit an assessment of success probabilities and resource costs. Comparisons between ideal simulations and noisy circuit implementations would clarify the robustness of the construction under imperfect state preparation and gate operations.

A central issue is the temporal resolution of evaporation on hyperbolic tessellations. In the row-removal dynamics studied by Ref.~\cite{ref16}, the coarse time step can exceed the Page time, preventing the transition in the dominant generalized entropy from being resolved. This limitation motivates circuit updates at the level of individual cells or links, together with a consistent prescription for the changing horizon and interior encoding. Such refinements could test whether the transition becomes resolvable while preserving the required gauge and encoding constraints. They would also distinguish the appearance of a nontrivial candidate quantum extremal surface from the later change in its global dominance.

\section*{Acknowledgments}
DMA thanks Hadyan L. Prihadi for helpful discussions on this topic. Research of FPZ is supported by PPMI FMIPA ITB and the
Ministry of Higher Education, Science, and Technology (Kemendiktisaintek).

\section*{Appendix A: Proof of Theorem 1}
\label{app:A}

This appendix provides a proof theorem 1 about the microscopic six-tetrahedron graph used to realize one pentagonal cell of the tessellation $\{5,4\}$, the microscopic transfer map
\begin{equation*}
M^{(j_l)}(\mathbf h_l):\mathbb C^{d_l^7}\longrightarrow\mathbb C^{d_l^5},
\qquad d_l=2j_l+1,
\end{equation*}
generically has full row rank,
\begin{equation*}
\operatorname{rank}M^{(j_l)}(\mathbf h_l)=d_l^5,
\end{equation*}
for Haar-almost every internal holonomy configuration $\mathbf h_l\in SU(2)^9$. The proof strategy is to choose one $d_l^5\times d_l^5$ maximal minor, reduce it by fixing two intertwiner channels to the singlet $i_0=i_2=0$, factorize the minor into one invertible one-leg transfer and one $K_4$ network, and then show that the determinant of the $K_4$ network is not the identically zero function by using complexification to $SL(2,\mathbb C)$, rank-one degeneration of the spin-$j_l$ representation, and a local triangular-intertwiner lemma with a diagonal that never vanishes. After generic surjectivity is proven, the right inverse of $M^{(j_l)}$ permits the synthesis of an arbitrary effective five-leg output tensor. In particular, when the local dimension $d_l$ supports a six-party perfect tensor, a coefficient tensor $\mathcal N$ can be chosen such that the effective six-leg tensor is proportional to a perfect tensor and simultaneously satisfies $\mathcal N^\dagger\mathcal N=I$. This yields an exact local HaPPY-like isometry and, under a radial contraction of a finite $\{5,4\}$ patch, an exact global bulk-to-boundary isometry after scalar normalization.

\subsection*{A.1. Statement to be proved}

We use the internal six-tetrahedron graph
\begin{equation*}
V=\{0,1,2,3,4,5\}
\end{equation*}
with nine internal links (see. Eq.~\eqref{eq:s3-src31} and Fig.~\ref{fig:microscopic realization})
\begin{align}
l_1&=(0,1),\qquad l_2=(0,2),\qquad l_3=(0,4),\label{eq:appA-src1}\\
l_4&=(1,2),\qquad l_5=(4,5),\qquad l_6=(1,5),\label{eq:appA-src2}\\
l_7&=(2,3),\qquad l_8=(3,4),\qquad l_9=(3,5).\label{eq:appA-src3}
\end{align}
For homogeneous spin,
\begin{equation}
j=j_l,
\qquad d_l\equiv d_{j_l}=2j_l+1.
\label{eq:appA-src4}
\end{equation}
The microscopic transfer map is defined by
\begin{equation}
M^{(j_l)}(\mathbf h_l):\mathbb C^{d_l^7}\longrightarrow\mathbb C^{d_l^5},
\label{eq:appA-src5}
\end{equation}
with rows labeled by the five open pentagonal magnetic legs
\begin{equation*}
\mathbf b=(b_1,b_2,b_3,b_4,b_5),
\end{equation*}
and columns labeled by
\begin{equation*}
A=(a,i_0,i_1,i_2,i_3,i_4,i_5).
\end{equation*}
Our goal is to prove
\begin{equation}
\operatorname{rank}M^{(j_l)}(\mathbf h_l)=d_l^5
\label{eq:appA-src6}
\end{equation}
for generic internal holonomies
\begin{equation*}
\mathbf h_l=(h_1,\ldots,h_9)\in SU(2)^9.
\end{equation*}
More strongly, we will show that the bad set
\begin{equation*}
Z_{j_l}\equiv\left\{\mathbf h_l\in SU(2)^9:\operatorname{rank}M^{(j_l)}(\mathbf h_l)<d_l^5\right\}
\end{equation*}
has Haar measure zero. Thus,
\begin{equation}
\operatorname{rank}M^{(j_l)}(\mathbf h_l)=d_l^5
\quad\text{for Haar-almost every }\mathbf h_l\in SU(2)^9.
\label{eq:appA-src7}
\end{equation}

\subsection*{A.2. Microscopic transfer tensor}

Use the normalized four-valent recoupling basis
\begin{equation}
\iota_{i}^{m_1m_2m_3m_4}
=
\sum_{M=-i}^{i}
\frac{(-1)^{i-M}}{\sqrt{2i+1}}
C^{iM}_{j_lm_1,j_lm_2}
C^{i,-M}_{j_lm_3,j_lm_4},
\label{eq:appA-src8}
\end{equation}
with
\begin{equation*}
i=0,1,\ldots,2j_l.
\end{equation*}
For equal spins,
\begin{equation*}
\dim\operatorname{Inv}_{SU(2)}\left(V_{j_l}^{\otimes4}\right)=2j_l+1=d_l,
\end{equation*}
so every intertwiner index $i_v$ has exactly $d_l$ values. The invariant metric on $V_{j_l}$ is written as
\begin{equation}
g^{(j_l)}_{mn}=(-1)^{j_l-m}\delta_{m,-n}.
\label{eq:appA-src9}
\end{equation}
For the rank calculation, the normalization factor $1/\sqrt{d_l}$ of the Bell state can be ignored because it is a nonzero scalar. We therefore use the unnormalized Bell-holonomy matrix
\begin{equation}
U_l\equiv D^{(j_l)}(h_l)g^{(j_l)}.
\label{eq:appA-src10}
\end{equation}
Since $D^{(j_l)}(h_l)$ and $g^{(j_l)}$ are invertible, every $U_l$ is invertible.

With the link orientations given above, the microscopic transfer matrix has components
\begin{align}
&\left(M^{(j_l)}\right)_{b_1b_2b_3b_4b_5;\,a,i_0i_1i_2i_3i_4i_5}
\nonumber\\
&=\sum_{\{m\}_{\rm int}}
\iota_{i_0}^{a,m_{01},m_{02},m_{03}}
\iota_{i_1}^{b_1,m_{14},m_{15},m_{16}}
\iota_{i_2}^{b_2,m_{21},m_{24},m_{27}}
\nonumber\\
&\quad\times
\iota_{i_3}^{b_3,m_{35},m_{38},m_{39}}
\iota_{i_4}^{b_4,m_{42},m_{47},m_{48}}
\iota_{i_5}^{b_5,m_{53},m_{56},m_{59}}
\nonumber\\
&\quad\times
(U_1)_{m_{21}m_{01}}(U_2)_{m_{42}m_{02}}(U_3)_{m_{53}m_{03}}
(U_4)_{m_{24}m_{14}}(U_5)_{m_{35}m_{15}}
\nonumber\\
&\quad\times
(U_6)_{m_{56}m_{16}}(U_7)_{m_{47}m_{27}}(U_8)_{m_{48}m_{38}}(U_9)_{m_{59}m_{39}}.
\label{eq:appA-src11}
\end{align}
All sums in Eq.~\eqref{eq:appA-src20} run over magnetic values $-j_l,-j_l+1,\ldots,j_l$.

\subsection*{A.3. Reduction to one maximal minor}

To prove full row rank of a $d_l^5\times d_l^7$ matrix, it is sufficient to find one $d_l^5\times d_l^5$ maximal minor with nonzero determinant. We choose the columns with
\begin{equation}
i_0=0,
\qquad i_2=0,
\label{eq:appA-src12}
\end{equation}
while the five indices
\begin{equation*}
(a,i_1,i_3,i_4,i_5)
\end{equation*}
are left free. Since each has $d_l$ values, the number of selected columns is exactly $d_l^5$. Define the square submatrix
\begin{equation}
\left(S^{(j_l)}\right)_{b_1b_2b_3b_4b_5;\,a,i_1,i_3,i_4,i_5}
=
\left(M^{(j_l)}\right)_{b_1b_2b_3b_4b_5;\,a,0,i_1,0,i_3,i_4,i_5}.
\label{eq:appA-src13}
\end{equation}
If
\begin{equation*}
\det S^{(j_l)}\neq0,
\end{equation*}
then the rank of $M^{(j_l)}$ is at least $d_l^5$, but it cannot exceed the number of rows $d_l^5$. Therefore,
\begin{equation*}
\det S^{(j_l)}\neq0
\quad\Longrightarrow\quad
\operatorname{rank}M^{(j_l)}=d_l^5.
\end{equation*}
Thus the rank problem is reduced to proving
\begin{equation}
\det S^{(j_l)}\not\equiv0
\label{eq:appA-src14}
\end{equation}
as a function of the internal holonomies.

\subsection*{A.4. Why the choice $i_0=i_2=0$ simplifies the network}

For the recoupling channel $i=0$, the Clebsch--Gordan coefficient is the singlet,
\begin{equation}
C^{00}_{j_lm,j_ln}
=
\frac{(-1)^{j_l-m}}{\sqrt{d_l}}\delta_{n,-m}.
\label{eq:appA-src15}
\end{equation}
Substitution into the intertwiner basis gives the factorized singlet intertwiner
\begin{equation}
\iota_0^{m_1m_2m_3m_4}
=
\frac{1}{d_l}g_{m_1m_2}g_{m_3m_4}
\label{eq:appA-src16}
\end{equation}
up to an overall convention-dependent phase that does not affect the rank. Thus, at vertex $0$,
\begin{equation}
\iota_0^{a,m_{01},m_{02},m_{03}}
=
\frac{1}{d_l}g_{a m_{01}}g_{m_{02}m_{03}},
\label{eq:appA-src17}
\end{equation}
and at vertex $2$,
\begin{equation}
\iota_0^{b_2,m_{21},m_{24},m_{27}}
=
\frac{1}{d_l}g_{b_2m_{21}}g_{m_{24}m_{27}}.
\label{eq:appA-src18}
\end{equation}
The interpretation is important. At vertex $0$, the pair $(a,l_1)$ separates algebraically from the pair $(l_2,l_3)$. At vertex $2$, the pair $(b_2,l_1)$ separates from $(l_4,l_7)$. Consequently, the maximal minor $S^{(j_l)}$ factorizes into one transfer matrix connecting $a$ with $b_2$ and one remaining four-vertex network.

\subsection*{A.5. Direct transfer $a\rightarrow b_2$}

The part passing through the link $l_1=(0,1)$ is
\begin{equation}
X_{b_2a}
=
\sum_{m_{21},m_{01}}
g_{b_2m_{21}}(U_1)_{m_{21}m_{01}}g_{a m_{01}}.
\label{eq:appA-src19}
\end{equation}
In matrix notation,
\begin{equation}
X=gU_1g^T,
\label{eq:appA-src20}
\end{equation}
up to harmless transpose/convention choices. Since $g$ and $U_1$ are invertible,
\begin{equation}
\det X\neq0.
\label{eq:appA-src21}
\end{equation}
Thus, the open microscopic index $a$ is mapped invertibly to the output index $b_2$.

\subsection*{A.6. Reduction of the remaining graph to $K_4$}

The singlet contraction through
\begin{equation*}
l_2-(v=0)-l_3
\end{equation*}
produces one effective edge connecting vertices $4$ and $5$. Likewise,
\begin{equation*}
l_4-(v=2)-l_7
\end{equation*}
produces one effective edge between vertices $1$ and $4$. The four internal edges already present are
\begin{equation*}
(1,3),\quad(1,5),\quad(3,4),\quad(3,5).
\end{equation*}
Thus, the subgraph on vertices
\begin{equation*}
\{1,3,4,5\}
\end{equation*}
becomes the complete graph
\begin{equation}
K_4.
\label{eq:appA-src22}
\end{equation}
Consequently, up to row/column permutations and an overall nonzero scalar,
\begin{equation}
S^{(j_l)}\simeq X\otimes K^{(j_l)},
\label{eq:appA-src23}
\end{equation}
where
\begin{equation}
K^{(j_l)}:(i_1,i_3,i_4,i_5)\longrightarrow(b_1,b_3,b_4,b_5)
\label{eq:appA-src24}
\end{equation}
is a square matrix
\begin{equation}
K^{(j_l)}\in\mathbb C^{d_l^4\times d_l^4}.
\label{eq:appA-src25}
\end{equation}
Since $X$ is invertible, it is enough to prove
\begin{equation}
\det K^{(j_l)}\not\equiv0.
\label{eq:appA-src26}
\end{equation}

\subsection*{A.7. Effective bonds on the $K_4$ graph}

The two effective edges generated by the singlet contractions have the form
\begin{equation}
U_{l_{14}}\sim U_7gU_4^T,
\label{eq:appA-src27}
\end{equation}
and
\begin{equation}
U_{l_{45}}\sim U_3gU_2^T,
\label{eq:appA-src28}
\end{equation}
while the other four edges are $U_5,U_6,U_8,U_9$. All factors $g$ are invertible. Moreover, the pair of holonomies $(h_4,h_7)$ enters only $U_{l_{14}}$, and $(h_2,h_3)$ enters only $U_{l_{45}}$. Hence, after complexification, the six effective edge matrices on $K_4$ can be varied independently on an open dense subset of parameter space. This independence is required so that the endpoint states in the rank-one degeneration below can be chosen freely at every half-edge.

\subsection*{A.8. Complexification to $SL(2,\mathbb C)$}

The spin-$j_l$ representation of $SU(2)$ extends algebraically to a representation of the complexification
\begin{equation*}
SU(2)_{\mathbb C}\simeq SL(2,\mathbb C).
\end{equation*}
If
\begin{equation*}
n=2j_l,
\end{equation*}
then
\begin{equation*}
V_{j_l}\simeq\operatorname{Sym}^{n}\mathbb C^2.
\end{equation*}
The matrices $D^{(j_l)}(G)$ for $G\in SL(2,\mathbb C)$ are therefore polynomial/rational-algebraic functions of the fundamental matrix entries and, after the determinant-one constraint is used, the determinant minor becomes a regular algebraic function on $SL(2,\mathbb C)^9$. Define
\begin{equation}
\Delta_{j_l}(G)\equiv\det S^{(j_l)}(G).
\label{eq:appA-src29}
\end{equation}
Our goal is now to show
\begin{equation*}
\Delta_{j_l}\not\equiv0
\end{equation*}
on $SL(2,\mathbb C)^9$. We do not interpret the complexified matrices as physical holonomies; complexification is used only as an algebraic tool to show that the determinant polynomial is not identically zero. We then return to the physical subgroup $SU(2)^9$.

\subsection*{A.9. Rank-one degeneration lemma for the spin-$j_l$ representation}

\textbf{Lemma 1 (Rank-one degeneration).} For every $j_l\in\frac12\mathbb N$, there exists a one-parameter family $G(t)\in SL(2,\mathbb C)$ such that, after multiplication by an appropriate scalar power, $D^{(j_l)}(G(t))$ converges to an arbitrary rank-one operator $|u\rangle\langle v|$, with $|u\rangle$ and $\langle v|$ lying on the coherent-state orbit of the highest-weight vector.

\textit{Proof.} Take
\begin{equation}
A(t)=
\begin{pmatrix}
t^{-1}&0\\
0&t
\end{pmatrix}
\in SL(2,\mathbb C),
\qquad t\neq0.
\label{eq:appA-src30}
\end{equation}
In the spin-$j_l$ representation,
\begin{equation}
D^{(j_l)}(A(t))|j_l,m\rangle=t^{-2m}|j_l,m\rangle.
\label{eq:appA-src31}
\end{equation}
Since $n=2j_l$,
\begin{equation}
t^{2j_l}D^{(j_l)}(A(t))
=
\sum_{m=-j_l}^{j_l}t^{2(j_l-m)}|j_l,m\rangle\langle j_l,m|.
\label{eq:appA-src32}
\end{equation}
As $t\rightarrow0$, all terms with $m<j_l$ vanish and only the highest-weight component remains:
\begin{equation}
\lim_{t\to0}t^{2j_l}D^{(j_l)}(A(t))=|j_l,j_l\rangle\langle j_l,j_l|.
\label{eq:appA-src33}
\end{equation}
Now take
\begin{equation*}
G(t)=LA(t)R^{-1},
\qquad L,R\in SL(2,\mathbb C).
\end{equation*}
Since the representation is a homomorphism,
\begin{equation}
t^{2j_l}D^{(j_l)}(G(t))
=
D^{(j_l)}(L)\left[t^{2j_l}D^{(j_l)}(A(t))\right]D^{(j_l)}(R^{-1})
\longrightarrow |u\rangle\langle v|,
\label{eq:appA-src34}
\end{equation}
with
\begin{equation*}
|u\rangle=D^{(j_l)}(L)|j_l,j_l\rangle,
\qquad
\langle v|=\langle j_l,j_l|D^{(j_l)}(R^{-1}).
\end{equation*}
Multiplication by the invariant metric $g$ only gives an additional invertible linear transformation, so the rank-one character is preserved. Thus, the leading edge operator can be made of the form
\begin{equation}
|u\rangle\langle v|.
\label{eq:appA-src35}
\end{equation}

\textbf{Remark 1.} This lemma does not state that an arbitrary vector in $V_{j_l}$ is a spin coherent state. What is needed in the proof is only one choice of endpoint states sufficient to make the local vertex matrices invertible. We will choose the endpoint states explicitly as $|j_l,j_l\rangle$, $|j_l,-j_l\rangle$, and one coherent state $|\xi\rangle$ with all magnetic components nonzero.

\subsection*{A.10. Local triangular-intertwiner lemma}

At every vertex of $K_4$, there is one open boundary index $b$, one intertwiner index $i$, and three internal legs. Choose the three incoming internal endpoint states as
\begin{equation*}
|j_l,j_l\rangle,\qquad |j_l,-j_l\rangle,\qquad |\xi\rangle.
\end{equation*}
Define the local matrix
\begin{equation}
A^{(j_l)}_{bi}
=
\sum_{m_1m_2m_3}
\iota_i^{b,m_1,m_2,m_3}
\delta_{m_1,j_l}\delta_{m_2,-j_l}\xi_{m_3}.
\label{eq:appA-src36}
\end{equation}
Thus,
\begin{equation}
A^{(j_l)}_{bi}
=
\sum_{m_3}\iota_i^{b,j_l,-j_l,m_3}\xi_{m_3}.
\label{eq:appA-src37}
\end{equation}

\textbf{Lemma 2 (Local invertibility).} There exists a coherent state $|\xi\rangle$ such that $A^{(j_l)}$ is invertible for every spin $j_l$.

\textit{Proof.} Substitution of the intertwiner basis gives
\begin{equation}
A^{(j_l)}_{bi}
=
\sum_{M,m_3}
\frac{(-1)^{i-M}}{\sqrt{2i+1}}
C^{iM}_{j_lb,j_lj_l}
C^{i,-M}_{j_l,-j_l,j_lm_3}
\xi_{m_3}.
\label{eq:appA-src38}
\end{equation}
The selection rule of the first Clebsch--Gordan coefficient requires
\begin{equation}
M=b+j_l.
\label{eq:appA-src39}
\end{equation}
The selection rule of the second coefficient requires
\begin{equation*}
-j_l+m_3=-M,
\end{equation*}
so that
\begin{equation}
m_3=-b.
\label{eq:appA-src40}
\end{equation}
All sums therefore collapse, and
\begin{equation}
A^{(j_l)}_{bi}
=
\frac{(-1)^{i-(b+j_l)}}{\sqrt{2i+1}}
C^{i,b+j_l}_{j_lb,j_lj_l}
C^{i,-(b+j_l)}_{j_l,-j_l,j_l,-b}
\xi_{-b}.
\label{eq:appA-src41}
\end{equation}
Define the integer
\begin{equation*}
r=b+j_l\in\{0,1,\ldots,2j_l\}.
\end{equation*}
Since $|M|\leq i$ is required for a nonzero Clebsch--Gordan coefficient and $M=r\geq0$,
\begin{equation}
A_{ri}=0\quad\text{for }i<r.
\label{eq:appA-src42}
\end{equation}
Hence $A$ is upper triangular when the rows and columns are ordered as $0,1,\ldots,n$, with $n=2j_l$. On the diagonal $i=r$, the required Clebsch--Gordan coefficients are
\begin{equation*}
C^{r,r}_{j_l,r-j_l;\,j_l,j_l}
\end{equation*}
and the symmetry-related coefficient
\begin{equation*}
C^{r,-r}_{j_l,-j_l;\,j_l,j_l-r}.
\end{equation*}
From the explicit Racah formula,
\begin{equation}
\left|C^{r,r}_{j_l,r-j_l;\,j_l,j_l}\right|^2
=
\frac{n!(2r+1)!}{r!(n+r+1)!},
\label{eq:appA-src43}
\end{equation}
and
\begin{equation}
\left|C^{r,-r}_{j_l,-j_l;\,j_l,j_l-r}\right|^2
=
\frac{n!(2r+1)!}{r!(n+r+1)!}.
\label{eq:appA-src44}
\end{equation}
All factorials are positive for $0\leq r\leq n$, so both Clebsch--Gordan coefficients never vanish. Choose now a spin coherent state along the $x$ axis,
\begin{equation}
|\xi\rangle=e^{-i\frac{\pi}{2}J_y}|j_l,j_l\rangle.
\label{eq:appA-src45}
\end{equation}
Its magnetic components are
\begin{equation}
|\xi_m|=2^{-j_l}\sqrt{\binom{2j_l}{j_l+m}},
\label{eq:appA-src46}
\end{equation}
which is nonzero for every $m=-j_l,\ldots,j_l$. Hence all diagonal entries $A_{rr}$ are nonzero. Explicitly,
\begin{equation}
|A_{rr}|
=
\frac{1}{\sqrt{2r+1}}
\frac{n!(2r+1)!}{r!(n+r+1)!}
|\xi_{j_l-r}|>0.
\label{eq:appA-src47}
\end{equation}
Since $A$ is triangular,
\begin{equation}
\det A^{(j_l)}=\prod_{r=0}^{n}A_{rr}\neq0.
\label{eq:appA-src48}
\end{equation}

\subsection*{A.11. Application to the four $K_4$ vertices}

At every vertex $v\in\{1,3,4,5\}$, use the rank-one degeneration on the three incident edges to choose endpoint states that, after an appropriate ordering of the internal legs, are $|j_l,j_l\rangle$, $|j_l,-j_l\rangle$, and $|\xi\rangle$. Since the left and right endpoint transformations on every complexified edge can be chosen independently through the matrices $L$ and $R$, these choices can be made consistently across all six edges of $K_4$.

Every edge carries the leading scaling $t^{-2j_l}$. Since $K_4$ has six edges, one matrix element of $K^{(j_l)}(t)$ has leading scaling $t^{-12j_l}$. Therefore,
\begin{equation}
\lim_{t\to0}t^{12j_l}K^{(j_l)}(t)
=
A_1^{(j_l)}\otimes A_3^{(j_l)}\otimes A_4^{(j_l)}\otimes A_5^{(j_l)},
\label{eq:appA-src49}
\end{equation}
up to a nonzero scalar and row/column permutations. The determinant of a tensor product satisfies
\begin{equation}
\det(A_1\otimes A_3\otimes A_4\otimes A_5)
=
\prod_{v\in\{1,3,4,5\}}(\det A_v)^{d_l^3}.
\label{eq:appA-src50}
\end{equation}
Since every $\det A_v\neq0$,
\begin{equation}
\det(A_1\otimes A_3\otimes A_4\otimes A_5)\neq0.
\label{eq:appA-src51}
\end{equation}
Therefore, the leading Laurent coefficient of $\det K^{(j_l)}(t)$ is nonzero. Hence the determinant of the $K_4$ network does not vanish identically as an algebraic function:
\begin{equation}
\det K^{(j_l)}\not\equiv0.
\label{eq:appA-src52}
\end{equation}

\subsection*{A.12. Return to the maximal minor $S^{(j_l)}$}

Since
\begin{equation*}
S^{(j_l)}\simeq X\otimes K^{(j_l)},
\end{equation*}
we have
\begin{equation}
\det S^{(j_l)}
=
(\det X)^{d_l^4}(\det K^{(j_l)})^{d_l}\times(\text{nonzero scalar}).
\label{eq:appA-src53}
\end{equation}
We have proved $\det X\neq0$ and $\det K^{(j_l)}\not\equiv0$. Thus,
\begin{equation}
\Delta_{j_l}(G)\equiv\det S^{(j_l)}(G)\not\equiv0
\label{eq:appA-src54}
\end{equation}
on $SL(2,\mathbb C)^9$ for every spin $j_l$.

\subsection*{A.13. Restriction back to physical $SU(2)$ holonomies}

We now use the Zariski density of the compact real form $SU(2)$ in $SL(2,\mathbb C)$. Intuitively, if a regular function on $SL(2,\mathbb C)$ vanishes on all of $SU(2)$, analytic continuation along complexified one-parameter subgroups forces the function to vanish identically on $SL(2,\mathbb C)$. The contrapositive gives
\begin{equation}
\Delta_{j_l}\not\equiv0\text{ on }SL(2,\mathbb C)^9
\Longrightarrow
\left.\Delta_{j_l}\right|_{SU(2)^9}\not\equiv0.
\label{eq:appA-src55}
\end{equation}
Hence there exists a physical holonomy configuration
\begin{equation}
\mathbf h_l^{\star}\in SU(2)^9
\qquad\text{with}\qquad
\det S^{(j_l)}(\mathbf h_l^{\star})\neq0.
\label{eq:appA-src56}
\end{equation}

\subsection*{A.14. Genericity and Haar-almost-everywhere full row rank}

The restriction $\left.\Delta_{j_l}\right|_{SU(2)^9}$ is a nonzero real-analytic function on the connected real-analytic manifold $SU(2)^9$. The zero set of a nontrivial real-analytic function has Haar measure zero. Define
\begin{equation}
Z_{j_l}=\left\{\mathbf h_l\in SU(2)^9:\Delta_{j_l}(\mathbf h_l)=0\right\}.
\label{eq:appA-src57}
\end{equation}
Then
\begin{equation}
\mu_{\rm Haar}(Z_{j_l})=0.
\label{eq:appA-src58}
\end{equation}
Outside $Z_{j_l}$, the maximal minor $S^{(j_l)}$ is invertible, so
\begin{equation}
\operatorname{rank}M^{(j_l)}(\mathbf h_l)=d_l^5
\qquad\text{for Haar-almost every }\mathbf h_l\in SU(2)^9.
\label{eq:appA-src59}
\end{equation}
In addition to having full measure, nonvanishing of the determinant is an open condition. Since the zero set of a nonzero analytic function has no interior, the full-rank set
\begin{equation*}
\mathcal G_{j_l}=SU(2)^9\setminus Z_{j_l}
\end{equation*}
is also open and dense.

\subsection*{A.15. Theorem on generic full-row rank}

\textbf{Theorem 1 (Generic full-row rank of the six-tetrahedron Bell-holonomy transfer map).} For every
\begin{equation*}
j_l\in\frac12\mathbb N,
\qquad d_l=2j_l+1,
\end{equation*}
consider the microscopic transfer map $M^{(j_l)}(\mathbf h_l)$ defined by Eq.~\eqref{eq:appA-src20} on the six-tetrahedron graph with nine internal links in Eqs.~\eqref{eq:appA-src4}--\eqref{eq:appA-src6}. There exists an open dense, full-Haar-measure subset
\begin{equation*}
\mathcal G_{j_l}\subset SU(2)^9
\end{equation*}
such that
\begin{equation}
\operatorname{rank}M^{(j_l)}(\mathbf h_l)=d_l^5,
\qquad
\forall\,\mathbf h_l\in\mathcal G_{j_l}.
\label{eq:appA-src60}
\end{equation}
In other words, the microscopic transfer map is generically surjective onto the effective five-leg magnetic output space.

\textit{Proof.} Choose the maximal minor $S^{(j_l)}$ with $i_0=i_2=0$. Singlet factorization at vertices $0$ and $2$ reduces the minor to $X\otimes K^{(j_l)}$, with $X$ invertible and $K^{(j_l)}$ a $K_4$ tensor-network transfer matrix. Complexification to $SL(2,\mathbb C)$ allows rank-one degeneration on every effective edge. With endpoint states $|j_l,j_l\rangle$, $|j_l,-j_l\rangle$, and the coherent state $|\xi\rangle$, the local vertex matrices become triangular with all diagonal entries nonzero; hence the limiting $K_4$ transfer matrix is invertible. Therefore, $\det S^{(j_l)}$ is not the identically zero algebraic function. The Zariski density of $SU(2)$ in $SL(2,\mathbb C)$ shows that the restriction to physical holonomies is also nontrivial. The zero set of the nonzero real-analytic determinant has Haar measure zero. Since a nonzero determinant is an open condition, the full-rank set is open, dense, and of full measure.

\subsection*{A.16. One generic connection works simultaneously for all spins}

The set of spin values
\begin{equation*}
\frac12,1,\frac32,2,\ldots
\end{equation*}
is countable. For each $j_l$, $\mathcal G_{j_l}$ has full Haar measure. Define
\begin{equation}
\mathcal G_{\infty}
=
\bigcap_{j_l\in\frac12\mathbb N}\mathcal G_{j_l}.
\label{eq:appA-src61}
\end{equation}
Its complement is the countable union
\begin{equation*}
\mathcal G_{\infty}^{c}
=
\bigcup_{j_l}\mathcal G_{j_l}^{c}.
\end{equation*}
Since every $\mathcal G_{j_l}^{c}$ has measure zero,
\begin{equation*}
\mu(\mathcal G_{\infty}^{c})=0.
\end{equation*}
Thus,
\begin{equation}
\mu(\mathcal G_{\infty})=1.
\label{eq:appA-src62}
\end{equation}
Therefore, a Haar-generic fundamental $SU(2)$ connection $\mathbf h_l$ has the property
\begin{equation}
\operatorname{rank}M^{(j_l)}(\mathbf h_l)=d_l^5
\label{eq:appA-src63}
\end{equation}
simultaneously for all spin representations.

\subsection*{A.17. Gauge redundancy does not change the rank}

The graph has
\begin{equation*}
E=9,
\qquad V=6,
\end{equation*}
so the cycle rank is
\begin{equation}
\beta_1=E-V+1=4.
\label{eq:appA-src64}
\end{equation}
Local gauge transformations can be used to set the holonomies on a spanning tree to the identity, leaving four independent loop holonomies. Gauge transformations act as invertible/unitary changes of basis on the half-edge indices and therefore do not change the matrix rank. Thus the generic full-rank theorem can be projected to the physical connection-moduli description based on four independent loop holonomies, modulo residual simultaneous conjugation.

This completes the proof of Theorem 1. $\blacksquare$

\phantomsection
\section*{Appendix B: Proof of Theorem 2}
\label{app:B}
\subsection*{B.1. Reduction of full row rank to a maximal minor}

The matrix $M_{p,j_l}$ has $d^p$ rows and $d^{p+2}$ columns. Full row rank is equivalent to the existence of at least one minor of size $d^p \times d^p$ with a non-zero determinant. Choose two adjacent vertices on the same cycle, say $u_0$ and $u_1$, such that there exists a link $l^* = (u_0, u_1)$. Restrict two intertwiner labels on columns to $i_{u_0} = i_{u_1} = 0$. Before restriction, column labels are $a$ and $2n = p+1$ intertwiner labels, totaling $p+2$ elementary indices. After freezing two labels, the remaining free column labels number $1 + p + 1 - 2 = p$. Since each free index has dimension $d$, the restricted column space has dimension $d^p$, equal to the row space. Thus, we obtain a square submatrix $S_{p,j_l} \in \mathbb{C}^{d^p \times d^p}$. It suffices to prove that $\det S_{p,j_l}$ is not identically the zero function of internal holonomies.

\medskip
\noindent\textbf{Step B: Structure of the $i=0$ intertwiner}

For $i=0$, the standard Clebsch-Gordan coefficient is
\begin{equation}
C_{j m, j n}^{0 0} = \frac{(-1)^{j-m}}{\sqrt{d_{j_l}}} \delta_{n,-m}.
\end{equation}
Substitution into the recoupling representation yields
\begin{equation}
(\iota_0)_{m_1 m_2 m_3 m_4} = \frac{1}{\sqrt{d_{j_l}}} g_{m_1 m_2}^{(j)} g_{m_3 m_4}^{(j)}
\end{equation}
up to an overall phase that does not affect the rank. Choose an ordering of legs at $u_0$ such that the open magnetic leg $a$ is paired by $\iota_0$ with the shared link $l^*$. At $u_1$, choose an ordering such that the open leg $b^*$ is paired with the same link $l^*$. The contraction through $l^*$ becomes
\begin{equation}
X_{b^* a} = \frac{1}{d_{j_l}^2} \sum_{r,s} g_{b^* r}^{(j)} U_{l^*}^{(j)}(r,s) g_{a s}^{(j)}.
\end{equation}
In matrix notation,
\begin{equation}
X = \frac{1}{d_{j_l}^2} g^{(j)} U_{l^*}^{(j)} (g^{(j)})^T.
\end{equation}
Since $g^{(j)}$ and $U_{l^*}^{(j)}$ are invertible, $\det X \neq 0$. The other two legs at $u_0$ are paired by the second factor $g^{(j)}$ of $\iota_0$, and likewise for the two remaining legs at $u_1$. After eliminating $u_0$ and $u_1$, those two pairs generate two effective bonds between the remaining neighboring vertices. The reduced graph has $2n-2 = p-1$ vertices. Every remaining vertex remains degree three: the loss of one incident link due to elimination is compensated by an effective link originating from singlet pairing at the frozen vertex. Therefore, the reduced graph, which we denote as $G_{n-1}$, remains cubic. For $n=3$, $G_2$ explicitly becomes $K_4$, so this step reduces precisely to the proof of the six-tetrahedron case. The restricted matrix can be factorized as
\begin{equation}
S_{p,j_l} \simeq X \otimes K_{p,j_l}
\end{equation}
up to row/column permutations and nonzero scalar factors, with $K_{p,j_l} \in \mathbb{C}^{d^{p-1} \times d^{p-1}}$. Since $X$ is invertible, it suffices to prove that $\det K_{p,j_l}$ is not identically zero.

\subsection*{B.2. Complexification as a proof device}

To prove the nonvanishing determinant, we temporarily complexify the group variables $\mathrm{SU}(2) \to \mathrm{SL}(2,\mathbb{C})$. This step does not alter the physical model; it merely provides a larger algebraic family in which a limiting configuration can be used to show that the determinant polynomial is not identically zero. First, take
\begin{equation}
A_t = \begin{pmatrix} t^{-1} & 0 \\ 0 & t \end{pmatrix} \in \mathrm{SL}(2,\mathbb{C}).
\end{equation}
In the spin-$j$ irreducible representation,
\begin{equation}
D^{(j)}(A_t) |j,m\rangle = t^{-2m} |j,m\rangle.
\end{equation}
Therefore,
\begin{equation}
t^{2j} D^{(j)}(A_t) = \sum_{m=-j}^j t^{2(j-m)} |j,m\rangle \langle j,m|.
\end{equation}
As $t \to 0$, all terms with $m < j$ vanish, leaving
\begin{equation}
\lim_{t \to 0} t^{2j} D^{(j)}(A_t) = |j,j\rangle \langle j,j|.
\end{equation}
More generally, for
\begin{equation}
G_l(t) = L_l A_t R_l^{-1},
\end{equation}
with $L_l, R_l \in \mathrm{SL}(2,\mathbb{C})$,
\begin{equation}
\lim_{t \to 0} t^{2j} D^{(j)}(G_l(t)) = |u_l\rangle \langle v_l|,
\end{equation}
where
\begin{equation}
|u_l\rangle = D^{(j)}(L_l) |j,j\rangle, \qquad \langle v_l| = \langle j,j| D^{(j)}(R_l^{-1}).
\end{equation}
Since $L_l$ and $R_l$ can be chosen independently, the endpoint coherent-state directions at the two ends of a link can be chosen independently. Multiplication by the fixed invariant metric $g^{(j)}$ does not alter the rank-one character. The effective links formed after the elimination of $u_0, u_1$ are finite products/contractions of the original link matrices and $g^{(j)}$; by choosing one factor along each effective path in rank-one degeneration and the other factors nonsingular, the effective bonds also possess independently tunable rank-one leading terms.

\subsection*{B.3. Local matrix at each remaining tetrahedron}

At each remaining cubic vertex, choose the three endpoint states from the incident rank-one bonds to be $|j,j\rangle, |j,-j\rangle, |\xi\rangle$, with $\xi_m \neq 0$ for all $m = -j, \dots, j$. Define the local matrix
\begin{equation}
A_{b i}^{(j)} = \sum_m (\iota_i)_{b, j, -j, m} \xi_m.
\end{equation}
Substituting the recoupling formula,
\begin{equation}
A_{b i}^{(j)} = \sum_{M,m} (-1)^{i-M} \sqrt{2i+1} C_{j b, j j}^{i M} C_{j,-j; j m}^{i,-M} \xi_m.
\end{equation}
The first Clebsch-Gordan selection rule requires $M = b+j$. The second selection rule requires $-j+m = -M$, so $m = -b$. Therefore, the sums collapse to
\begin{equation}
A_{b i}^{(j)} = (-1)^{i-b-j} \sqrt{2i+1} C_{j b, j j}^{i, b+j} C_{j,-j; j,-b}^{i, -b-j} \xi_{-b}.
\end{equation}
Define the integer $r = b+j$, so $r = 0, 1, \dots, 2j$. Since the Clebsch-Gordan coefficient is nonzero only when $M \le i$, while $M = r \ge 0$, we obtain
\begin{equation}
A_{r i}^{(j)} = 0 \qquad \text{for } i < r.
\end{equation}
With ordering of rows and columns $0, 1, \dots, 2j$, the matrix $A^{(j)}$ is therefore upper triangular. On the diagonal $i = r$, the magnitude of the first coefficient is
\begin{equation}
\left| C_{j, r-j; j, j}^{r, r} \right|^2 = \frac{(2j)! (2r+1)!}{r! (2j+r+1)!} > 0,
\end{equation}
and the second coefficient has the same magnitude,
\begin{equation}
\left| C_{j,-j; j, j-r}^{r, -r} \right|^2 = \frac{(2j)! (2r+1)!}{r! (2j+r+1)!} > 0.
\end{equation}
We can choose, for instance, the coherent state
\begin{equation}
|\xi\rangle = e^{-i \pi J_y / 2} |j,j\rangle,
\end{equation}
which has components
\begin{equation}
\xi_m = 2^{-j} \sqrt{\binom{2j}{j+m}} > 0
\end{equation}
for all allowed $m$. Hence, every diagonal element $A_{rr}^{(j)}$ is nonzero and
\begin{equation}
\det A^{(j)} = \prod_{r=0}^{2j} A_{rr}^{(j)} \neq 0.
\end{equation}

\subsection*{B.4. Determinant of the reduced network}

The reduced cubic graph has $p-1 = 2n-2$ vertices. Since the graph is cubic, the number of reduced links is
\begin{equation}
L' = \frac{3(p-1)}{2} = 3n - 3.
\end{equation}
Each complexified link at leading order carries a factor $t^{-2j_l}$. Thus, after multiplying by $t^{2j_l L'}$ and taking $t \to 0$, the reduced transfer matrix factorizes into a tensor product of local matrices,
\begin{equation}
\lim_{t \to 0} t^{2j_l L'} K_{p,j_l}(t) = \bigotimes_{v=1}^{p-1} A_v^{(j_l)},
\end{equation}
up to a fixed nonzero scalar and permutations of tensor factors. For a tensor product of $p-1$ matrices of size $d \times d$, we have
\begin{equation}
\det \left( \bigotimes_{v=1}^{p-1} A_v \right) = \prod_{v=1}^{p-1} (\det A_v)^{d^{p-2}}.
\end{equation}
Since each $\det A_v \neq 0$,
\begin{equation}
\det \left( \bigotimes_{v=1}^{p-1} A_v \right) \neq 0.
\end{equation}
Consequently, the leading coefficient of $\det K_{p,j_l}(t)$ is non-zero, so
\begin{equation}
\det K_{p,j_l} \not\equiv 0.
\end{equation}
Since $X$ is invertible, this implies
\begin{equation}
\det S_{p,j_l} \not\equiv 0.
\end{equation}

\subsection*{B.5. Return to physical $\mathrm{SU}(2)$ holonomies}

The spin-$j$ matrix coefficients on $\mathrm{SL}(2,\mathbb{C})$ can be realized as matrix coefficients of the symmetric-power representation and are therefore regular algebraic functions of the fundamental matrix entries. Thus,
\begin{equation}
\Delta_{p,j}(G) \equiv \det S_{p,j}(G)
\end{equation}
is a regular function on $\mathrm{SL}(2,\mathbb{C})^{3n}$ and, based on the previous step, $\Delta_{p,j} \not\equiv 0$. The compact real form $\mathrm{SU}(2)^{3n}$ is Zariski dense in its complexification. Therefore, the restriction of $\Delta_{p,j}$ to the physical submanifold $\mathrm{SU}(2)^{3n}$ is also not identically zero. On the connected real-analytic manifold $\mathrm{SU}(2)^{3n}$, the zero set of a non-trivial real-analytic function has Haar measure zero and an empty interior. By defining
\begin{equation}
\mathcal{Z}_{p,j} = \{ h \in \mathrm{SU}(2)^{3n} \mid \Delta_{p,j}(h) = 0 \},
\end{equation}
we have
\begin{equation}
\mu_{\mathrm{Haar}}(\mathcal{Z}_{p,j}) = 0.
\end{equation}
The complement
\begin{equation}
\mathcal{G}_{p,j} = \{ h \in \mathrm{SU}(2)^{3n} \mid \Delta_{p,j}(h) \neq 0 \}
\end{equation}
is open, dense, and of full measure. For all $h \in \mathcal{G}_{p,j}$, the maximal minor $S_{p,j}$ is non-zero, so that
\begin{equation}
\mathrm{Rank}\, M_{p,j} = d^p.
\end{equation}
This completes the proof of Theorem 2. $\blacksquare$

\phantomsection
\section*{Appendix C: Proof of Theorem 3}
\label{app:C}
The matrix element of $V_\Gamma^\dagger V_\Gamma$ is decomposed as follows:
\begin{equation}
(V_\Gamma^\dagger V_\Gamma)_{\mu'_1 \dots \mu'_N; \mu_1 \dots \mu_N} = \sum_{\mathcal{L}_L} V_\Gamma^*(\mathcal{L}_L; \mu'_1 \dots \mu'_N) V_\Gamma(\mathcal{L}_L; \mu_1 \dots \mu_N).
\end{equation}
Substituting the radial decomposition $(\widetilde{K})$, the sum over the outer boundary indices $\mathcal{L}_L$ acts only on the pair $(\widetilde{K}^{(L)})^\dagger \widetilde{K}^{(L)}$. Using the local shell isometry,
\begin{equation}
\sum_{\mathcal{L}_n} (\widetilde{K}^{(n)})^\dagger(\mathcal{L}_n; x_{n-1}', \mu_n') \widetilde{K}^{(n)}(\mathcal{L}_n; x_{n-1}, \mu_n) = \delta_{x_{n-1}', x_{n-1}} \delta_{\mu_n', \mu_n}.
\end{equation}
The frontier delta identifies the bra and ket indices at layer $n-1$, so the next contraction becomes $(\widetilde{K}^{(n-1)})^\dagger \widetilde{K}^{(n-1)}$. Repeating this step recursively yields
\begin{equation}
(V_\Gamma^\dagger V_\Gamma)_{\mu'; \mu} = \prod_{l=1}^L \delta_{\mu_l', \mu_l}.
\end{equation}
The right-hand side is the matrix element of the identity operator on
\begin{equation}
\mathcal{H}_{\mathrm{bulk}} = \bigotimes_{f \in F_\Gamma} \mathcal{H}_{\mu_f}.
\end{equation}
Thus,
\begin{equation}
V_\Gamma^\dagger V_\Gamma = \mathbb{I}_{\mathcal{H}_{\mathrm{bulk}}}.
\end{equation}
If raw physical kernels $L_l = U_l / \sqrt{d_{j_l}}$ and unnormalized $T_f$ are used instead, the same contraction yields
\begin{equation}
(V_\Gamma^{\mathrm{raw}})^\dagger V_\Gamma^{\mathrm{raw}} = \kappa_\Gamma^2 \mathbb{I},
\end{equation}
for some non-zero scalar $\kappa_\Gamma$. Defining $V_\Gamma = \kappa_\Gamma^{-1} V_\Gamma^{\mathrm{raw}}$ restores exact isometry. 
This completes the proof of Theorem 3. $\blacksquare$

\end{document}